\documentclass[aps,prx,twocolumn,superscriptaddress]{revtex4-2} 
\usepackage{amsmath,amssymb}
\usepackage[pdftex]{graphicx}
\usepackage{xcolor}
\usepackage{bm}
\usepackage{booktabs}
\usepackage{multirow}
\usepackage{colortbl}
\usepackage{here}

\begin{document}
\title{Computation time and thermodynamic uncertainty relation of Brownian circuits}

\author{Yasuhiro Utsumi}
\address{Department of Physics Engineering, Faculty of Engineering, Mie University, Tsu, Mie, 514-8507, Japan}

\author{Yasuchika Ito}
\address{Department of Physics Engineering, Faculty of Engineering, Mie University, Tsu, Mie, 514-8507, Japan}

\author{Dimitry Golubev}
\address{QTF Centre of Excellence, Department of Applied Physics, Aalto University, P.O. Box 15100, FI-00076 Aalto, Finland}

\author{Ferdinand Peper}
\address{National Institute of Information and Communications Technology, Iwaoka-588-2 Iwaokacho, Nishi Ward, Kobe, Hyogo 651-2492, Japan}

\begin{abstract}
We analyze a token-based Brownian circuit in which Brownian particles, coined `tokens,' move randomly by exploiting thermal fluctuations, searching for a path in multi-token state space corresponding to the solution of a given problem. 
The circuit can evaluate a Boolean function with a unique solution. 
However, its computation time varies with each run. 
We numerically calculate the probability distributions of Brownian adders' computation time, given by the first-passage time, and analyze the thermodynamic uncertainty relation and the thermodynamic cost based on stochastic thermodynamics. 
The computation can be completed in finite time without environment entropy production, i.e., without wasting heat to the environment. 
The thermodynamics cost is paid through error-free output detection and the resets of computation cycles. 
The signal-to-noise ratio quantifies the computation time's predictability, and it is well estimated by the mixed bound, which is approximated by the square root of the number of token detections. 
The thermodynamic cost tends to play a minor role in token-based Brownian circuits in computation cycles. 
This contrasts with the logically reversible Brownian Turing machine, in which the entropy production increases logarithmically with the size of the state space, and thus worsens the mixed bound.
\end{abstract}

\date{\today}
\maketitle

\newcommand{\mat}[1]{\mbox{\boldmath$#1$}}

\section{Introduction}

Computers can be regarded as heat engines transforming free energy into waste heat and mathematical work~\cite{Bennett1982,Bennett1985}. 
The fundamental physical limit of computation has attracted attention for more than four decades since the infancy of present-day computers~\cite{Landauer1961,Bremermann1967,Landauer1988,Landauer1996}. 
A notable example is the thermodynamic cost of information erasure: Landauer's bound~\cite{Landauer1961} states that the minimum energy dissipation in erasing one bit of memory at temperature $T$ is $k_{\rm B}T \ln 2$, 
where $k_{\rm B}$ is the Boltzmann constant. 
On the other hand, there is no unavoidable minimal energy requirement per transmitted information unit~\cite{Landauer1988,Landauer1996}. 
From the early days, it has been speculated that there is a possible connection between thermodynamic reversibility and logical reversibility. 
In logically reversible computation, each computation step is carried out without discarding information. 
In common understanding, logically reversible computation can be performed without a thermodynamic cost if it is carried out quasi-statically~\cite{Bennett1982,Feynamn1996}.
This idea has been the driving force for in-depth studies on logical reversibility, and it has led to a fertile research field revolving around {\it reversible computing}~\cite{Bennett1973,Morita2017} in the computer science community.

Historically, the origin of the thermodynamics of computation lies in Maxwell's demon gedankenexperiment~\cite{Leff2003}, 
which shed light on the role of information in thermodynamics. 
In the last decade, the physics community has succeeded in incorporating information into the framework of thermodynamics~\cite{Sagawa2014,Parrondo2015}. 
This research field, termed information thermodynamics, has as its primary focus the conversion of information to work.
Though in the same framework, several recent works have emphasized the role played by computation~\cite{Sagawa2019,Wolpert2019,Wolpert2020,Kolchinsky2020}, based on the information thermodynamics of Bayesian nets~\cite{Ito2013,Ito2016}, resulting in a transparent description of the relations between logical reversibility and thermodynamic reversibility~\cite{Sagawa2019}.

%

A basis of information thermodynamics is stochastic thermodynamics~\cite{Seifert2012,VandenBroeck2013,VandenBroeck2015}, which introduces quantities associated with a single stochastic trajectory, such as stochastic entropy, energy, heat, and work. 
This view naturally applies to small systems where thermal fluctuations are relatively prominent. 
Moreover, thermodynamics at the trajectory level plays an essential role in discovering universal relations valid in regimes far from equilibrium, such as the fluctuation relation~\cite{Esposito2009,Campisi2011,Seifert2012,Klages2012,Pekola2015}. 
In recent decades, solid-state quantum devices such as single-electron transistors~\cite{Pekola2015} have revealed themselves as powerful experimental tools for demonstrating the fluctuation relations~\cite{Utsumi2010,Kueng2012} as well as the Jarzynski equality~\cite{Saira2012,Hofmann2016}. 
Precise feedback control techniques realize the gedankenexperiments on information-to-work conversion, Sizlard's engine, and Maxwell's demon~\cite{Koski2014,Chida2017}. 
These experimental achievements have helped the paradigm of stochastic thermodynamics to pervade the device physics community.

Recently, a new research direction has emerged on the device physics side: the fabrication of functional circuits by integrated solid-state quantum devices. 
Examples include circuits consisting of magnetic tunnel junctions that realize neuromorphic computing, such as physical reservoir computing~\cite{Torrejon2019} and probabilistic computing~\cite{Borders2019}. 
Although integrated quantum-device circuit technology is still in its infancy, it has been a motivation to analyze the stochastic thermodynamics of the whole computation process involving realistic small-scale integrated quantum-device circuits. 

%
%

A standard theoretical model of the thermodynamics of computation is the Brownian computer~\cite{Bennett1982}. 
It utilizes random thermal transitions for searching through a labyrinth of the configuration space isomorphic to the desired computation. 
A single-stranded RNA and an RNA polymerase hypothetically consist of a Brownian Turing machine: the former and latter are interpreted as a ``tape" and a ``tape head'' respectively~\cite{Bennett1982}. 
In the present paper, we discuss a specific computation model of a Brownian computer, the so-called \emph{token-based Brownian circuit} proposed in computer science~\cite{Lee2010,Peper2013,Lee2016}. 
In this type of circuit, Brownian particles, i.e., tokens subject to thermal fluctuations, perform a random search in multi-token state space. 
Brownian circuits can be constructed from three primitive circuit elements: the Hub, the Conservative Join (CJoin), and the Ratchet~\cite{Peper2013,Lee2016}. 
The circuits use so-called \emph{dual-rail encoding} to represent binary information, and they have been proven to be universal if stochastic backtracking is not disabled. 
Although the token-based Brownian circuit was introduced as a theoretical model for asynchronous computation, several efforts have been conducted to realize them by integrated quantum devices. 
For example, primitive circuit elements of Brownian circuits implemented by single-electron-tunneling circuits have been proposed~\cite{Safiruddin2008,Agbo2009}. 
In another technology, one of the circuit elements, the Hub, was experimentally realized in a ferromagnetic thin film in which magnetic skyrmions represent tokens~\cite{Jibiki2020}. 
Theoretical study for the actual implementation of skyrmion token-based Brownian circuits is also in progress~\cite{Brems2021}. 

%
%

Brownian circuits are asynchronous, and the computation time varies with each run. 
Regardless of its intrinsic randomness necessary for stochastic backtracking, a token-based Brownian circuit always outputs the intended solution. 
A natural definition of the computation time $\tau$ would be the first-passage time~\cite{Redner2001,Saito2017,Singh2019}, which corresponds to the duration of time to reach a final state (output state) for the first time starting from an initial state (input state). 
The computation time can be shortened by increasing the intrinsic operating frequency of each circuit element. 
In other words, the computation time is system parameter dependent. 
A proper figure of merit of a Brownian computer is the signal-to-noise ratio (SNR): 
\begin{align}
\frac{S}{N}= \frac{ \langle \! \langle \tau \rangle \! \rangle_F }{ \sqrt{ \langle \! \langle \tau^2 \rangle \! \rangle_F } }  \, ,
\label{eqn:snr}
\end{align}
where $\langle \! \langle \tau \rangle \! \rangle_F$ and $\langle \! \langle \tau^2 \rangle \! \rangle_F$, respectively, are the average and the variance of the computation time. 
The subscript $F$ indicates the first-passage ensemble. 
The computation time is deterministic in the limit of $S/N \to \infty$. 
On the other hand, the computation time is unpredictable in the limit of $S/N \to 0$.

%
%

To analyze the computation time based on an analogy between a computer and a thermodynamic engine, we need to go beyond traditional thermodynamics. 
A heat engine achieves its maximum efficiency, the Carnot limit, when it operates quasi-statically, i.e., infinitely slow. 
Recently, a trade-off between efficiency and power consumption has been discovered~\cite{Shiraishi2016}. 
Since a finite power consumption implies a finite operating time, this trade-off relation would apply to our problem. 
In the last few years, it has been recognized that there exists a family of trade-off relations applicable beyond a quasi-static regime, which is exemplified by the thermodynamic uncertainty relation (TUR)~\cite{Horowitz2020,Gingrich2016,Shiraishi2016,Garrahan2017,Gingrich2017,Shiraishi2021,Hiura2021,Pal2021,Pal2021_1,Wolpert2020_2}. 
Several earlier works~\cite{Gingrich2016,Garrahan2017,Gingrich2017} are based on a stochastic trajectory ensemble. 
In practice, the TUR for the first-passage time~\cite{Garrahan2017,Gingrich2017} and the TUR with unidirectional transitions~\cite{Pal2021,Pal2021_1} have provided an upper bound of the SNR in Eq.~(\ref{eqn:snr}) [see Eqs.~(\ref{eqn:tur_activity}) and (\ref{eqn:tur_mix})]. 
In the present paper, we use these results to estimate the performance of Brownian computers.

%
%

The thermodynamic cost of a Brownian computer has been a matter of argument~\cite{Norton2013}. 
Brownian computation is argued to be thermodynamically irreversible from the analogy of one `Brownian particle' gas expanding irreversibly into a box, which is the analog of the state space $\Omega$ of a computation model. 
The cost is the system entropy production due to the Brownian particle exploring the state space $k_{\rm B} \ln |\Omega|$~\cite{Norton2013}, where $|\Omega|$ is the size of the state space~\cite{set_theory_1}. 
The entropy $k_{\rm B} \ln |\Omega|$ emerges since the Brownian particle position is uncertain. 
The cost is the amount of external work necessary to reset to the initial zero entropy state in a computation cycle. 
This idea has been analyzed further for stochastic Turing machines (TMs)~\cite{Strasberg2015}. 
However, there remain ambiguities since, at the end of a computation, the Brownian particle is at a unique position corresponding to a solution: what is uncertain is the time to reach this unique position. 
In the present paper, we analyze the cost in the framework of the stochastic thermodynamics of resetting~\cite{Fuchs2016}, which applies to the problem of the first-passage time and to its extension of first-passage resetting~\cite{DeBruyne2020}. 

%
%

There are several recent works relevant to this research. 
Stochastic thermodynamics of lumped circuits~\cite{Freitas2020,Freitas2021} has been used to specify the information-bearing degrees of freedom in terms of high and low voltage states. 
In SET technology, the fundamental energy limits of Brownian circuits have been analyzed~\cite{Ercan2018,Ercan2021} based on physical information theory~\cite{Ercan2013}. 
In contrast with these works, we do not discuss the practical implementation of circuit elements: we rather focus on the whole computation process of small integrated circuits described by large-scale stochastic Petri nets (SPN)~\cite{Murata1989}. 
Technically, we adopt the full-counting statistics (FCS) of a classical master equation~\cite{Bagrets2003} with a large and sparse transition rate matrix to perform the stochastic thermodynamics analysis.

%
%

The structure of this paper is as follows. 
Section~\ref{sec:FPTTUR} introduces our theoretical framework, the FCS of resetting, and summarizes the TUR and the thermodynamic cost of resetting. 
We also apply them to an example of a logically reversible Brownian Turing machine. 
Section~\ref{sec:tbbc} explains token-based Brownian circuits and their transition rate matrix. 
In Sec.~\ref{sec:cal_tim}, we present numerical results on the computation time of binary adders. 
In Sec.~\ref{sec:summary}, we summarize our results after the discussion in Sec.~\ref{sec:discussion}.

\section{computation time, TUR and cost}
\label{sec:FPTTUR}

The physical limit of computation speed is the primary interest of the thermodynamics of computation. 
Already around the 1980s, it has been commonly known that the following trade-off relations exist between energy cost versus speed~\cite{Feynamn_3eqs}:
\begin{align}
k_{\rm B} T \log r = U_1-U_2 \, , \label{eqn:feynmann1} \\
k_{\rm B} T \log r = k_{\rm B} T (\log D_2 -\log D_1) = (S_2-S_1) T \, , \label{eqn:feynmann2} \\
{energy \; loss \; per \; step} \approx  k_{\rm B} T \frac{\gamma^+ - \gamma^-}{(\gamma^+ + \gamma^-)/2} \, .
\label{eqn:feynmann3}
\end{align}
Here, $r=\gamma^+/\gamma^-$ is the ratio between the forward and backward transition rates, $U_{1(2)}$ is the energy of the initial (final) state, $D_{1(2)}$ is the number of available states, and $S_{1(2)}=k_{\rm B} \ln D_{1(2)}$ is the entropy.

The first two relations, (\ref{eqn:feynmann1}) and (\ref{eqn:feynmann2}), are unified as $k_{\rm B} T \log r = F_1-F_2$ by using the free energy $F=U-T S $. 
It is the \emph{detailed balance relation} or, stated in modern language, the \emph{fluctuation relation}~\cite{Esposito2009,Campisi2011,Seifert2012,Klages2012,Pekola2015}. 
In this section, we summarize stochastic thermodynamics of computation. 
Using a logically reversible TM (RTM) as an example, we demonstrate that the third relation (\ref{eqn:feynmann3}) can be regarded as the TUR. 
New results are given in Sec.~\ref{sec:Finite_RTM}.
In the following, we set $k_{\rm B}=1$.

\subsection{Logically reversible Brownian Turing machine}

\begin{figure}[ht]
\begin{center}
\includegraphics[width=0.5 \columnwidth]{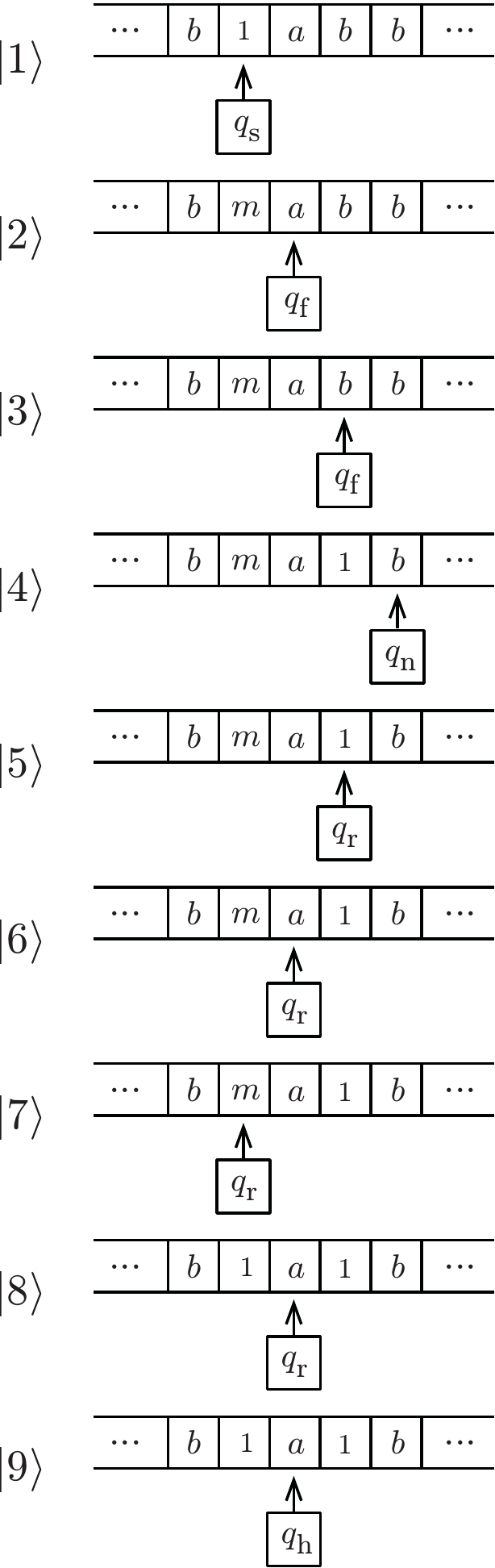}
\caption{
Step-by-step snapshots of computational configurations of a logically reversible Turing machine, $T_{\rm copy}$. 
It copies a unary number at the left of the delimiter $a$ to the right of the delimiter~\cite{Morita2017}. 
$b$ represents a blank symbol. 
}
\label{fig:RTMTcopy}
\end{center}
\end{figure}

As an example, we take the Turing machine~\cite{Bennett1973, Morita2017}, which is the basic model of computation. 
The precise definition is given in Appendix \ref{sec:RTM}. 
The TM consists of a read/write head (arrow) plus a finite control (box) with a finite number of internal states and a two-way infinite tape (Fig.~\ref{fig:RTMTcopy}). 
The tape consists of an infinite sequence of tape cells. 
The TM operates according to prescribed rules expressed as the set of, e.g., quintuples~(\ref{eqn:quintuple}). 
In each step, there is only one applicable rule. 
Accordingly, the TM reads and overwrites a symbol on a cell at the tape head, changes the internal state $q$ of the finite control, and shifts the position of the tape head. 
Figure \ref{fig:RTMTcopy} shows an example of step-by-step snapshots of the computational configurations of a TM, $T_{\rm copy}$, whose program is given by the set~(\ref{eqn:T_copy})~\cite{Morita2017}. 
Initially, the finite control is in the start state $q_{\rm s}$. 
At the end of the computation, the internal state is in the halt state $q_{\rm h}$. 
The start and end of the computation are identified only by the internal state of the finite control. 
The input, reflected by the initial tape content, can be arbitrary. 
The computational configuration specifies the `many-body state' of the composite system consisting of the tape-head and the tape. 
In Fig.~\ref{fig:RTMTcopy}, the computational configuration evolves as $|1 \rangle \to |2 \rangle \to \cdots \to |9 \rangle$.

The RTM is a standard model for analyzing the connection between logical reversibility and thermodynamic reversibility~\cite{Bennett1973, Morita2017}. 
In an RTM, there is a one-to-one correspondence between the present computational configuration and the next computational configuration obtained after employing a unique applicable rule. 
In other words, any possible computational configuration, except for the start state, possesses a unique predecessor. 
The TM in~\cite{Morita2017}, $T_{\rm copy}$, is an example of an RTM. 
The RTM possesses the inverse RTM $T_{\rm copy}^{-1}$ undoing the computation performed by the RTM $T_{\rm copy}$. 
The precise definition of an RTM is explained in Appendix \ref{sec:RTM}.

The dynamics of the RTM are implemented by introducing stochastic transitions between computational configurations. 
The stochastic transitions are formally defined by a continuous-time Markov chain, or the master equation~\cite{Strasberg2015}. 
In the bra-ket notation~\cite{Bagrets2003} it reads as follows:
\begin{align}
\frac{d}{dt} | p(t) \rangle = \hat{L}^{\rm f} | p(t) \rangle \, . 
\end{align}
Each state of the master equation corresponds to a computational configuration. 
The state space, i.e., the set of computational configurations, is 
$ \Omega \cup \{ |{\rm f}\rangle \}$, 
where we separate the singleton of the final state $\{ |{\rm f}\rangle \}$ from the set of the other states $ \Omega = \{ |1 \rangle, \cdots, ||\Omega| \rangle \}$. 
The initial state is included in the state space $| {\rm i} \rangle \in \Omega$. 
The distribution probability to find the $n$th state is $\langle n|p(t) \rangle = p_n(t)$, ($n=1,\cdots, |\Omega|, {\rm f}$). 
The matrix element of the transition rate matrix is, 
\begin{align}
\langle n' | \hat{L}^{\rm f} | n \rangle &= L^{\rm f}_{n',n}  = \left \{ \begin{array}{cc} \Gamma^{\rm f}_{n',n} & (n' \neq n) \\ - \sum_{n'' (\neq n)} \Gamma^{\rm f}_{n'',n} & (n' = n) \end{array} \right. \, . \label{eqn:full_Liouvillian}
\end{align}
We assume that there is no direct transition from the initial state to the final state, i.e.,  $\Gamma^{\rm f}_{{\rm i},{\rm f}} =0$. 

\begin{figure}[ht]
\begin{center}
\includegraphics[width=0.9 \columnwidth]{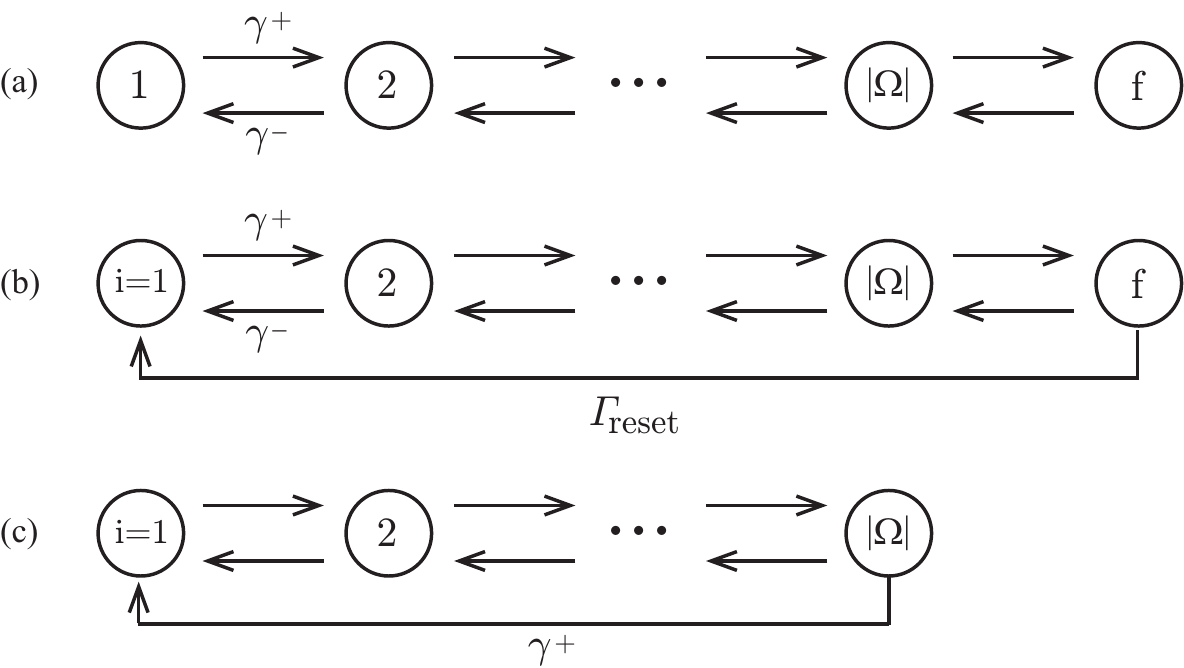}
\caption{
(a) Transition state diagram for a logically reversible Turing machine. 
Except for the start state $|{\rm i} \rangle = |1 \rangle$ and the halting state $| {\rm f} \rangle$, each state posses a unique predecessor and a unique successor. 
The states constitute a finite one-dimensional chain. 
(b) Transition state diagram for the RTM with resetting: 
if the final state $| {\rm  f} \rangle$ is reached, the system is reset to the initial state $| {\rm i} \rangle$ with a resetting rate $\Gamma_{\rm reset}$. 
(c) Transition state diagram for the RTM with error-free resetting, $\Gamma_{\rm reset} \to \infty$. 
It is derived from (b) as follows.
The directed edges with tail $|{\rm f} \rangle$ in (b) are removed. 
The directed edge with tail $\left| |\Omega| \right \rangle$ and head $|{\rm f} \rangle$ in (b) is transformed to the directed edge with tail $\left| |\Omega| \right \rangle$ and head $|{\rm i} \rangle = |1 \rangle$. 
}
\label{fig:ctmarkovchain_RTM}
\end{center}
\end{figure}

Figure \ref{fig:ctmarkovchain_RTM} (a) shows the transition state diagram of the RTM. 
Here, the transition state diagram is a directed graph~\cite{graph_theory_1} without self-loop: each node represents a state $|n \rangle \in \Omega$ (In the sequel, we often write $|n \rangle$ as $n$, when the meaning is clear from the context). 
A directed edge, which we often denote as an ordered pair $(|n' \rangle \leftarrow |n \rangle) \in E$ of two states $|n \rangle , \, |n' \rangle \in \Omega$, represents a transition with non-zero transition rate $\Gamma^{\rm f}_{n',n} > 0 \, \wedge \, n' \neq n$. 
We assume the transition rates associated with all rules, i.e., all quintuples of the set (\ref{eqn:T_copy}), are the same; the forward and backward transition rates are $\gamma^+$ and $\gamma^-$, respectively. 
Then, the transition rate matrix of the RTM is, 
\begin{align}
\hat{L}^{\rm f}(\xi,\eta) =& \sum_{n=2}^{|\Omega|} \biggl( - (\gamma^+ + \gamma^-) |n \rangle \langle n| +\gamma^+ e^{i \Delta \Sigma_{\rm bi}^{\rm env} \xi + i \eta} \nonumber \\
& \times |n \rangle \langle n-1| + \gamma^- e^{-i \Delta \Sigma_{\rm bi}^{\rm env} \xi + i \eta} |n \rangle \langle n+1| \biggl)
\nonumber \\
& - \gamma^+ |1 \rangle \langle 1 | + \gamma^- e^{-i \Delta \Sigma_{\rm bi}^{\rm env} \xi + i \eta} |1 \rangle \langle 2 | 
\nonumber \\
& - \gamma^- |{\rm f} \rangle \langle {\rm f} | + \gamma^+ e^{i \Delta \Sigma_{\rm bi}^{\rm env} \xi + i \eta} |{\rm f} \rangle \langle |\Omega|| \, , \label{eqn:revTM_liouvill}
\end{align}
where, $|{\rm i} \rangle=|1 \rangle$, $|{\rm f} \rangle=||\Omega|+1 \rangle$ and $\langle m | n \rangle = \delta_{m,n}$. 
The environment entropy production accompanies a stochastic transition induced by thermal fluctuation: 
\begin{align}
\Delta \Sigma_{\rm bi}^{\rm env} = \ln  \frac{\gamma^+}{\gamma^-} \, . \label{eqn:env_ent}
\end{align}
This is the detailed balance relation corresponding to Eq.~(\ref{eqn:feynmann2}). 
We call Eq.~(\ref{eqn:revTM_liouvill}) the \emph{modified} transition rate matrix since it is modified with the counting fields $\xi$ and $\eta$ to count the amount of environment entropy production and the number of transitions~\cite{Bagrets2003}, i.e. the activity. 
The standard form is recovered for $\xi=\eta=0$.

In general, the stochastic dynamics of a Brownian RTM is equivalent to a discrete random walk on a finite one-dimensional chain~\cite{Strasberg2015}. 
In addition, any one-tape TM can be transformed into an RTM, such as the three-tape RTM in Refs.~\cite{Bennett1973,Morita2017,Strasberg2015}, whose Brownian version is also a discrete random walk on a finite one-dimensional chain~\cite{Strasberg2015}.

\subsection{Thermodynamic uncertainty relation and the stochastic thermodynamics of resetting}

\subsubsection{Resetting as a periodic boundary condition}

The natural definition of the computation time is the first-passage time~\cite{Redner2001}, which is the time taken for the system to reach the final state for the first time starting from a given initial state. 
For the theoretical analysis, it is convenient to use the concept of resetting: if the final state $| {\rm  f} \rangle$ is reached, the system is reset to the initial state $| {\rm i} \rangle$ with a resetting rate $\Gamma_{\rm reset}$, see Fig.~\ref{fig:ctmarkovchain_RTM} (b). 
The transition rate matrix is modified as $\hat{L}^{\rm f} \to \hat{L}^{\rm f}+\hat{V}$, where the unidirectional reset is described by, 
\begin{align}
\hat{V} = \Gamma_{\rm reset} (|{\rm i} \rangle \langle {\rm f}| - |{\rm f} \rangle \langle {\rm f}|) \, . \label{eqn:reset_ham}
\end{align}
We can formulate the first-passage time of $W$ resets with this extension. 

The reset hypothetically describes a computation cycle. 
Although the reset would be realized by externally controlling the system, we do not discuss how to implement it in practice. 
The reset imposes a periodic boundary condition, which hypothetically realizes a steady-state, and enjoys the following merits in the theoretical analysis: 
(i) The thermodynamic cost of resets in the steady-state can be estimated in the framework of the stochastic thermodynamics of resetting~\cite{Fuchs2016}. 
(ii) The number of resets $W$ is time-extensive observable under the periodic boundary condition. 
Then, TURs for the first-passage time when a time-extensive observable first reaches a certain threshold value~\cite{Garrahan2017,Gingrich2017} are available. 
In addition, there exist simple relations between the fluctuations of the number of resets over trajectories with fixed measurement-time and the fluctuations of first-passage time over trajectories with a fixed number of resets~\cite{Garrahan2017,Gingrich2017}. 
For the analysis of fluctuations with fixed measurement-time, one can utilize systematic approaches developed in the theory of FCS~\cite{Bagrets2003,FlindtPRB2010}.

In the present paper, we focus on error-free detection, i.e., the resetting rate is taken to be $\Gamma_{\rm reset} \to \infty$. 
In this limit, the final state is identified with the initial state, $|{\rm f} \rangle = |{\rm i} \rangle$. 
In other words, the final state is removed, and the state space is reduced to $\Omega$. 
Consequently, the off-diagonal elements of the transition rate matrix are modified as, 
\begin{align}
\Gamma_{i,j} = \left \{ \begin{array}{cc} \Gamma^{\rm f}_{i,j} & (i, j \in \Omega \, \wedge \, i \neq j ) \\  \Gamma^{\rm f}_{ {\rm i} , j} + \Gamma^{\rm f}_{ {\rm f} , j} & (i={\rm i} \, \wedge \, j \in \Omega) \end{array} \right.  \, . 
\end{align}
The diagonal component should be modified to satisfy probability conservation. 

In the transition state diagram, the above modifications correspond to (i) the removal of all directed edges with tail $|{\rm f} \rangle$, (ii) the replacement of each direct edge whose head node is $|{\rm f} \rangle$ with a directed edge whose head node is $|{\rm i} \rangle$ while keeping the tail node unchanged, and (iii) the removal of the node $|{\rm f} \rangle$. 
Consequently, the transition state diagram of the RTM in Fig.~\ref{fig:ctmarkovchain_RTM} (b) is modified to Fig.~\ref{fig:ctmarkovchain_RTM} (c). 
An accurate and general description in a graph-theoretical representation is as follows.
In the transition state diagram, directed edges with tail $|{\rm f}\rangle$, 
\begin{align}
\delta^+ {\rm f} = \left \{  ( |i \rangle \leftarrow |{\rm f} \rangle) \middle | \Gamma^{\rm f}_{ i, {\rm f}} >0  \wedge |i \rangle \in \Omega  \right \} \, , 
\end{align}
are removed and nodes being tails of directed edges with head $|{\rm f}\rangle$, 
\begin{align}
{\it \Gamma}^- {\rm f} = \left \{  |j \rangle \in \Omega \middle | \Gamma^{\rm f}_{ {\rm f},j} >0 \right \} \, , 
\end{align}
transform into nodes being tails of directed edges with head $|{\rm i} \rangle$.

\subsubsection{Kinetic and mixed bounds}

Any transition process from $|j \rangle$ to $|i \rangle$ is either unidirectional or bidirectional. 
For a token-based Brownian circuit, unidirectional processes occur when tokens are detected at output terminals.
That is, when the final state is reached after the detection of the last token, it is reset to the initial state. 
In graph-theoretical representation~\cite{graph_theory_1}, a transition state diagram is a directed graph comprising the set of nodes $\Omega$ corresponding to the states, as well as the set of directed edges $E$ corresponding to the possible transitions. 
Let the sets of directed edges for unidirectional and bidirectional transitions be $E_{\rm uni}$ and $E_{\rm bi}$, respectively:  
\begin{align}
E_{\rm uni} =& \left \{  (i \leftarrow j) \middle | \Gamma_{i,j} >0 \, \wedge \,  \Gamma_{j,i} = 0 \wedge i,j \in \Omega \right \} \, , \label{eqn:E_uni} \\
E_{\rm bi} =& \left \{ (i \leftarrow j) \middle | \Gamma_{i,j} > 0 \, \wedge \, \Gamma_{j,i} >0 \wedge i,j \in \Omega  \right \} \, . \label{eqn:E_bi} 
\end{align}
They are mutually disjoint $E_{\rm uni} \cap E_{\rm bi} = \emptyset$ and $E = E_{\rm uni} \cup E_{\rm bi}$. 

In stochastic thermodynamics, an observable $O$ is defined for each stochastic trajectory~\cite{Seifert2012,VandenBroeck2015}. 
We write the average of the stochastic observable $O$ for a fixed number of resets $W$ reached for the first time as $\langle \! \langle O \rangle \! \rangle_F $ and the average for a fixed measurement time $\tau$ as $\langle \! \langle O \rangle \! \rangle$. 
As we explain in Sec. \ref{sec:fcs}, in the limit of a long measurement time $\tau$, the two ensemble averages are related as 
$\langle \! \langle O \rangle \! \rangle_F \approx \langle \! \langle o \rangle \! \rangle \langle \! \langle \tau \rangle \! \rangle_F$. 
We use a small letter to represent the rate of the observable 
$\langle \! \langle O \rangle \! \rangle/\tau = \langle \! \langle o \rangle \! \rangle$. 
The average first-passage time for $W$ resets $\langle \! \langle \tau \rangle \! \rangle_F$ is connected to the `reset current' $\langle \! \langle w \rangle \! \rangle$ as, 
\begin{align}
\langle \! \langle w \rangle \! \rangle \langle \! \langle \tau \rangle \! \rangle_F  \approx W \, . \label{eqn:c_t} 
\end{align}

We discuss two kinds of TURs, which bound the SNR of the first-passage time from above~\cite{Garrahan2017,Hiura2021, Pal2021}. 
The first TUR is the kinetic bound~\cite{Garrahan2017,Hiura2021}, 
\begin{align}
r_A =& \frac{S}{N} \frac{1}{ \sqrt{ \langle \! \langle A \rangle \! \rangle_F } } \leq 1 \label{eqn:tur_activity} \, . 
\end{align}
Here, the activity $A$ is the total number of transitions. 
The activity rate is
\begin{align}
\langle \! \langle a \rangle \! \rangle = \sum_{ i \neq j \, \wedge \, i,j \in \Omega} \Gamma_{i,j} p_{j}^{\rm st} \, , \label{eqn:act_rat}
\end{align}
whereby $p_{j}^{\rm st}$ is the steady-state distribution probability satisfying $\sum_{j \in \Omega} \Gamma_{i,j} p_{j}^{\rm st} = 0$. 
The second TUR is the mixed bound, expressed as the total entropy production of the bidirectional processes and the activity of the unidirectional processes~\cite{Pal2021} (Appendix \ref{sec:tur_oneway}): 
\begin{align}
r_\Sigma =& \frac{S}{N} \frac{1}{ \sqrt{ \langle \! \langle \Sigma^{{\rm tot}}_{{\rm bi}} \rangle \! \rangle_F/2 + \langle \! \langle A_{{\rm uni}} \rangle \! \rangle_F } } \leq 1 \, . \label{eqn:tur_mix} 
\end{align}
The activity rate is taken over the unidirectional transitions as, 
\begin{align}
\langle \! \langle a_{{\rm uni}} \rangle \! \rangle = \sum_{ (i \leftarrow j) \in E_{{\rm uni}}} \Gamma_{i,j} p_{j}^{\rm st}  \, . 
\end{align}

The environment entropy production is associated with the bidirectional transitions. 
The average total entropy production is the sum of the environment entropy production and the system entropy production: 
\begin{align}
\langle \! \langle \Sigma^{{\rm tot}}_{{\rm bi}} \rangle \! \rangle_F = \langle \! \langle \Sigma^{{\rm env}}_{{\rm bi}} \rangle \! \rangle_F + \langle \! \langle \Sigma^{{\rm sys}}_{{\rm bi}} \rangle \! \rangle_F \, . 
\end{align}
The entropy production rates are, 
\begin{align}
\langle \! \langle \sigma^{{\rm env}}_{{\rm bi}} \rangle \! \rangle = \frac{1}{2} \sum_{ (i \leftarrow j) \in E_{\rm bi}}  j_{i,j}^{\rm st} \, \Delta \Sigma_{ {\rm bi} , i \leftarrow j}^{\rm env} \, , \label{eqn:sig_env_bi}
\\
\langle \! \langle \sigma^{{\rm sys}}_{{\rm bi}} \rangle \! \rangle = \frac{1}{2} \sum_{ (i \leftarrow j) \in E_{\rm bi}} j_{i,j}^{\rm st}  \ln \left( \frac{ p_{j}^{\rm st} }{ p_{i}^{\rm st} } \right) \,   \, , \label{eqn:sig_sys_bi}
\end{align}
where $j_{i,j}^{\rm st}$ is the flux, 
\begin{align}
\tilde{j}_{i,j} =\Gamma_{i,j} p_{j} -\Gamma_{j,i} p_{i} \, , 
\end{align}
defined in the steady-state, whereby $p_{j}=p^{\rm st}_{j}$. 
Corresponding to Eq.~(\ref{eqn:env_ent}), 
\begin{align}
\Delta \Sigma_{ {\rm bi} , i \leftarrow j}^{\rm env} = \ln \frac{ \Gamma_{i,j} }{ \Gamma_{j,i} } \, , \label{eqn:env_ent_general}
\end{align}
is the environment entropy production associated with the transition $|j \rangle \to |i \rangle $.

\subsubsection{Thermodynamic cost of Brownian computation}

In the framework of the stochastic thermodynamics of resetting~\cite{Fuchs2016}, the thermodynamic cost of maintaining the reset state is introduced. 
We assume the theory is applicable to any unidirectional process, including the token-detection process. 
The system entropy production rate associated with unidirectional processes, 
\begin{align}
\langle \! \langle \sigma^{{\rm sys}}_{{\rm uni}} \rangle \! \rangle = \sum_{ (i \leftarrow j) \in E_{\rm uni}} \Gamma_{i,j} p^{\rm st}_{j} \ln \frac{ p_{j}^{\rm st} }{  p_{i}^{\rm st} }  = \langle \! \langle \sigma^{{\rm abs}}_{{\rm uni}} \rangle \! \rangle + \langle \! \langle \sigma^{{\rm ins}}_{{\rm uni}} \rangle \! \rangle \, , \label{eqn:sig_sys_uni}
\end{align}
is separated into the entropy production rates due to absorption and insertion: 
\begin{align}
\langle \! \langle \sigma^{{\rm abs}}_{{\rm uni}} \rangle \! \rangle =& \sum_{ (i \leftarrow j) \in E_{\rm uni}} \Gamma_{i,j} p^{\rm st}_{j} \ln p_{j}^{\rm st} \, , \label{eqn:s_abs} \\
\langle \! \langle \sigma^{{\rm ins}}_{{\rm uni}} \rangle \! \rangle =& - \sum_{ (i \leftarrow j) \in E_{\rm uni}} \Gamma_{i,j} p^{\rm st}_{j} \ln p_{i}^{\rm st} \, .  \label{eqn:s_ins} 
\end{align}
In the steady-state, rates of the stochastic observable $O$ associated with bidirectional and unidirectional processes compensate each other, which can be deduced from the following equality, 
\begin{align}
\sum_{i \in \Omega} O_i \dot{p}_i =& \frac{1}{2} \sum_{(i \leftarrow j) \in E_{\rm bi} } ( O_i - O_j ) \tilde{j}_{i,j} \nonumber \\ & + \sum_{(i \leftarrow j) \in E_{\rm uni} } ( O_i - O_j ) \Gamma_{i,j} p_j \, . \label{eqn:bi_uni_rate_balance}
\end{align}
For example, in the steady-state, the time derivative of the Shannon entropy, 
\begin{align}
H( \{ p_j \} )= - \sum_{j \in \Omega} p_j \ln p_j \, , 
\end{align}
vanishes, which results in $\langle \! \langle \sigma^{{\rm sys}}_{{\rm bi}} \rangle \! \rangle + \langle \! \langle \sigma^{{\rm sys}}_{{\rm uni}} \rangle \! \rangle =0$.

The thermodynamic cost is the amount of external work ${\mathcal W}^{\rm ext}$ provided to maintain the steady-state. 
From the second-law of non-equilibrium thermodynamics~\cite{Parrondo2015}, the following inequality is obtained~\cite{Fuchs2016}: 
\begin{align}
\dot{\mathcal W}^{\rm ext} \geq \sum_{ (i \leftarrow j) \in E_{\rm uni}} (F_i - F_j) \Gamma_{i,j} p_j^{\rm st} = {k_{\rm B} T} \langle \! \langle \sigma^{{\rm tot}}_{{\rm bi}} \rangle \! \rangle \, , \label{eqn:Fuchs_bound}
\end{align}
where we exploited Eq.~(\ref{eqn:bi_uni_rate_balance}). 
Here the stochastic free energy is, $F_{i} = U_i + {k_{\rm B} T} \ln p_i^{\rm st}$. 
The energy difference satisfies $U_i-U_j = {k_{\rm B} T} \Delta \Sigma^{{\rm env}}_{{\rm bi}, j \leftarrow i}$ for $(j \leftarrow i ) \in E_{\rm bi}$. 
The inequality (\ref{eqn:Fuchs_bound}) corresponds to Landauer's principle in the steady-state in the stochastic thermodynamics of resetting~\cite{Fuchs2016}. 
The mixed bound (\ref{eqn:tur_mix}) can be replaced with a looser upper bound written as, 
\begin{align*}
\sqrt{ \langle \! \langle \Sigma^{{\rm tot}}_{{\rm bi}} \rangle \! \rangle_F/2 + \langle \! \langle A_{{\rm uni}} \rangle \! \rangle_F } 
\leq 
\sqrt{ {\mathcal W}^{\rm ext}/2 + \langle \! \langle A_{{\rm uni}} \rangle \! \rangle_F } 
\, ,
\end{align*}
where ${\mathcal W}^{\rm ext} = \dot{\mathcal W}^{\rm ext} \langle \! \langle \tau \rangle \! \rangle_F \geq {k_{\rm B} T} \langle \! \langle \Sigma^{{\rm tot}}_{{\rm bi}} \rangle \! \rangle_F $.

\subsection{Full-counting statistics of resetting}
\label{sec:fcs}

We will summarize the theoretical framework adopted in the present work. 
Detailed calculations are relegated to Appendix~\ref{sec:First_passage_resettings}. 
The probability distribution of the first-passage time $\tau$ for $W$ resets is~\cite{Garrahan2017},  
\begin{widetext}
\begin{align}
F_W(\tau) = \int_0^\tau dt_{W-1} \int_0^{t_{W-1}} dt_{W-2} \cdots \int_0^{t_{3}} dt_{2}  \int_0^{t_{2}} dt_{1} \langle {\rm i} | \hat{V} e^{\hat{L}^{\rm f} (\tau -t_{W-1})} \hat{V} e^{\hat{L}^{\rm f} (t_{W-1} -t_{W-2})} \cdots \hat{V} e^{\hat{L}^{\rm f} (t_2 -t_1)} \hat{V} e^{\hat{L}^{\rm f} t_1}| {\rm i} \rangle \, . \label{eqn:fw_def}
\end{align}
\end{widetext}
Here we suppress the counting fields for the activity and the environment entropy production, $\eta$ and $\xi$. 
The characteristic function is the Laplace transform of Eq.~(\ref{eqn:fw_def}) expressed as 
\begin{align}
F_W(s)=\int_0^\infty d \tau e^{-s \tau} F_W(\tau) = F_1(s)^W \, .
\end{align}
In the limit of $\Gamma_{\rm reset} \to \infty$ and $W=1$, it reproduces the well-known form~\cite{Redner2001}: 
\begin{align}
F_1(s) = \frac{\langle {\rm f}|( s \hat{I} - \hat{L}^{\rm f})^{-1} |{\rm i} \rangle }{\langle {\rm f}|( s \hat{I} - \hat{L}^{\rm f})^{-1} |{\rm f} \rangle } \, . \label{eqn:cff1}
\end{align}

The probability distribution of the first-passage time for $W$ resets, $F_W(\tau)$, and that of the number of resets $W$ during the measurement time $\tau$, $P_\tau(W)$, are related to each other according to Eq.~(\ref{eqn:pw})~\cite{Redner2001,Saito2017,Singh2019}. 
In the limit of $\tau \to \infty$, where the steady-state is reached, the characteristic function becomes~\cite{Saito2017,Singh2019}, 
\begin{align}
F_W(s) =& \frac{P_s(W)}{P_s(0)} \, , \label{eqn:cf_fpt} \\ P_s(W) \approx& \int_{-\pi}^\pi \frac{d \chi}{2 \pi} \frac{e^{-i \chi W}}{s-\Lambda_0(\chi)} \, . \label{eqn:pws}
\end{align}
Here, $\Lambda_0(\chi)$ is the scaled cumulant generating function (CGF), 
\begin{align}
\Lambda_0(\chi) = \lim_{\tau \to \infty} \frac{1}{\tau} \ln \sum_{W=0}^\infty e^{i \chi W} P_\tau(W) \, . \label{eqn:cf_long_tau}
\end{align}
In the limit of $\Gamma_{\rm reset} \to \infty$, it is the eigenvalue of the coarse-grained modified transition rate matrix, 
\begin{align}
\hat{ L }(\chi) = \hat{P} \hat{L}^{\rm f} \hat{P} + e^{i \chi} |{\rm i} \rangle  \langle {\rm f}| \hat{L}^{\rm f} \hat{P} \, , \label{eqn:eff_liouvill}
\end{align}
with maximum real part. 
Here the projection operator $\hat{P}=\hat{I}-|{\rm f} \rangle \langle {\rm f}|$ projects out the final state. 

An approximate probability distribution is obtained by expanding the scaled cumulant generating function up to the second order in the counting field, 
$\Lambda_0(\chi) \approx \langle \! \langle w \rangle \! \rangle (i \chi)+\langle \! \langle w^2 \rangle \! \rangle (i \chi)^2/2$. 
Here the first cumulant, 
\begin{align}
\langle \! \langle w \rangle \! \rangle =  \sum_{ j \in {\it \Gamma}^- {\rm  f} } \Gamma_{ {\rm i},j} p_j^{\rm st}  \, ,
\end{align}
is the average reset current. 
By substituting the second-order expansion to Eq.~(\ref{eqn:pws}), we obtain the probability distribution, which is the inverse Gaussian distribution, also known as the Wald distribution~\cite{Singh2019},
\begin{align}
F_W(\tau) \approx \frac{W}{ \tau \, \sqrt{ 2 \pi \tau \langle \! \langle w^2 \rangle \! \rangle  } } \exp \left( - \frac{ \left( W - \tau \langle \! \langle w \rangle \! \rangle \right)^2}{2 \tau \langle \! \langle w^2 \rangle \! \rangle} \right) \, , \label{eqn:inverse_gauss_tau}
\end{align}
after some calculations (Appendix~\ref{sec:First_passage_resettings}). 
From Eq.~(\ref{eqn:inverse_gauss_tau}), we can derive Eq.~(\ref{eqn:c_t}).

The SNR (\ref{eqn:snr}) for $W$ resets can be re-expressed by the Fano factor of the reset current, $\langle \! \langle w^2 \rangle \! \rangle/\langle \! \langle w \rangle \! \rangle$~\cite{Blanter2000,Belzig2005}: 
\begin{align}
\frac{S}{N} = \sqrt{ \frac{W \langle \! \langle w \rangle \! \rangle}{\langle \! \langle w^2 \rangle \! \rangle} } \, . \label{eqn:snr_wald}
\end{align}
This relation connects the fluctuations of the computation time with the current fluctuations in the steady-state transport. 
The approximation explained above can be extended for the joint probability distribution of the first-passage time and the activity, see Eq.~(\ref{eqn:inverse_gauss_tau_a}).

\subsection{Reversible TM with semi-infinite state space}
\label{sec:semi-infinite_reversible_TM}

In this section, we discuss speed limits, noting that Eqs.~(\ref{eqn:feynmann1}), (\ref{eqn:feynmann2}) and (\ref{eqn:feynmann3}) are applicable to the intermediate steps of computation in an infinite state-space. 
The RTM with infinite state-space was considered in previous work~\cite{Strasberg2015} focusing on position fluctuations in the state-space with fixed measurement time.
Here, we analyze the fluctuations of computation time with a semi-infinite state-space ($|\Omega| \to \infty$) close to the final state $|{\rm f} \rangle = ||\Omega|+1 \rangle$. 
Precisely, we start to measure the duration of time at an intermediate state, i.e., the state $m$ steps before the final state $| |\Omega|+1 - m \rangle$. 
Technically, we calculate the CGF (\ref{eqn:cff1}) with the replacement $|{\rm i} \rangle$ by $| |\Omega|+1 - m \rangle$ (Appendix \ref{sec:suppl_semi-infinite_reversible_TM}). 
For $\gamma^+ > \gamma^-$, the CGF is, 
\begin{align}
F_1(s,\xi,\eta) =& \left( \frac{s + \gamma - \sqrt{ (s + \gamma)^2-4 \gamma^+ \gamma^- e^{i 2 \eta} } }{2 \gamma^- e^{i \eta}} \right)^m
\nonumber \\ & \times e^{i m \Delta \Sigma_{\rm bi}^{\rm env} \xi} \, , \label{eqn:cgf_infinite_TM} 
\end{align}
where $\gamma=\gamma^+ + \gamma^-$. 
The joint cumulant is calculated from the derivatives as, 
\begin{align}
\langle \! \langle \tau^j A^k \Sigma_{\rm bi}^{ {\rm env} \, \ell } \rangle \! \rangle_F = \left. \frac{\partial^{j+k+\ell} \ln F_W(s,\xi,\eta)}{\partial (-s)^j \partial (i \xi)^k \partial (i \eta)^\ell}\right|_{s=\xi=\eta=0} \, . \label{eqn:joint_cumulants}
\end{align}
Equation~(\ref{eqn:cgf_infinite_TM}) indicates that the entropy production does not fluctuate and its higher cumulants vanish:  
\begin{align}
\langle \! \langle \Sigma_{\rm bi}^{ {\rm env} \, \ell}  \rangle \! \rangle_F =&\left \{ \begin{array}{cc} m \Delta \Sigma_{\rm bi}^{\rm env} & (\ell=1) \\
0  & (\ell>1) \end{array} \right. \, . \label{eqn:envent_infiniRTM}
\end{align}
This property would be universal for any Brownian computer processing a problem with a unique solution since the correct computation path, and associated stochastic environment entropy production are unique. 
Obviously, Eq.~(\ref{eqn:envent_infiniRTM}) corresponds to Eq.~(\ref{eqn:feynmann1}). 

The average first-passage time is, 
\begin{align}
\langle \! \langle \tau \rangle \! \rangle_F =& \frac{m}{\gamma^+ - \gamma^-} \, . \label{eqn:infiniteTM_ave_tau} 
\end{align}
In the limit of $\gamma^+/\gamma^- \to 1$, the environment entropy production (\ref{eqn:envent_infiniRTM}) vanishes and the average computation time (\ref{eqn:infiniteTM_ave_tau}) goes to infinity. 
Therefore, when we focus on a portion of a large RTM, i.e., on a  scale much smaller than the state space, the computation can be done without the environment entropy production if it is carried out with speed approaching zero.

The first and second cumulants other than Eq.~(\ref{eqn:infiniteTM_ave_tau}) are summarized in Eqs.~(\ref{eqn:infiniteTM_ave_A}), (\ref{eqn:infiniteTM_var_tau}), (\ref{eqn:infiniteTM_var_A}), and (\ref{eqn:infiniteTM_cross}). 
Using these relations, we check the kinetic bound (\ref{eqn:tur_activity}): 
\begin{align*}
\frac{S}{N} = \sqrt{ m \frac{\gamma^+ - \gamma^-}{\gamma^+ + \gamma^-} } \leq \sqrt{ \langle \! \langle A \rangle \! \rangle_F } = \sqrt{ m \frac{\gamma^+ + \gamma^-}{\gamma^+ - \gamma^-} }\, . 
\end{align*}
By applying the inequality~\cite{Shiraishi2016} 
$\ln (\gamma^+/\gamma^-) \geq 2(\gamma^+-\gamma^-)/(\gamma^++\gamma^-)$, 
to Eq.~(\ref{eqn:envent_infiniRTM}), we obtain
\begin{align}
\frac{ \langle \! \langle \Sigma_{\rm bi}^{\rm env} \rangle \! \rangle_F }{m} \geq \frac{\gamma^+ - \gamma^-}{(\gamma^+ + \gamma^-)/2} \, . \label{eqn:ref_em}
\end{align}
Since the left-hand side of the inequality is the energy loss per step, the inequality would be regarded as a refinement of Eq.~(\ref{eqn:feynmann3}). 
The inequality (\ref{eqn:ref_em}) can be written in a form similar to Eq.~(\ref{eqn:tur_mix}) as, 
\begin{align*}
\frac{S}{N} = \sqrt{ m \frac{\gamma^+ - \gamma^-}{\gamma^+ + \gamma^-} } \leq \sqrt{ \frac{ \langle \! \langle \Sigma_{\rm bi}^{\rm env} \rangle \! \rangle_F }{2} } \, , 
\end{align*}
which provides the reasoning that Eq.~(\ref{eqn:feynmann3}) would be understood as a TUR~(\ref{eqn:tur_mix}). 

The correlation coefficient between the first-passage time and activity is,
\begin{align}
r=\frac{ \langle \! \langle \tau A \rangle \! \rangle_F }{ \sqrt{ \langle \! \langle \tau^2 \rangle \! \rangle_F \langle \! \langle A^2 \rangle \! \rangle_F } } = 2 \frac{ \sqrt{ \gamma^+ \gamma^- } }{\gamma^+ + \gamma^-} \, . \label{eqn:cor_coe_seminfTM}
\end{align}
A perfect linear correlation is realized when $\gamma^+ = \gamma^-$. 
In this case, the activity is always $A = \gamma \tau$.

\subsection{Reversible TM}
\label{sec:Finite_RTM}

If the state space is finite, the computation is done within a finite time even when the environment entropy production is zero, $\gamma^+ = \gamma^-=\gamma/2$. 
Here we present novel consequences obtained by integrating theories explained in the previous sections. 
The reduced transition rate matrix (\ref{eqn:eff_liouvill}) corresponding to Fig. \ref{fig:ctmarkovchain_RTM}~(c) is 
\begin{align}
\hat{ {L} }(\chi) =& \sum_{j=2}^{|\Omega|-1} - \gamma |j \rangle \langle j| + \gamma^+ |j \rangle \langle j-1| + \gamma^- |j \rangle \langle j+1| \nonumber \\ &
- \gamma ||\Omega| \rangle \langle |\Omega|| + \gamma^+ ||\Omega| \rangle \langle |\Omega|-1| \nonumber \\ &
- \gamma^+ |1 \rangle \langle 1| + \gamma^- |1 \rangle \langle 2| + \gamma^+ e^{i \chi} |1 \rangle \langle |\Omega|| \, , \label{eqn:mod_Liouvill_FiniteRevTM}
\end{align}
where $|{\rm i} \rangle = |1 \rangle$. 
The steady-state solution satisfying $\hat{ {L} }(0) |p^{\rm st} \rangle =0$ is, 
\begin{align}
|p^{\rm st} \rangle = \sum_{n=1}^{|\Omega|} p^{\rm st}_n |n \rangle \, , \;\;\;\; p^{\rm st}_n = \frac{2 (|\Omega|+1-n)}{|\Omega|(|\Omega|+1)} \, . \label{eqn:dist_prob_finite_TM}
\end{align}
The probability decreases linearly with the distance from the initial state. 
We emphasize that the error-free resets drive the system out of equilibrium. 
Then the average reset current is, 
\begin{align*}
\langle \! \langle w \rangle \! \rangle =& \frac{\gamma}{2} p^{\rm st}_{|\Omega|} = \frac{\gamma}{|\Omega|(|\Omega|+1)} \, , \label{eqn:reset_curr}
\end{align*}
where $\langle {\rm e}|j \rangle=1$. 
By exploiting Eq.~(\ref{eqn:c_t}), the average first-passage time for $W$ resets is,
\begin{align}
\langle \! \langle \tau \rangle \! \rangle_F = \frac{W}{\langle \! \langle w \rangle \! \rangle} = \frac{W |\Omega|(|\Omega|+1)}{\gamma} \, .
\end{align}
The result indicates that the computation time is finite and increases quadratically with the size of the state space $|\Omega|$. 

The activity rate associated with the unidirectional transition is equal to the reset current, 
\begin{align*}
\langle \! \langle a_{\rm uni} \rangle \! \rangle = \Gamma_{1,|\Omega|} p^{\rm st}_{|\Omega|}  = \langle \! \langle w \rangle \! \rangle \, .
\end{align*}
Therefore the corresponding activity is simply the number of resets, 
\begin{align}
\langle \! \langle A_{\rm uni} \rangle \! \rangle_F = \langle \! \langle a_{\rm uni} \rangle \! \rangle \langle \! \langle \tau \rangle \! \rangle_F = W \, ,
\end{align}
where we exploited Eq.~(\ref{eqn:c_t}). 
The system entropy production rates due to absorption and insertion are, 
\begin{align}
\langle \! \langle \sigma^{{\rm abs}}_{{\rm uni}} \rangle \! \rangle =& \Gamma_{1,|\Omega|} p^{\rm st}_{|\Omega|} \ln { p^{\rm st}_{|\Omega|} } = - \langle \! \langle w \rangle \! \rangle \ln \frac{ |\Omega| ( |\Omega|+1 )}{2}  \, , \label{eqn:s_abs_RTM} \\
\langle \! \langle \sigma^{{\rm ins}}_{{\rm uni}} \rangle \! \rangle =& - \Gamma_{1,|\Omega|} p^{\rm st}_{|\Omega|} \ln { p^{\rm st}_{1} } = \langle \! \langle w \rangle \! \rangle \ln \frac{|\Omega|+1}{2}  \, . \label{eqn:s_ins_RTM} 
\end{align}
Then the system entropy production due to the unidirectional resets is, 
\begin{align}
\langle \! \langle \Sigma^{\rm sys}_{\rm uni} \rangle \! \rangle_F = \langle \! \langle \sigma^{\rm sys}_{\rm uni} \rangle \! \rangle \langle \! \langle \tau \rangle \! \rangle_F = -W \ln |\Omega| \, , \label{eqn:sys_uni_finite_TM}
\end{align}
which means that the resets decrease the system entropy. 
The total entropy production associated with the bidirectional transition processes is, 
\begin{align}
\langle \! \langle \Sigma^{\rm tot}_{\rm bi} \rangle \! \rangle_F = \langle \! \langle \Sigma^{\rm sys}_{\rm bi} \rangle \! \rangle_F = - \langle \! \langle \Sigma^{\rm sys}_{\rm uni} \rangle \! \rangle_F = W \ln |\Omega| \, . 
\end{align}
By comparing this with Eq.~(\ref{eqn:Fuchs_bound}), we conclude that the resetting cost increases logarithmically in the size of the state space. 
At first glance, our results seem consistent with Refs.~\onlinecite{Norton2013,Strasberg2015}. 
Sec.~\ref{sec:discussion} discusses this interpretation in more detail.

Summarizing the above, we obtain the mixed bound as, 
\begin{align}
\sqrt{ \langle \! \langle \Sigma^{\rm tot}_{\rm bi} \rangle \! \rangle_F/2 + \langle \! \langle A_{\rm uni} \rangle \! \rangle_F } = \sqrt{ W \left( 1 + \frac{ \ln |\Omega| }{2} \right) } \, . \label{eqn:mixed_bound_finite_TM}
\end{align}

\begin{figure}[ht]
\begin{center}
\includegraphics[width=0.8 \columnwidth]{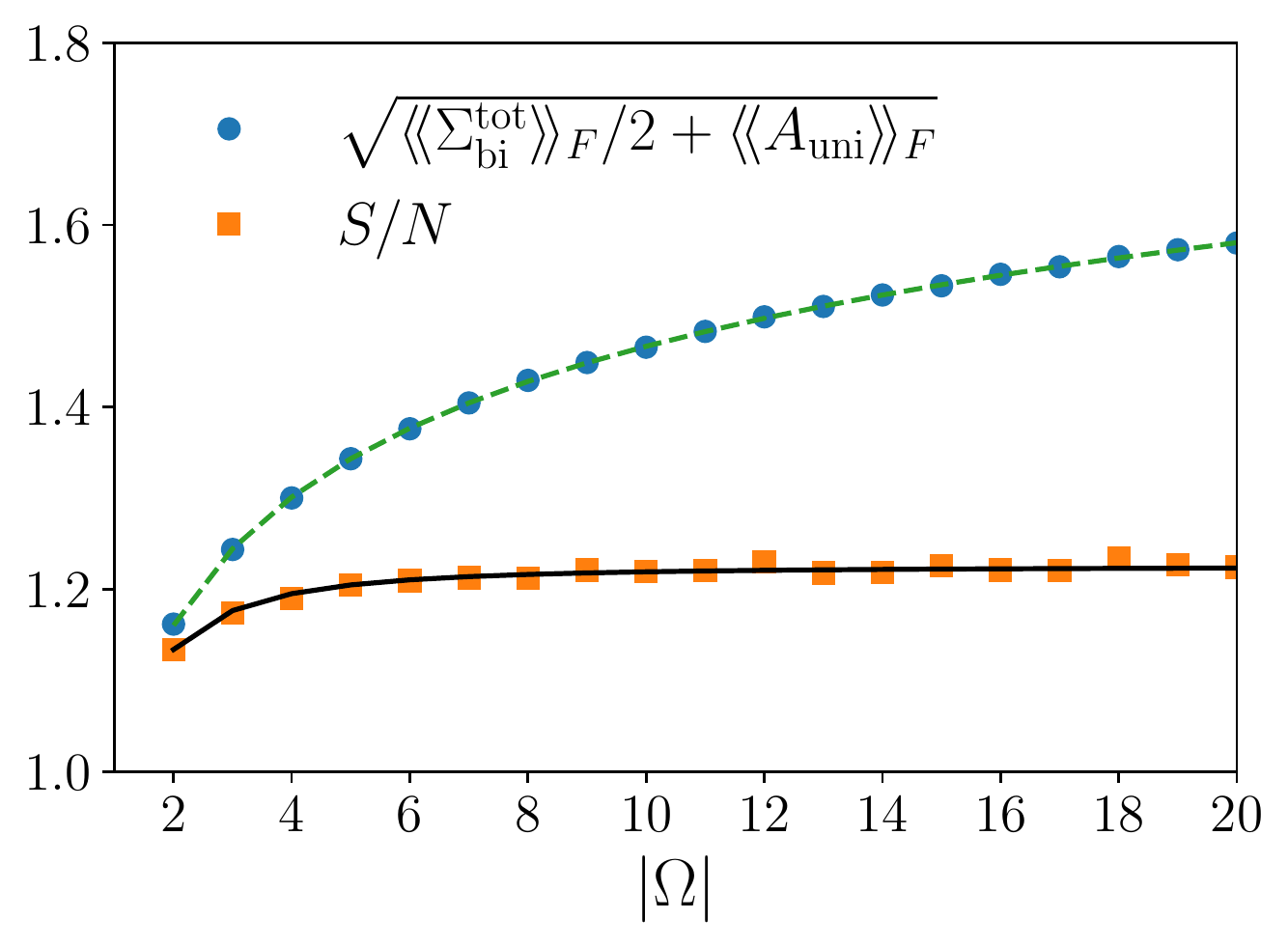}
\caption{
The signal-to-noise ratio and the mixed bound versus state space size for $W=1$. 
The number of stochastic trajectories taken is $10^5$. 
The solid line indicates the SNR for one reset in Eq.~(\ref{eqn:snr_wald}). 
The dashed line indicates the mixed bound in Eq.~(\ref{eqn:mixed_bound_finite_TM}). 
}
\label{fig:FinRevTMTUR}
\end{center}
\end{figure}

Figure \ref{fig:FinRevTMTUR} shows the SNR and the mixed bound versus the size of the state space $|\Omega|$ calculated by Gillespie's algorithm (Appendix \ref{sec:Gillespies_algorithm}). 
The mixed bound corresponds well with the analytic result according to Eq.~(\ref{eqn:mixed_bound_finite_TM}) (dashed line). 
The solid line is the approximate SNR, Eq.~(\ref{eqn:snr_wald}), which is expressed by the reset current $\langle \! \langle w \rangle \! \rangle$ and its noise $\langle \! \langle w^2 \rangle \! \rangle$ calculated by perturbatively solving the eigenvalue equation of the modified transition rate matrix (\ref{eqn:mod_Liouvill_FiniteRevTM}) (Appendix \ref{sec:RSPT}). 
The mixed bound increases because of the logarithmic increase in the system entropy production. 
On the other hand, the SNR saturates: 
the Fano factor $\langle \! \langle w^2 \rangle \! \rangle/\langle \! \langle w \rangle \! \rangle$ is independent of the length of the chain once it becomes longer than the characteristic length. 
Therefore, the resetting entropy production cannot be used as the cost for improving the SNR for the RTM.

\section{Token-based Brownian circuits}
\label{sec:tbbc}

A Brownian RTM's stochastic dynamics are simple, but its rules (programs) are highly abstract. 
This section analyzes the token-based Brownian circuit, which consists of modules of logic gates constructed from three primitive circuit elements. 
The computation process of the token-based Brownian circuit is intuitively straightforward, but its stochastic dynamics are involved.

\subsection{Circuit elements: Stochastic Petri net in Doi-Peliti formalism}

\begin{figure}[ht]
\begin{center}
\includegraphics[width=0.9 \columnwidth]{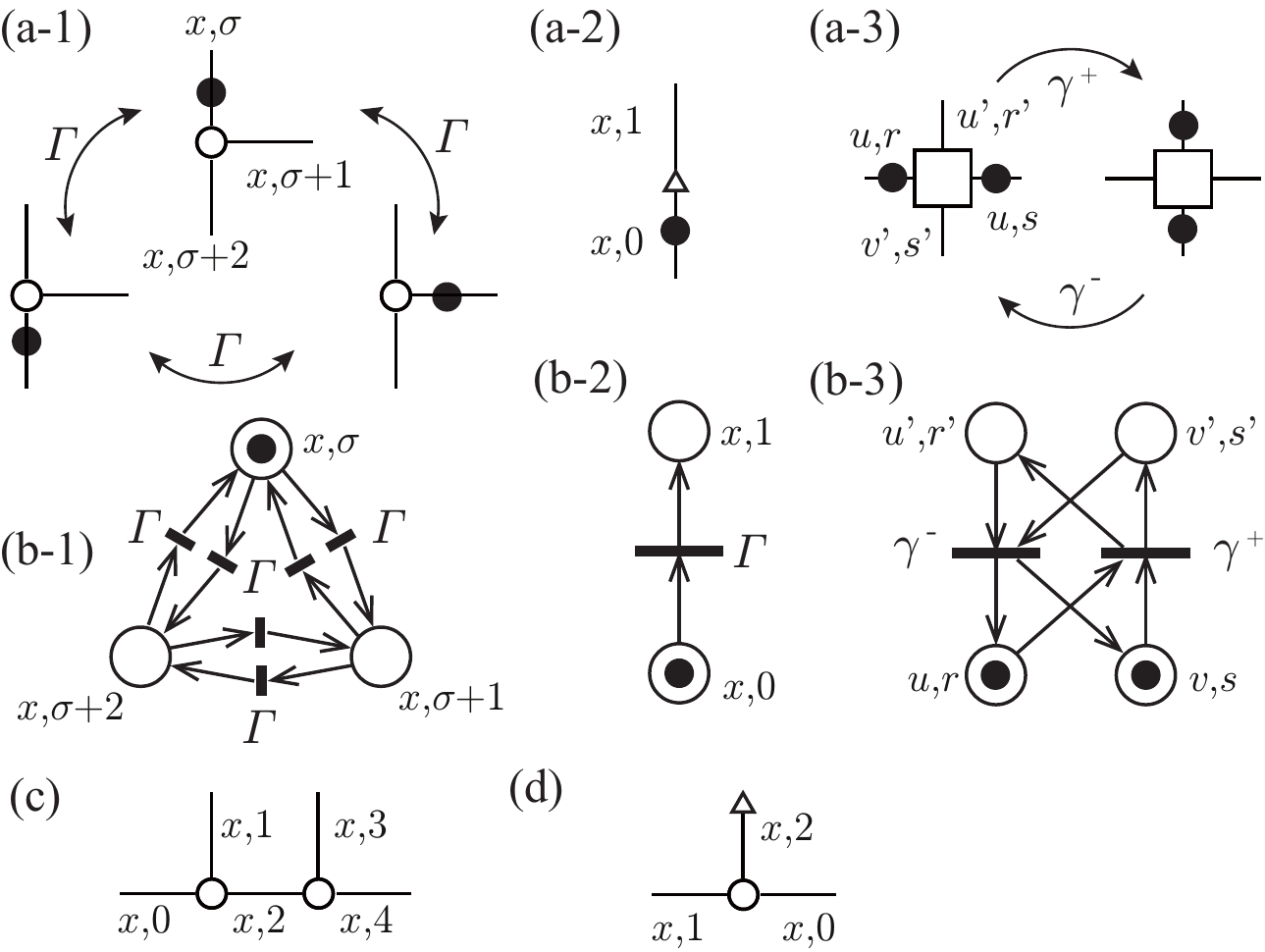}
\caption{
Circuit elements of the token-based Brownian circuit, (a-1) Hub (open circle), (a-2) Ratchet (open triangle), and (a-3) Cjoin (open squire). 
(b-1), (b-2) and (b-3) are corresponding stochastic Petri nets. 
Circles represent {\it places} and bars represent {\it transitions}. 
Tokens (black dots) move along arcs. 
(c) Two Hubs connected in series. 
(d) A Hub with a Ratchet attached to an output terminal. 
}
\label{fig:Circuit_elements}
\end{center}
\end{figure}

A token-based Brownian circuit diagram~\cite{Peper2013,Lee2016} (see Fig.~\ref{fig:Circuit_elements}) is a graph whose nodes are circuit elements and whose edges are wires through which tokens transmit. 
A Boolean variable and its truth value are encoded by the index of the wire on which a token resides (dual-rail encoding). 
Figure~\ref{fig:Circuit_elements} shows the circuit diagrams of three circuit elements, the Hub [panel (a-1)], the Ratchet [panel (a-2)] and the Conservative Join (CJoin) [panel (a-3)]. 
The Hub [panel (a-1)] is denoted by an open circle with three wires. 
It forks into two branches corresponding to two possible computation paths with the rate $\Gamma$. 
However, a token can also return on its tracks. 
In this way, a token searches for a computational path. 
A triangle denotes the Ratchet [panel (a-2)]. 
It works as a kind of `diode' representing the unidirectional transition corresponding to the readout process, i.e., the detection of a token. 
The CJoin [panel (a-3)] is denoted by an open square with one wire attached to each of its four sides. 
If one token each is injected into the left side and the right side, one of them is emitted from the top side and the other from the bottom side. 
The forward and backward transition rates are $\gamma^+$ and $\gamma^-$, respectively. 
The CJoin synchronizes a process~\cite{Murata1989} through two-token scattering. 
The CJoin is the only circuit element producing environment entropy. 
It pushes the computation according to the setting $\gamma^+ > \gamma^-$ at the cost of environment entropy production [Eq.~(\ref{eqn:env_ent})], see also Eq.~(\ref{eqn:env_ent_general}). 
Note that a Ratchet is a measurement device whose thermodynamic cost is not counted as the environment entropy production: the Ratchet changes the system entropy, Eq.~(\ref{eqn:sig_sys_uni}), when it detects a token. 
Any terminal of a circuit element can be connected to a terminal of another circuit element. 
A terminal in the circuit diagram that is not connected to any node is an input or output terminal.

The formal representation of the token-based Brownian circuit~\cite{Peper2013,Lee2016} is the stochastic Petri net (SPN) (Appendix~\ref{sec:SPN})~\cite{Murata1989}, which is a directed bipartite graph. 
Panels (b-1), (b-2), and (b-3) show the SPNs corresponding to the above circuit elements. 
Graphically, two kinds of nodes, {\it places} and {\it transitions} are drawn as circles and bars, respectively. 
Places of the SPN correspond to wires of the token-based Brownian circuit diagram. 
Directed edges, or arcs, represents the direction of the token transition. 
In Panels (b-1), (b-2), and (b-3), a transition is enabled if each input place, which is the tail of an arc whose head is the transition, contains at least one token. 
When the enabled transition is fired, a token is removed from each input place and added to each output place, which is the head of an arc whose tail is the transition. 

The SPN is uniquely defined once sets of places, transitions, flow relations, and transition rates are provided (Appendix \ref{sec:SPN})~\cite{Murata1989}. 
However, this formal definition is cumbersome for the token-based Brownian circuit, which becomes a large-scale SPN possessing a modular structure. 
In the present paper, we introduce the Doi-Peliti formalism for SPN (Appendix~\ref{sec:DPH}). 
In this formalism, the transition rate matrix is given in the second quantized form $\hat{H}(\hat{a}^\dagger,\hat{a})$ expressed with the boson creation and annihilation operators satisfying the commutation relation $[\hat{a},\hat{a}^\dagger]=1$. 
This technique enables us to write down the transition rate matrix compactly and to utilize field theory approaches~\cite{Kamenev2002,Kamenevbook2011,Baez2018}.

In a token-based Brownian circuit, the information-bearing degrees of freedom~\cite{Bennett1982} are defined by the locations of the tokens. 
More precisely, a logical state $x$, which we define as a Boolean variable and its truth value, is encoded as the index of the place where a token resides. 
A group of places connected by Hub(s), which we call a \emph{cell}, form a single logical state $x$. 
It is convenient to specify a place as a pair $(x,\sigma)$ of a cell index $x$ and an index $\sigma$ that specifies its fine-grained location in the cell. 
The state with a single token at $(x,\sigma)$ is denoted as $|x \rangle |\sigma \rangle = \hat{a}_{x,\sigma}^\dagger |0 \rangle$, where $|0 \rangle$ is the vacuum state. 

In the token-based Brownian circuit, a single token carries the information of a Boolean variable and its truth value. 
Consequently, each cell can host at most one token. 
Therefore to specify the $N$-token state, it is convenient to consider the set of occupied places $\{ (x_1,\sigma_1), (x_2,\sigma_2), \cdots, (x_N,\sigma_N)\}$ and denote the $N$-token fine-grained state as, 
\begin{align}
|x_1, x_2, \cdots x_N \rangle |\sigma_1, \sigma_2, \cdots \sigma_N \rangle = \left( \prod_{\ell = 1}^N \hat{a}_{x_\ell,\sigma_\ell}^\dagger \right) |0 \rangle  \, . \label{eqn:Ntokenstate}
\end{align}
We exclude $N$-token states with more than one token in any cell since they are erroneous states.

The general form of the modified Doi-Peliti Hamiltonian of SPN is given in Eq.~(\ref{eqn:Doi_Peliti_Ham}). 
For an isolated Hub, it is
\begin{align}
\hat{H}_{ {\rm Hub}, \sigma } (x;\eta) &= \Gamma (\hat{a}_{x,\sigma+1}^\dagger e^{i \eta} - \hat{a}_{x,\sigma}^\dagger) \hat{a}_{x,\sigma} \nonumber \\ &+ \Gamma (\hat{a}_{x,\sigma+2}^\dagger e^{i \eta} - \hat{a}_{x,\sigma+1}^\dagger) \hat{a}_{x,\sigma+1} \nonumber \\ &+ \Gamma (\hat{a}_{x,\sigma}^\dagger e^{i \eta} - \hat{a}_{x,\sigma+2}^\dagger) \hat{a}_{x,\sigma+2} + {\rm H.c.} \, . \label{eqn:1hub}
\end{align}
The Hub does not produce environment entropy, although it always racks up computation time $\sim 1/\Gamma$ to search for a computation path that leads to an output state. 
The argument $x$ and the subscript $\sigma$, respectively, specify the Hub's logical state and the place of the input terminal. 
The logical state is encoded by a cell, which consists of $n$ Hubs ($n \in {\mathbb Z}^+$, ${\mathbb Z}^+$ being the set of positive integers) connected in series [Fig.~\ref{fig:Circuit_elements} (c)]. 
For $n=1$, $\hat{H}_{1}(x;\eta) = \hat{H}_{ {\rm Hub}, 0 } (x;\eta)$ and for $n \geq 2$, 
\begin{align}
\hat{H}_{n}(x;\eta) = \hat{H}_{n-1}(x;\eta) + \hat{H}_{{\rm Hub},2n-2}(x;\eta) \, .  \label{eqn:Ham_nhub}
\end{align}
For $n=0$, we define,  
\begin{align}
\hat{H}_{0}(x) = \Gamma (\hat{a}_{x,1}^\dagger - \hat{a}_{x,0}^\dagger) \hat{a}_{x,0} + \Gamma (\hat{a}_{x,0}^\dagger - \hat{a}_{x,1}^\dagger) \hat{a}_{x,1}\, . \label{eqn:Ham_0hub}
\end{align}

The Ratchet [Fig.~\ref{fig:Circuit_elements} (a-2)] represents a unidirectional transition. 
For simplicity, when the Ratchet is attached to a terminal of a Hub [Fig.~\ref{fig:Circuit_elements} (d)], we remove the backward transition(s): 
\begin{align}
\hat{H}_{0}^{\rm R}(x) =& \hat{H}_{0}(x) - \Gamma (\hat{a}_{x,0}^\dagger - \hat{a}_{x,1}^\dagger) \hat{a}_{x,1} \, , \label{eqn:Ham_Ratchet} \\
\hat{H}_n^{\rm R}(x) =& \hat{H}_n(x) - \Gamma \sum_{j=1,2} (\hat{a}_{x,2n-j}^\dagger - \hat{a}_{x,2n}^\dagger) \hat{a}_{x,2n}  \, , \label{eqn:terminal_Ratchet}
\end{align}
where $n \in \mathbb{Z}^{+}$. 

Let $X_n$ ($X_n^{\rm R}$) be the set of logical states encoded by $n$ Hub cells without (with) Ratchet. 
Let $P(x)$ and $T(x)$ be the sets of places and transitions, respectively, in the cell $x$. 
Then the sizes of these sets are, 
\begin{align}
|P(x)| =& \left \{ \begin{array}{cc} 2 & (x \in {X}_0 \cup {X}_0^{\rm R}) \\ 2n+1  & (x \in {X}_n \cup {X}_n^{\rm R} \, , n \in {\mathbb Z}_{+}) \end{array} \right.  \, , \label{eqn:size_logical_state}
\\
|T(x)| =& \left \{ \begin{array}{cc} 2 & (x \in {X}_0) \\ 6n & (x \in {X}_n \, , n \in {\mathbb Z}_{+}) 
\end{array} \right. \, .
\end{align}

The modified CJoin Doi-Peliti Hamiltonian [Fig.~\ref{fig:Circuit_elements} (b-3)] annihilating two tokens at input places, $(u,r)$ and $(v,s)$, and creating two tokens at output places, $(u',r')$ and $(v',s')$, is
\begin{align}
& \hat{V}_{r',s'|r,s}(u',v'|u,v;\xi,\eta) \nonumber \\ =& \gamma^+  (\hat{a}_{u',r'}^\dagger a_{v',s'}^\dagger \hat{a}_{u,r} \hat{a}_{v,s} e^{i \eta + i \xi \Delta \Sigma_{\rm bi}^{\rm env}} - \hat{a}_{u,r}^\dagger \hat{a}_{v,s}^\dagger \hat{a}_{u,r} \hat{a}_{v,s}) \nonumber \\ &+ \gamma^-  (\hat{a}_{u,r}^\dagger \hat{a}_{v,s}^\dagger \hat{a}_{u',r'} \hat{a}_{v',s'} e^{i \eta - i \xi \Delta \Sigma_{\rm bi}^{\rm env}} \nonumber \\ & - \hat{a}_{u',r'}^\dagger \hat{a}_{v',s'}^\dagger \hat{a}_{u',r'} \hat{a}_{v',s'}) \, . \label{ham_c_join}
\end{align}
When the ratio $\gamma^+/\gamma^-$ increases, the computation time becomes shorter at the cost of environment entropy production, see Eq.~(\ref{eqn:env_ent}).  

Since the extension from the ordinary Doi-Peliti Hamiltonian to the modified one is straightforward, in the following we sometimes leave out the counting fields.

\subsection{Plan of this section}

By connecting circuit elements, we construct circuits evaluating Boolean functions. 
We will focus on the computation processes of two simple arithmetic operations: (i) the addition of two single-digit binary numbers, $x$ and $y$, and (ii) the addition of two two-digit binary numbers, $x_1x_0$ and $y_1y_0$. 

\begin{figure}[ht]
\begin{center}
\includegraphics[width=0.9 \columnwidth]{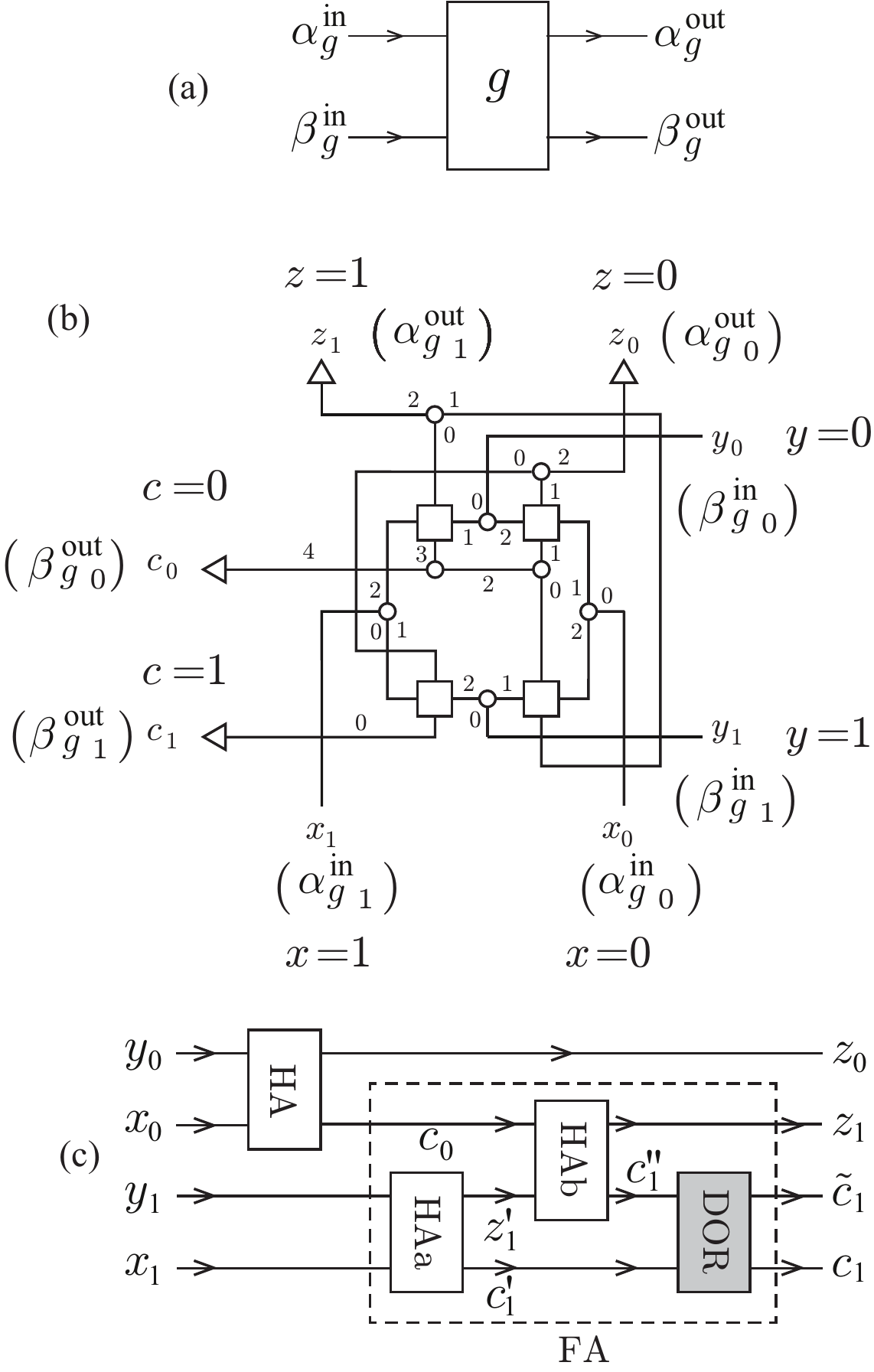}
\caption{ 
(a) A logical gate $g$ corresponding to a Boolean function $f_g(\alpha^{\rm in}_g,\beta^{\rm in}_g) = (\alpha^{\rm out}_g,\beta^{\rm out}_g)$. 
(b) Circuit diagram of a logical gate $g$ (The half-adder $g={\rm HA}$). 
In the dual-rail encoding scheme the Boolean variable $\alpha^{\rm in}_g =x$ with truth value 0(1) is encoded by the cell ${ \alpha^{\rm in}_g }_{0(1)}=x_{0(1)}$. 
Numbers assigned to wires are indexes of places inside each cell. 
(c) Block diagram of the circuit for adding 2-digit binary numbers. 
The half-Adder (HA), taking inputs $x_0$ and $y_0$, is connected as a cascade to the full-adder (FA) in the dashed box. 
The FA module consists of two HA modules, HAa and HAb, and one duplicate-output OR (DOR) gate module (shaded box).
} 
\label{fig:blockdiag2in2out_module}
\end{center}
\end{figure}

The circuit for (i) is the half-adder (HA). 
Explicitly, its arithmetic operation is, 
\begin{align}
x + y = (c)z \, , \label{eqn:arth_HA}
\end{align}
where $z$ is the sum and $c$ is the carry. 
Formally, it is a $2$-input and $2$-output Boolean function, 
$f_{\rm HA}: \{1,0\}^2 \to \{1,0\}^2 $, $f_{\rm HA}(x,y)=(z,c)$, 
whose truth table is given in Table \ref{tab:half_adder}. 
Figure~\ref{fig:blockdiag2in2out_module} (a) shows a block-diagram of the 2-input 2-output logic gate, which is a basic constituent of token-based Brownian circuits. 
The HA is constructed by connecting circuit elements as shown in Fig.~\ref{fig:blockdiag2in2out_module} (b). 

\begin{table}[htb]
\begin{tabular}{cc|cc l} 
$x$ & $y$ & $z$ & $c$ & \\ 
\cline{1-4}
0 & 0 & 0 & 0 & (a) \\ 
\rowcolor{gray!30}%
0 & 1 & 1 & 0 & \\
\rowcolor{gray!30}%
1 & 0 & 1 & 0 &  \multirow{-2}{*}{(b)} \\ 
1 & 1 & 0 & 1 & (c) \\
\end{tabular}
\caption{Truth table of $f_{\rm HA}$.}
\label{tab:half_adder}
\end{table}

To perform the addition of multi-digit binary numbers, we need the full-adder (FA), 
which also adds carry-in $c_{\rm in}$ from the previous stage: 
$x + y +c_{\rm in} = (c_{\rm out})z$. 
The FA consists of two HAs and one duplicate-output OR (DOR) gate module (see Sec.~\ref{sec:DOR}), the dashed box in Fig.~\ref{fig:blockdiag2in2out_module} (c). 
The addition of two-digit binary numbers (ii), 
\begin{align}
x_1 x_0+y_1 y_0=(c_1)z_1 z_0 \, ,  \label{eqn:arth_HAFA}
\end{align}
is realized by a cascade connection of HA and FA, Fig.~\ref{fig:blockdiag2in2out_module} (c). 
The circuit realizes a 4-input 4-output Boolean function, $f_{\rm HA+FA}:\{0,1\}^4 \to \{0,1\}^4$, defined as 
$f_{\rm HA+FA}(x_0,x_1,y_0,y_1) = (z_0,z_1,c_1,\tilde{c}_1)$, 
where $\tilde{c}_1$ is a copy of $c_1$. 
The truth table is given in Table \ref{tab:full_truth_table_two_binary_digit_sum}.

\begin{table}[htb]
\begin{tabular}{cccc|ccc|cccc|c l}
{$x_1$} &  $x_0$ &  $y_1$ &  $y_0$ &  $z_1$ &  $z_0$ & $c_1(\tilde{c}_1)$ &  $z_1^\prime$ &  $c_0$ &  $c_1^\prime$ &  $c_1^{\prime \prime}$ & $c_0+c_1'+c_1''$ & \\
\cline{1-12}
0 & 0 & 0 & 0 & 0 & 0 & 0 & 0 & 0 & 0 & 0 & 0 & (a) \\
\rowcolor{gray!30}%
1 & 1 & 1 & 1 & 1 & 0 & 1 & 0 & 1 & 1 & 0 & 2 & (b)  \\
0 & 0 & 0 & 1 & 0 & 1 & 0 & 0 & 0 & 0 & 0 & 0 & \multirow{2}{*}{(c)} \\
0 & 1 & 0 & 0 & 0 & 1 & 0 & 0 & 0 & 0 & 0 & 0 & \\
\rowcolor{gray!30}%
1 & 0 & 1 & 1 & 0 & 1 & 1 & 0 & 0 & 1 & 0 & 1 &  \\
\rowcolor{gray!30}%
1 & 1 & 1 & 0 & 0 & 1 & 1 & 0 & 0 & 1 & 0 & 1 & \multirow{-2}{*}{(d)} \\
0 & 0 & 1 & 0 & 1 & 0 & 0 & 1 & 0 & 0 & 0 & 0 & \multirow{3}{*}{(e)} \\
0 & 1 & 0 & 1 & 1 & 0 & 0 & 0 & 1 & 0 & 0 & 1 & \\
1 & 0 & 0 & 0 & 1 & 0 & 0 & 1 & 0 & 0 & 0 & 0 & \\
\rowcolor{gray!30}%
0 & 1 & 1 & 1 & 0 & 0 & 1 & 1 & 1 & 0 & 1 & 2 &  \\
\rowcolor{gray!30}%
1 & 0 & 1 & 0 & 0 & 0 & 1 & 0 & 0 & 1 & 0 & 1 & \\
\rowcolor{gray!30}%
1 & 1 & 0 & 1 & 0 & 0 & 1 & 1 & 1 & 0 & 1 & 2 & \multirow{-3}{*}{(f)}  \\
0 & 0 & 1 & 1 & 1 & 1 & 0 & 1 & 0 & 0 & 0 & 0 & \multirow{4}{*}{(g)}  \\
0 & 1 & 1 & 0 & 1 & 1 & 0 & 1 & 0 & 0 & 0 & 0 & \\
1 & 0 & 0 & 1 & 1 & 1 & 0 & 1 & 0 & 0 & 0 & 0 & \\
1 & 1 & 0 & 0 & 1 & 1 & 0 & 1 & 0 & 0 & 0 & 0 & \\
\end{tabular}
\caption{Truth table of $f_{\rm HA+FA}$. 
There are 16 inputs and seven different outputs. 
The truth values of internal Boolean variables, $z_1'$, $c_0$, $c_1'$, and $c_1''$, and the number of carries transmitting 1, $c_0+c_1'+c_1''$, are also presented. }
\label{tab:full_truth_table_two_binary_digit_sum}
\end{table}

An $N$-input and $N$-output Boolean function is calculated by an $N$-token Brownian circuit. 
The rest of this section will explain the circuit design and the rate matrix describing transitions among multi-token states. 
In Sec.~\ref{sec:modules_circuits} we explain the 2-input 2-output logic gate modules and the Doi-Peliti Hamiltonians of circuits. 
The Doi-Peliti Hamiltonian contains two-body interactions associated with CJoins. 
The circuit is designed so that a unique solution becomes a `bright state' reachable from a given input state, while wrong solutions are unreachable `dark states.' 
In Sec.~\ref{sec:reachability_graph}, we analyze the reachability among multi-token logical states. 
In Sec.~\ref{sec:trm_tb_b_c}, we sketch how to construct the transition rate matrix of token-based Brownian circuits, relegating details to Appendices~\ref{sec:Ser_num_for_fin_gra_mul_tok_sta} and \ref{sec:transition_rate_matrix}.

\subsection{2-input 2-output logic gate modules and circuits}
\label{sec:modules_circuits}

The basic constituent is a circuit module that is a 2-input 2-output logic gate $g$ realizing a Boolean function, $f_g:\{0,1\}^2 \to \{0,1 \}^2$, 
\begin{align*}
f_{g}(\alpha_g^{\rm in},\beta_g^{\rm in})=(\alpha_g^{\rm out},\beta_g^{\rm out}) \, .
\end{align*}
Figure \ref{fig:blockdiag2in2out_module} (a) shows the block diagram~\cite{block_diagram}. 
A node (box) represents the logic gate $g$. 
Two directed edges representing horizontal wires, which have as the head (tail) node $g$, carry the truth values of the input (output) Boolean variables $\alpha_g^{\rm in \, (out)}$ and $\beta_g^{\rm in \, (out)}$. 
In the block diagram, we abuse notation and label each directed edge with its Boolean variable. 
Since the three circuit elements conserve the number of tokens, the number of wires is the same at the input and output sides. 

The function accepts all possible inputs, and thus, the size of the domain is $|\{0,1\}^2|=2^2$. 
If the Boolean function is not injective, the image is smaller than the domain, $f(\{0,1\}^2) \subset \{0,1\}^2 $. 
The size of a fiber, as defined in~\cite{set_theory_1}, 
\begin{align}
|f_{g}^{-1}(\alpha_g^{\rm out},\beta_g^{\rm out})|
\end{align}
is compatible with the number of paths merging in the corresponding step of the reachability graph, as we will see in Sec.~\ref{sec:reachability_graph}. 

Figure \ref{fig:blockdiag2in2out_module} (b) shows an example of the circuit diagram of the 2-input 2-output logic gate $g$. 
The truth value of a Boolean variable is encoded in the dual-rail encoding scheme~\cite{Lee2010}. 
Namely, for each Boolean variable, two wires are prepared to express two different truth values, $0$ and $1$. 
On the right and bottom (left and top) sides, we attach four input (four output) terminals corresponding to the two truth values of two input (two output) Boolean variables. 
To each wire, we assign a logical state, the Boolean variable, and its truth value denoted by a subscript.
For example, ${\alpha_{g}^{\rm in} }_{0(1)}$, expresses the Boolean variable and its truth value as ${\alpha_{g}^{\rm in} }=0(1)$. 

In fact, Fig.~\ref{fig:blockdiag2in2out_module} (b) is the circuit diagram of HA,
whereby numbers assigned to wires are indexes of places inside each cell. 
The input terminals are labeled as $x_{0(1)}$ (${\alpha^{\rm in}_g}_{0(1)}$) and $y_{0(1)}$ (${\beta^{\rm in}_g}_{0(1)}$). 
The output terminals for the sum and the carry are labeled as $z_{0(1)}$ (${\alpha^{\rm out}_g}_{0(1)}$) and $c_{0(1)}$ (${\beta^{\rm out}_g}_{0(1)}$). 
For example, when the truth values of Boolean input variables are $x=0$ and $y=0$, two tokens are injected from input terminals of cells $x_{0}$ and $y_{0}$.

The 2-input 2-output logic gate $g$ consists of four CJoins corresponding to $2^2=4$ different inputs. 
Each input cell contains one Hub, which enables a token in the cell ${\alpha^{\rm in}_{g} }_{0(1)}$ to interact with a token in the cell ${\beta^{\rm in}_{g} }_{0}$ or ${\beta^{\rm in}_{g} }_{1}$ by a CJoin and visa versa. 
On the other hand, the size of an output cell depends on preimages~\cite{set_theory_1}. 
Let the set of outputs as tuples with $\alpha_g^{\rm out}$ ($\beta_g^{\rm out}$) be $(\alpha_g^{\rm out},\cup)= \{ (\alpha_g^{\rm out},0), (\alpha_g^{\rm out},1) \} \cap f_g(\{ 0,1\}^2)$ ($(\cup,\beta_g^{\rm out}) = \{ (0,\beta_g^{\rm out}), (1,\beta_g^{\rm out}) \} \cap f_g(\{ 0,1\}^2)$). 
Then the output cell ${\alpha_g^{\rm out}}_*$ (${\beta_g^{\rm out}}_*$), where the wildcard $*$ stands for 0 or 1, consists of a total of
\begin{align}
\left |f_{g}^{-1}(\alpha_g^{\rm out},\cup) \right|-1 \, \biggl( \left |f_{g}^{-1}(\cup,\beta_g^{\rm out}) \right|-1 \biggl) \,  \label{eqn:nhub_output_cell}
\end{align}
Hubs connected in series. 
More explicitly, the preimages~\cite{set_theory_1} are, 
\begin{align}
f_{g}^{-1}(\alpha_g^{\rm out},\cup ) =& \bigcup_{ \substack{ \beta_g^{\rm out} \in \{ 0,1\}  \\ \, \wedge \, (\alpha_g^{\rm out},\beta_g^{\rm out}) \in f_g(\{ 0,1\}^2) } } f_{g}^{-1}(\alpha_g^{\rm out},\beta_g^{\rm out}) \,  \nonumber \\
f_{g}^{-1}(\cup,\beta_g^{\rm out}) =& \bigcup_{ \substack{ \alpha_g^{\rm out} \in \{ 0,1\} \\ \, \wedge \, (\alpha_g^{\rm out},\beta_g^{\rm out}) \in f_g(\{ 0,1\}^2) } } f_{g}^{-1}(\alpha_g^{\rm out},\beta_g^{\rm out}) \, .
\label{eqn:preimages}
\end{align}

The output-cell size quantifies the resolution of a signal carried by a token. 
In this sense, the token-based Brownian circuit meets the physical prerequisite, i.e., \emph{the amount of information that can be encoded in the state of a finite system is bounded}~\cite{Fredkin1982}. 
Since the number of Hubs is roughly the size of the cell [see Eq.~(\ref{eqn:size_logical_state})], it affects the computation time, as we discuss in Sec.~\ref{sec:cscs}.

One can connect the input wires of a gate module to the output wires of other gate modules. 
In this way, an $N$-input $N$-output logic gate realizing a Boolean function, $f:\{0,1\}^N \to \{0,1 \}^N$ where $N \in {\mathbb Z}^{+}$, is constructed. 
By definition, the size of the domain is $|\{0,1\}^N|=2^N$. 
The image is equal to or smaller than the domain, $f(\{0,1\}^N) \subseteq \{0,1\}^N$. 
Figure~\ref{fig:blockdiag2in2out_module} (c), shows such a circuit. 
The horizontal direction along wires does not correspond tightly with time, as our circuit is asynchronous. 
Thus, the gate HA can fire before or after the gate HAa fires. 

In Fig.~\ref{fig:blockdiag2in2out_module} (c), there are four internal directed edges, $z_1'$, $c_0$, $c_1'$ and $c_1''$, which are not input or output terminals of the circuit. 
Such an internal directed edge is constructed by connecting an output terminal of one gate module to an input terminal of another gate module. 
For example, an internal directed edge is constructed by connecting $\alpha_g^{\rm out}$ ($\beta_g^{\rm out}$) of the gate $g$ with an input terminal of another gate, $\partial^- \alpha_g^{\rm out}$ ($\partial^- \beta_g^{\rm out}$). 
Such an internal cell consists of a total of
\begin{align}
\left |f_{g}^{-1}(\alpha_g^{\rm out},\cup) \right| \, \biggl( \left |f_{g}^{-1}(\cup,\beta_g^{\rm out}) \right| \biggl) \,  \label{eqn:nhub_inter_cell}
\end{align}
Hubs connected in series.

\subsubsection{Half adder module}

Figure~\ref{fig:blockdiag2in2out_module} (b) shows the circuit diagram of HA. 
Depending on the size of the preimages, as determined based on Eq.~(\ref{eqn:nhub_output_cell}) and Appendix \ref{sec:preima}, the set of output cells $\{z_{0}, z_{1} , c_{0} , c_{1} \}$ is partitioned into, 
${X}_1^{\rm R} = \{ z_{0}, z_{1} \}$, 
${X}_2^{\rm R} = \{ c_{0} \}$, and 
${X}_0^{\rm R} = \{ c_{1} \}$. 
The Doi-Peliti Hamiltonian is, 
\begin{align}
\hat{H}_{\rm HA} =& \sum_{a \in {X}_1 } \hat{H}_{1}(a) + \sum_{n=0}^2 \sum_{a \in {X}_n^{\rm R} } \hat{H}_{n}^{\rm R}(a) + \hat{V}_{\rm HA}(z,c|x,y) \, , \label{eqn:DPHamiltonianHA}
\end{align}
where the set of input cells is, ${X}_1 = \{ x_{0}, x_{1}, y_{0}, y_{1} \}$. 
The CJoin part is, 
\begin{align*}
\hat{V}_{\rm HA}(z,c|x,y) =& 
\hat{V}_{1,1||P(x_{0})|-2,|P(y_{0})|-1}(z_{0},c_{0}|x_{0},y_{0}) \nonumber \\ & + \hat{V}_{1,0| |P(y_{1})|-2,|P(x_{0})|-1}(z_{1},c_{0}|y_{1},x_{0}) \nonumber \\ & + \hat{V}_{0,0| |P(x_{1})|-2,|P(y_{1})|-1}(z_{0},c_{1}|x_{1},y_{1}) \nonumber \\ & + \hat{V}_{0,3| |P(y_{0})|-2,|P(x_{1})|-1}(z_{1},c_{0}|y_{0},x_{1}) \, . 
\end{align*}
Although the Doi-Peliti Hamiltonian (\ref{eqn:DPHamiltonianHA}) is compact, the SPN is large: in total, there are $|P|=25$ places and $|T|=51$ transitions. 

\begin{figure}[ht]
\begin{center}
\includegraphics[width=0.8 \columnwidth]{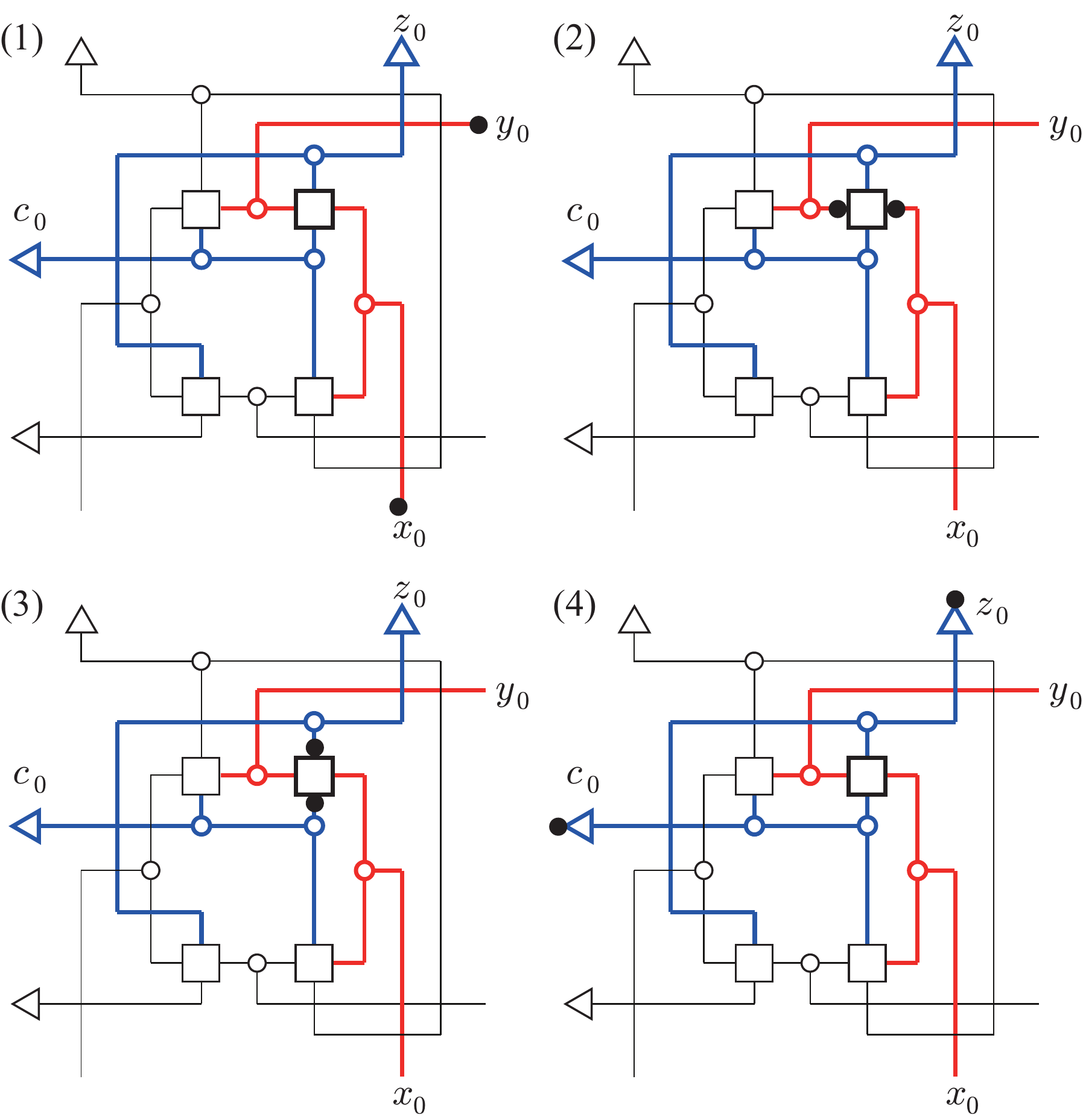}
\caption{
Computation process of $0+0=(0)0$: ($\mathrm{i}$) One token each is input at two input terminals of cells $x_{0}$ and $y_{0}$ [panel (1)]. 
The initial state is, 
$|x_{0}, y_{0} \rangle |0,0 \rangle  = \hat{a}_{x_0,0}^\dagger \hat{a}_{y_0,0}^\dagger |0 \rangle$. 
($\mathrm{ii}$) Each token randomly jumps between places (wires) in the cells $x_0$ and $y_0$ (in red). 
At some point, the two tokens simultaneously reach the right and left sides of the CJoin [box with a thick line in panel (2)]. 
($\mathrm{iii}$) The CJoin fires with the forward rate $\gamma^+$. 
If this happens, one token each is emitted from the top and the bottom sides [panel (3)]. 
($\mathrm{iv}$) Each token randomly jumps between places of the cells $c_0$ and $z_0$ (in blue). 
If the two tokens come back to the configuration in panel (3), the CJoin fires with the backward rate $\gamma^-$, causing the two tokens to return to the state in panel (2).  
($\mathrm{v}$) If one token comes to the output terminal of the cell, $c_{0}$ or $z_{0}$, it is detected by a Ratchet. 
When Ratchets detect both tokens at the two output terminals, the final state, 
$|z_0, c_0 \rangle |2,4 \rangle  = \hat{a}_{z_0,2}^\dagger \hat{a}_{c_0,4}^\dagger |0 \rangle$ 
is reached. 
This is the end of the computation, and we obtain the solution, the sum $z=0$ and the carry $c=0$.
}
\label{fig:halfadder4}
\end{center}
\end{figure}

Figure~\ref{fig:halfadder4} illustrates the computation process of $0+0=(0)0$. 
In the computation process, only one CJoin fires, and a limited number of two-token states participate.

\subsubsection{DOR gate module}
\label{sec:DOR}

The logical OR has two Boolean input variables, $x$, and $y$,
and one Boolean output variable $z$, 
\begin{align*}
x \vee y = z \, .
\end{align*}
It is realized by a $2$-input $1$-output logic gate. 
To construct it from token-conserving circuit elements, we introduce a $2$-input $2$-output logic gate and then `embed' the logical OR. 
One such a logical gate is called the duplicate-output OR (DOR) gate~\cite{Lee2010}. 
The DOR gate outputs $z$ and its copy $\tilde{z}$. 
Formally, it is a 2-input and 2-output Boolean function, 
$f_{\rm DOR}:\{0,1\}^2 \to \{0,1\}^2$, $f_{\rm DOR}(x,y)=(z,\tilde{z})$. 
Its truth table is given in Table~\ref{tab:dor}.

\begin{table}[h]
\begin{tabular}{cc|cc} 
$x$  & $y$ & $z$ & $\tilde{z}$ \\ 
\hline 
0  & 0 & 0 & 0 \\ 
\rowcolor{gray!30}%
0  & 1 & 1 & 1 \\ 
\rowcolor{gray!30}%
1  & 0 & 1 & 1 \\ 
\rowcolor{gray!30}%
1  & 1 & 1 & 1 \\
\end{tabular}
\caption{Truth table of $f_{\rm DOR}$.}
\label{tab:dor}
\end{table}

\begin{figure*}[ht]
\begin{center}
\includegraphics[width=1.8 \columnwidth]{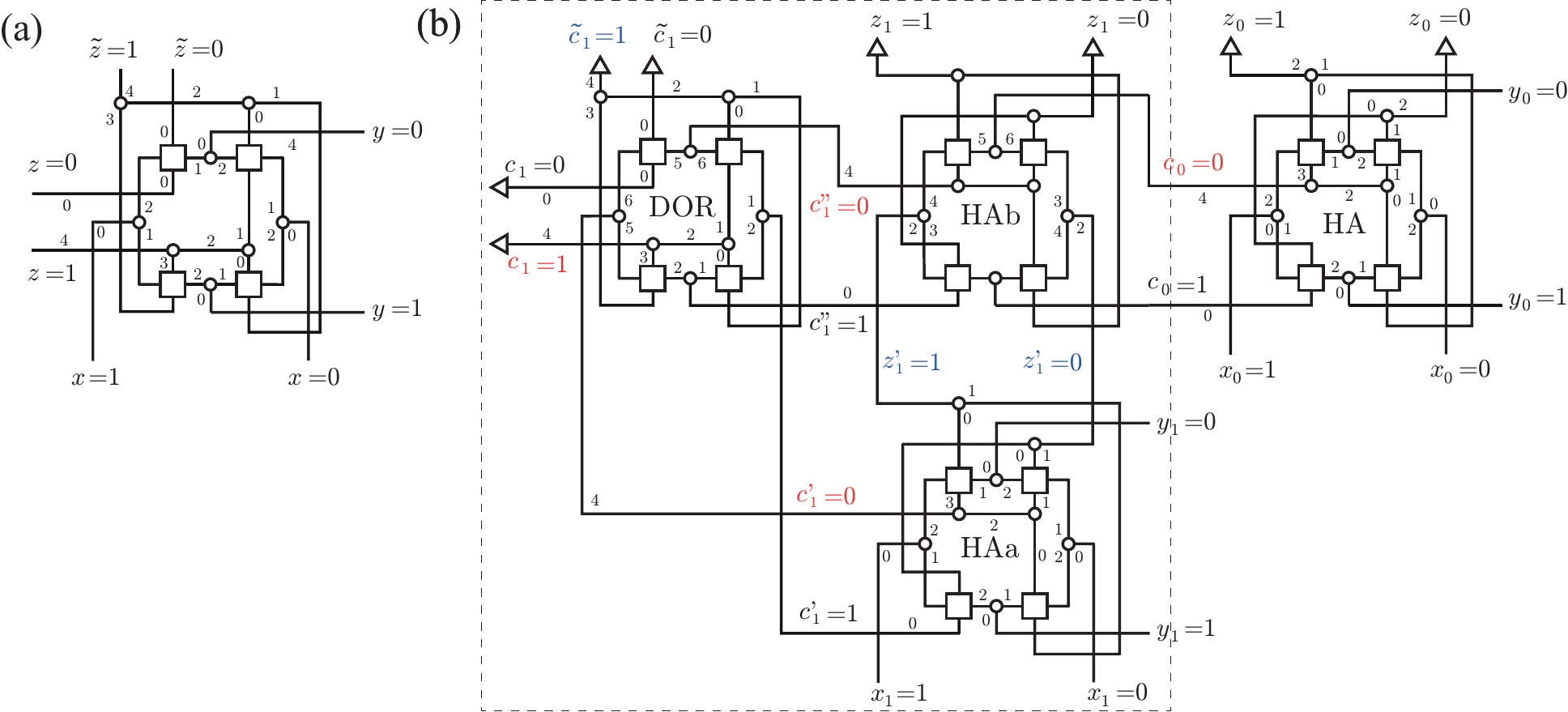}
\caption{
(a) Circuit diagram of a duplicate-output OR gate. 
(b) Circuit diagram of a cascade of a half-adder and a full-adder [Fig.~\ref{fig:blockdiag2in2out_module} (c)]. 
The dashed box indicates the full-adder module. 
Numbers assigned to wires are indexes of places inside cells. 
}
\label{fig:DOR_HAFA}
\end{center}
\end{figure*}

Figure \ref{fig:DOR_HAFA} (a) shows the circuit diagram of a DOR. 
The set of output cells $\{z_{0}, z_{1} , \tilde{z}_{0} , \tilde{z}_{1} \}$ is partitioned into, 
${X}_0 = \{ z_{0}, \tilde{z}_{0} \}$ and ${X}_2 = \{ z_{1}, \tilde{z}_{1} \}$. 
The Doi-Peliti Hamiltonian is, 
\begin{align}
\hat{H}_{\rm DOR} =& \sum_{n=0}^2 \sum_{a \in {X}_n } \hat{H}_{n}(a) + \hat{V}_{\rm DOR}(z,\tilde{z}|x,y) \, , 
\end{align}
where the set of input cells is ${X}_1 = \{ x_{0}, x_{1}, y_{0}, y_{1} \}$ and, 
\begin{align}
\hat{V}_{\rm DOR}(z,\tilde{z}|x,y) =& \hat{V}_{1,0| |P(x_{1})|-2,|P(y_{0})|-1}(z_{1},\tilde{z}_{1}|x_{1},y_{0}) \nonumber \\ & + \hat{V}_{0,1| |P(y_{1})|-2,|P(x_{1})|-1}(z_{1},\tilde{z}_{1}|y_{1},x_{1}) \nonumber \\ & + \hat{V}_{3,3| |P(x_{0})|-2,|P(y_{1})|-1}(z_{1},\tilde{z}_{1}|x_{0},y_{1}) \nonumber \\ & + \hat{V}_{0,0| |P(y_{0})|-2, |P(x_{0})|-1}(z_{0},\tilde{z}_{0}|y_{0},x_{0}) \, . 
\end{align}
The redundant token encoding the output $\tilde{z}$ is `garbage' and is a cost of computation.

\subsubsection{Cascade connection of half-adder and full-adder}

Figure~\ref{fig:blockdiag2in2out_module} (c) shows the block diagram of the cascade connection of a HA and a FA. 
The leftmost HA module (box) adds the second digits, $x_0$ and $y_0$, and outputs the sum $z_0$ and the carry $c_0$. 
The FA (dashed box) adds the first digits, $x_1$ and $y_1$, and the carry input $c_0$. 
It outputs the sum $z_1$, the carry output $c_1$ and its copy $\tilde{c}_1$. 
Table \ref{tab:full_truth_table_two_binary_digit_sum} indicates that there are $2^4=16$ different inputs and seven different outputs. 
It also includes the truth values of internal Boolean variables, $z_1'$, $c_0$, $c_1'$, and $c_1''$. 

Figure~\ref{fig:DOR_HAFA} (b) shows the circuit diagram corresponding to Fig.~\ref{fig:blockdiag2in2out_module} (c). 
The dashed box indicates the FA module. 
The Boolean variables $x_1$ and $x_0$ in the string $x_1 x_0$ indicate a two-digit binary number's first and second digits. 
To represent a truth value, we introduce another subscript separated by a comma: e.g. $x_{0}=0$ is expressed as $x_{0,0}$
and $x_{0}=1$ is expressed as $x_{0,1}$.
The set of output cells is partitioned into, 
${X}_0^{\rm R} = \{ c_{1,0}, \tilde{c}_{1,0} \}$. 
${X}_1^{\rm R} = \{ z_{0,1}, z_{0,1}, z_{1,1}, z_{1,1} \}$, 
and ${X}_2^{\rm R} = \{ c_{1,1}, \tilde{c}_{1,1} \}$. 
The set of internal cells is partitioned into,  
${X}_{1 , \, {\rm int}} = \{ c_{0,1}, c'_{1,1}, c''_{1,1} \}$, 
${X}_2 = \{ z'_{1,0}, z'_{1,1} \}$, and 
${X}_3 = \{ c_{0,0}, c'_{1,0}, c''_{1,0} \}$.
The Doi-Peliti Hamiltonian is, 
\begin{align}
\hat{H}_{\rm HA+FA} =&  \sum_{n=1}^3 \sum_{a \in {X}_n } \hat{H}_{n}(a) + \sum_{n=0}^2 \sum_{a \in {X}_n^{\rm R} } \hat{H}_{n}^{\rm R}(a)
\nonumber \\ & + \hat{V}_{\rm HA}(z_0,c_0|x_0,y_0) + \hat{V}_{\rm HA}(z_1,c''_1|z'_1,c_0)
\nonumber \\ & + \hat{V}_{\rm HA}(z'_1,c'_1| x_1,y_1) + \hat{V}_{\rm DOR}(c_1,\tilde{c}_1| c'_1,c''_1) \, , \label{eqn:ham_hafa}
\end{align}
where 
$X_1 = \{ x_{0,0}, x_{0,1}, y_{0,0}, y_{0,1}, x_{1,0}, x_{1,1}, y_{1,0}, y_{1,1} \} \cup X_{1, \, {\rm int}}$. 
The corresponding SPN consists of $|P|=90$ places and $|T|=214$ transitions.

\subsection{Reachability graph}
\label{sec:reachability_graph}

The {\it reachability} is a fundamental basis for the dynamical properties of Petri nets~\cite{Murata1989}. 
The reachable states are visualized by the reachability graph, whose nodes are states and whose arcs describe possible transitions. 
Since all places inside a particular cell are reachable, it is sufficient to analyze the reachability among the logical states and neglect the indexes of fine-grained places. 
In other words, we perform the coarse-graining for the Doi-Peliti Hamiltonian as, 
$\hat{a}_{x,s} \to \hat{a}_{x}$. 
In the Petri net description, the coarse-graining corresponds to replacing each cell with a single place while keeping arcs connected to transitions of CJoins. 
Such a coarse-grained graph is obtained by short-circuiting all arcs of each Hub in a way that removes all transitions of the Hub.

\subsubsection{Addition of two 1-digit binary numbers: 2-token circuit}

After the coarse-graining, the Doi-Peliti Hamiltonians of the Hubs vanish, and only the CJoin part remains: 
\begin{align}
\hat{ \tilde{H} }_{\rm HA} =& \hat{V}(z_{0},c_{0}|x_{0},y_{0}) + \hat{V}(z_{1},c_{0}|y_{1},x_{0}) \nonumber \\ & + \hat{V}(z_{0},c_{1}|x_{1},y_{1}) + \hat{V}(z_{1},c_{0}|y_{0},x_{1}) \, ,  \label{cg_ha}
\end{align}
where $V(x,y|u,v)$ is the CJoin Hamiltonian (\ref{ham_c_join}) without indexes of fine-grained places. 

The reachability graph is constructed systematically~\cite{Murata1989} by starting from the final 2-token logical states, 
\begin{align*} 
| z_{0},c_{0} \rangle \, ,  | z_{1},c_{0} \rangle \, ,  | z_{0},c_{1} \rangle \, ,
\end{align*}
each of which becomes the {\it root} of the graph by repeatedly applying the Doi-Peliti Hamiltonian (\ref{cg_ha}) and constructing descendant 2-token logical states until no descendant can be constructed.
In the present paper, we denote the final state as the root and the initial states as {\it leaves}.

\begin{figure}[ht]
\begin{center}
\includegraphics[width=0.9 \columnwidth]{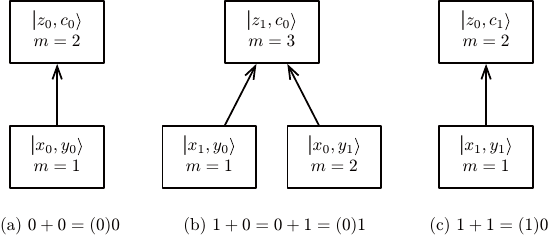}
\caption{Reachability graphs of HA. 
Nodes and arcs correspond to 2-token logical states and transitions by CJoins, respectively. 
Three different reachability graphs correspond to three different outputs in Table~\ref{tab:half_adder}. 
The function $f_{\rm HA}$ is non-injective, as there is a merging of paths in panel (b).
}
\label{fig:reachabilityHA}
\end{center}
\end{figure}

Figure~\ref{fig:reachabilityHA} shows the three reachability graphs. 
Each node (box) represents a 2-token logical state. 
The bottom nodes are initial states, and the top nodes are final states. 
The computation proceeds from the bottom layer to the top layer. 
Each directed edge indicates a possible forward transition between 2-token logical states induced by the firing of a CJoin. 

For each reachability graph, we assign serial numbers $m$ and express the multi-token logical states as $|m \rangle$. 
The sets of nodes ${\mathcal R}$ and arcs ${\mathcal E}$ are as follows:

For Fig.~\ref{fig:reachabilityHA} (a):
\begin{align*}
{\mathcal R}(| 1 \rangle) =& \{ | 1 \rangle \, , |2 \rangle \} \, , 
\;\;\;\;
{\mathcal E}(| 1 \rangle) = \{ ( | 2 \rangle \leftarrow |1 \rangle ) \} \, , 
\end{align*}
where $( |1 \rangle,  |2 \rangle ) = ( |x_{0},y_{0} \rangle, |z_{0}, c_{0} \rangle)$. 

For Fig.~\ref{fig:reachabilityHA} (b):
\begin{align*}
{\mathcal R}(| 1 \rangle) =& {\mathcal R}(| 2 \rangle) = \{ | 1 \rangle \, , | 2 \rangle \, , | 3 \rangle \} \, , \\
{\mathcal E}(| 1 \rangle) =& {\mathcal E}(| 2 \rangle) = \{ ( |3 \rangle \leftarrow |1 \rangle ),  ( | 3 \rangle \leftarrow |2 \rangle ) \} \, , 
\end{align*}
where $( |1 \rangle, |2 \rangle, |3 \rangle ) = ( |x_{1},y_{0} \rangle, |x_{0},y_{1} \rangle, |z_{1},c_{0} \rangle)$. 

For Fig.~\ref{fig:reachabilityHA} (c):
\begin{align*}
{\mathcal R}(| 1 \rangle) =& \{ | 1 \rangle \, , | 2 \rangle \} \, , 
\;\;\;\;
{\mathcal E}(| 1 \rangle) = \{ ( | 2 \rangle \leftarrow |1 \rangle ) \} \, , 
\end{align*}
where $( |1 \rangle,  |2 \rangle ) = ( |x_{1},y_{1} \rangle, |z_{0},c_{1} \rangle)$. 

For panels (a) and (c) of Fig.~\ref{fig:reachabilityHA}, there is a one-to-one correspondence between the initial 2-token input logical state and the final 2-token output logical state, while in panel (b), the paths from two input states merge into one path to the output state. 
This is because $f_{\rm HA}$ is not an injective function~\cite{Landauer1961,Bennett1982}.

\subsubsection{Addition of 2-digit binary numbers : 4-token circuit}

\begin{figure*}[ht]
\begin{center}
\includegraphics[width=1.8 \columnwidth]{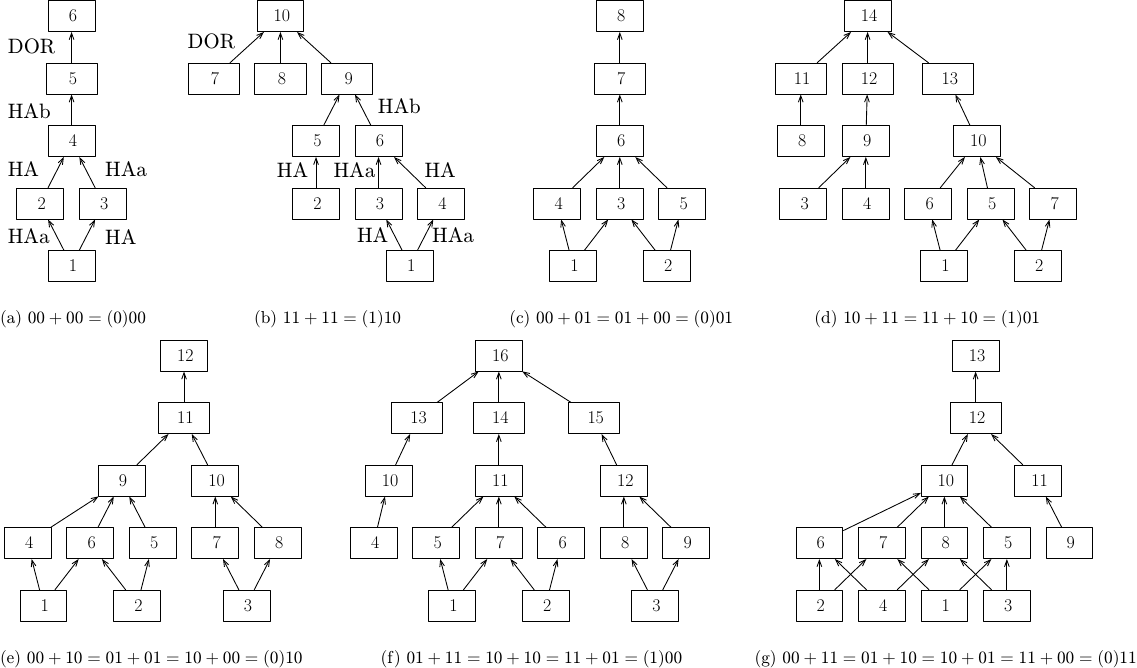}
\caption{Reachability graphs for the circuit in Fig.~\ref{fig:DOR_HAFA} (b). 
In each node, a serial number is assigned to a 4-token logical state $m$. 
The explicit expressions of 4-token logical states are summarized in the tables in Fig.~\ref{tab:HAFAtable} in Appendix~\ref{sec:Ser_num_for_fin_gra_mul_tok_sta}. 
There are seven distinct reachability graphs corresponding to the seven distinct outputs in Table~\ref{tab:full_truth_table_two_binary_digit_sum}. 
}
\label{fig:reachabilityHAFA}
\end{center}
\end{figure*}

Figure~\ref{fig:reachabilityHAFA} shows the reachability graphs for the circuit Fig.~\ref{fig:DOR_HAFA} (b) or Eq.~(\ref{eqn:ham_hafa}). 
There are seven distinct reachability graphs corresponding to seven distinct outputs in Table~\ref{tab:full_truth_table_two_binary_digit_sum}. 
We assign serial numbers to reachable 4-token logical states of each reachability graph. 
The corresponding 4-token logical states are summarized in the tables in Fig.~\ref{tab:HAFAtable} in Appendix~\ref{sec:Ser_num_for_fin_gra_mul_tok_sta}. 


The length of each path from the initial state, represented by a node at the bottom layer, to the final state, represented by a node at the top layer, is 4. 
For example, in Fig.~\ref{fig:reachabilityHAFA} (a), one such a path is $|1 \rangle \to |2 \rangle \to |4 \rangle \to |5 \rangle \to |6 \rangle$, 
as there are four forward transitions caused by firings of four CJoins in three HAs and one DOR [see Fig.~\ref{fig:blockdiag2in2out_module} (c)]. 
The following Boolean functions are computed in the course of the path: 
\begin{align*}
\begin{array}{ll}
{\rm 4th \; step} & f_{\rm DOR}(c_1',c_1'')=(c_1,\tilde{c}_1) \, , \\
{\rm 3rd \; step} & f_{\rm HA}(z_1',c_0)=(z_1,c_1'') \, , \\
{\rm 2nd \; step} & f_{\rm HA}(x_1,y_1)=(z'_1,{c}'_1)  \, , \\
{\rm 1st \; step} & f_{\rm HA}(x_0,y_0)=(z_0,{c}_0)  \, .
\end{array}
\end{align*}

It is easily verified in Fig.~\ref{fig:reachabilityHAFA} (a) that there are two paths connecting the initial state to the final state in each reachability graph. 
The two paths are associated with branching and merging in the first and second steps. 
They appear because a CJoin of HA and a CJoin of HAa are concurrent: in the first step, the branching emerges since either HA or HAa in Fig.~\ref{fig:blockdiag2in2out_module} (c) is available. 
If the CJoin of HA fires in the first step, the CJoin of HAa should fire in the second step before the HAb is available, and visa versa. 
In the third and fourth steps, the DOR is available only after the CJoin of HAb fires. 
In this way, the computation process is regulated autonomously.


Some reachability graphs in Fig.~\ref{fig:reachabilityHAFA}, have merging of paths at the fourth and third steps, which originates from the non-injectivity of DOR and HAb. 
For example, the reachability graphs (b), (d), and (f) possess the merging of three paths at the 4th step. 
In these reachability graphs, the output 2-tuple of DOR is $(c_1,\tilde{c}_1)=(1,1)$. 
The number of paths is the size of the fiber, 
\begin{align*}
\left | f_{\rm DOR}^{-1}(1,1) \right| = \left |\left \{ (0,1), (1,0), (1,1) \right \} \right| =3 \, .
\end{align*}
Likewise, in the reachability graphs (b), (e), and (g) two paths are merged at the 3rd step. 
In these reachability graphs, the output 2-tuple of HAb is $(z_1,c''_1)=(1,0)$, whose fiber size is 
\begin{align*}
\left | f_{\rm HA}^{-1}(1,0) \right| = \left | \{ (0,1) , (1,0) \} \right| =2 \, .
\end{align*}

Additionally, in panels (b), (d), (f), and (g), there exist nodes without predecessors consistent with any computation path (so-called `garden-of-Eden' states~\cite{Bennett1982}). 
For example, in panel (b), $| 2 \rangle$, $| 7 \rangle$ and $| 8 \rangle$ are such states. 
In the fourth step of panel (b), the merging of paths to
\begin{align*}
|10 \rangle =|z_{0,0},z_{1,1},c_{1,1},\tilde{c}_{1,1} \rangle \,
\end{align*}
occurs from the 4-token logical states satisfying, 
$(c'_1,c''_1) \in f^{-1}_{\rm DOR}(1,1)$: 
\begin{align*}
|7 \rangle =& |z_{0,0},z_{1,1},c'_{1,1},c''_{1,1} \rangle \, , \\
|8 \rangle =& |z_{0,0},z_{1,1},c'_{1,0},c''_{1,1} \rangle \, , \\
|9 \rangle =& |z_{0,0},z_{1,1},c'_{1,1},c''_{1,0} \rangle \, . 
\end{align*}
Among these three states, $|7 \rangle$ and $|8 \rangle$ are not on the intended computation path since $(z_1,c''_1)=(1,1) \notin f_{\rm HA}( \{0,1\}^2)$ is not consistent with the output of HAb. 
In the third step of the panel (b), the merging of paths to $|9 \rangle$ by firing HAb originates from the 4-token logical states satisfying $(c_0,z'_1) \in f^{-1}_{\rm HA}(1,0)$: 
\begin{align*}
|5 \rangle =|z_{0,0},z'_{1,1},c_{0,0},c'_{1,1} \rangle \, , \\
|6 \rangle =|z_{0,0},z'_{1,0},c_{0,1},c'_{1,1} \rangle \, . 
\end{align*}
However, the state $|5 \rangle$ is not on the intended computation path, since $(z'_1,c'_1)=(1,1) \notin f_{\rm HA}( \{0,1\}^2)$ is not consistent with the output of HAa. 
On the other hand, $(z_0,c_0)=(0,0) = f_{\rm HA}(0,0)$ is consistent with the output of HA. 
Therefore, $|5 \rangle$ has a predecessor, 
\begin{align*}
|2 \rangle = |x_{0,0},y_{0,0}, z'_{1,1},c'_{1,1} \rangle \, . 
\end{align*}
In general, the extraneous predecessors slow down the computation time due to temporary detours via the extraneous paths~\cite{Bennett1982}.

\subsection{Transition rate matrix}
\label{sec:trm_tb_b_c}

The full transition rate matrix is decomposed into independent blocks corresponding to different reachability graphs. 
In each reachability graph, we assign serial numbers (\ref{eqn:serial_number}) to all reachable fine-grained multi-token states satisfying the condition that the SPN is safe and token-conserving (Appendix~\ref{sec:Ser_num_for_fin_gra_mul_tok_sta}). 
Details on how to construct the transition rate matrix are relegated to Appendix~\ref{sec:transition_rate_matrix}. 
The transition rate matrix is further divided into submatrices according to multi-token logical-states, $|m \rangle$, i.e., nodes of the reachability graphs in Figs.~\ref{fig:reachabilityHA} and \ref{fig:reachabilityHAFA}. 
The block-diagonal components (\ref{eqn:block_diagonal}) represent transitions of tokens inside cells. 
For an $N$-token circuit, the size of the $m$th block-diagonal submatrix is $NU_{N}^{(m)} \times NU_{N}^{(m)}$ [see Eq.~(\ref{eqn:num_l_tokenstates_of_mth_state})]. 
The block-off-diagonal components (\ref{eqn:tra_rat_blo_off_f}) and (\ref{eqn:tra_rat_blo_off_b}) represent transitions between different multi-token logical states induced by CJoins. 
The transition rate matrix is sparse, and its size increases exponentially in the number of tokens.

\section{Computation time distribution}
\label{sec:cal_tim}

In the following, we set $\Gamma=1$. 
To obtain histograms, we construct $10^5$ trajectories by using Gillespie's algorithm (Appendix \ref{sec:Gillespies_algorithm}).

\subsection{Half adder}

\begin{figure}[ht]
\begin{center}
\includegraphics[width=0.9 \columnwidth]{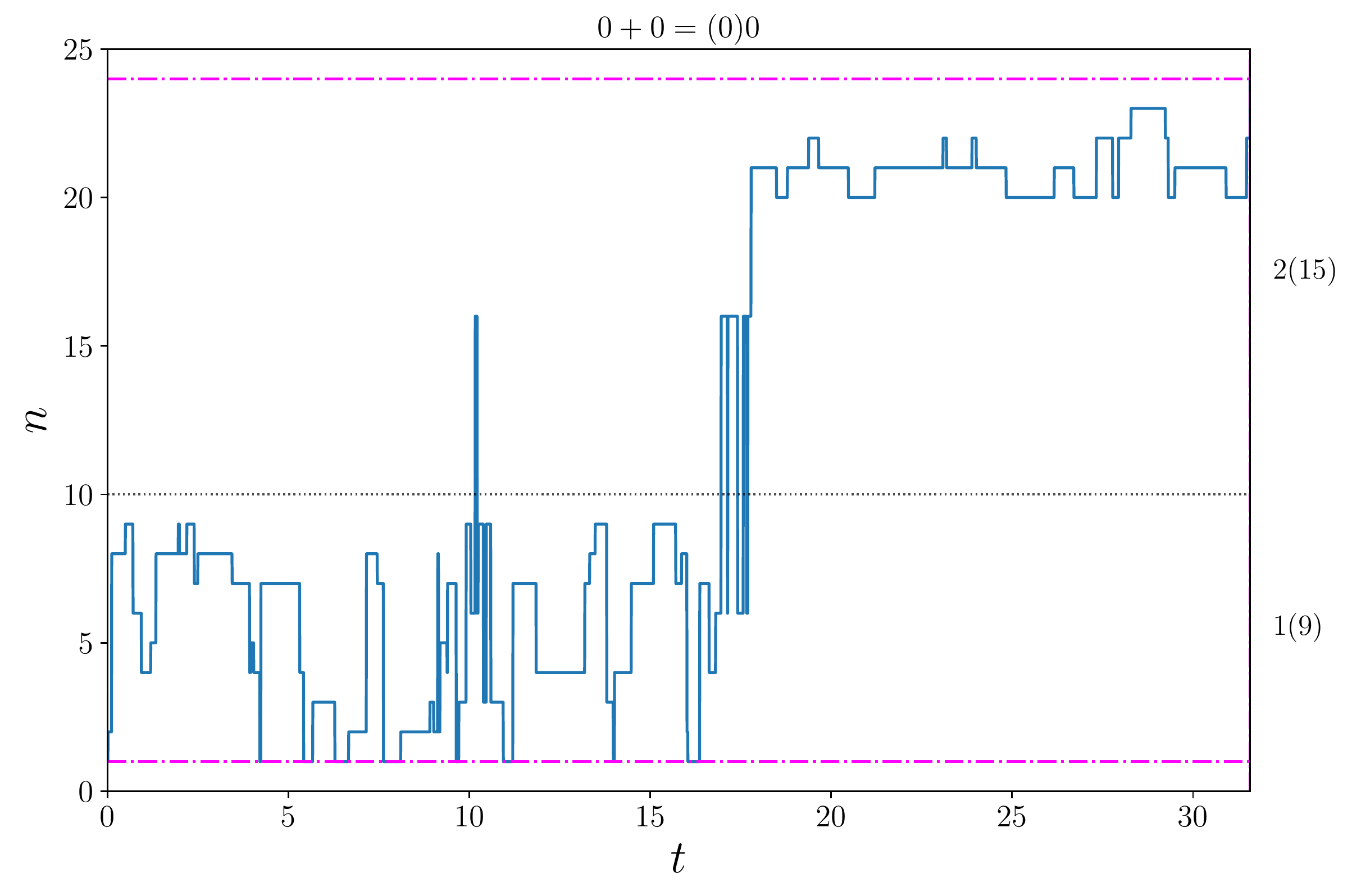}
\caption{Stochastic trajectory for the computation process $0+0=(0)0$ with $\gamma^\pm=10$. 
The left vertical axis indicates the serial numbers of the states, Eq.~(\ref{eqn:serial_number}). 
The horizontal dot-dashed lines indicate the initial and final states. 
The right vertical axis denotes the index of the $2$-token logical states, $|m \rangle$, and the number of $2$-token fine grained states, $NU_{2}^{(m)}$, as $m(NU_{2}^{(m)})$. 
The horizontal dotted lines separate different $2$-token logical states. 
}
\label{fig:lis_dat_HA0p0f10b10_path}
\end{center}
\end{figure}

Figure~\ref{fig:lis_dat_HA0p0f10b10_path} shows a single stochastic trajectory versus time for the computation process $0+0=(0)0$ [Fig.~\ref{fig:reachabilityHA} (a)]. 
The left vertical axis indicates the serial numbers of fine-grained 2-token states (\ref{eqn:serial_number}). 
The CJoin forward and backward transition rates are fixed to be the same, $\gamma^\pm=10$, and thus there is no environment entropy production by the CJoin. 
The horizontal dot-dashed lines indicate the initial and final states, $|n_{\rm i} \rangle =|1 \rangle $ and $|n_{\rm f} \rangle =|24 \rangle$ (Table \ref{tab:ha_initial_final}). 
The right vertical axis denotes the serial numbers of $2$-token logical states, $|m \rangle$, and the number of fine grained states, $NU_{2}^{(m)}$, as $m(NU_{2}^{(m)})$, according to Eq.~(\ref{eqn:num_l_tokenstates_of_mth_state}). 
The horizontal dotted line separates different $2$-token logical states. 
The 2 tokens are in the initial state $|n_{\rm i} \rangle$ at $t=0$. 
Then after $A=92$ jumps, the 2 tokens reach the final state $|n_{\rm f} \rangle$, for the first time at $\tau=31.6$, which is the computation time of this specific trajectory.

\begin{table}[hb]
\begin{tabular}{cccc}
  & $|n_{\rm i} \rangle $ & $ |n_{\rm f} \rangle$ & \\ \cmidrule{1-3} 
$0+0=(0)0$  & $|1 \rangle = |x_{0},y_{0} \rangle |0,0 \rangle$ & $|24 \rangle = |c_{0},z_{0} \rangle |4,2 \rangle$ & (a) \\ 
\rowcolor{gray!30}
$1+0=(0)1$  & $|1 \rangle = |x_{1},y_{0} \rangle |0,0 \rangle$ & & \\ 
\rowcolor{gray!30}
$0+1=(0)1$  & $|10 \rangle = |x_{0},y_{1} \rangle |0,0 \rangle$ & \multirow{-2}{*}{$|33 \rangle = |c_{0},z_{1} \rangle |4,2 \rangle$} & \multirow{-2}{*}{(b)} 
\\ 
$1+1=(1)0$  & $|1 \rangle = |x_{1},y_{1} \rangle |0,0 \rangle$ & $|15 \rangle = |c_{1},z_{0} \rangle |0,2 \rangle$ & (c) \\ 
\end{tabular}
\caption{Initial and final 2-token fine grained-states of the 4 computation processes of $f_{\rm HA}$. }
\label{tab:ha_initial_final}
\end{table}

\begin{figure}[ht]
\begin{center}
\includegraphics[width=0.8 \columnwidth]{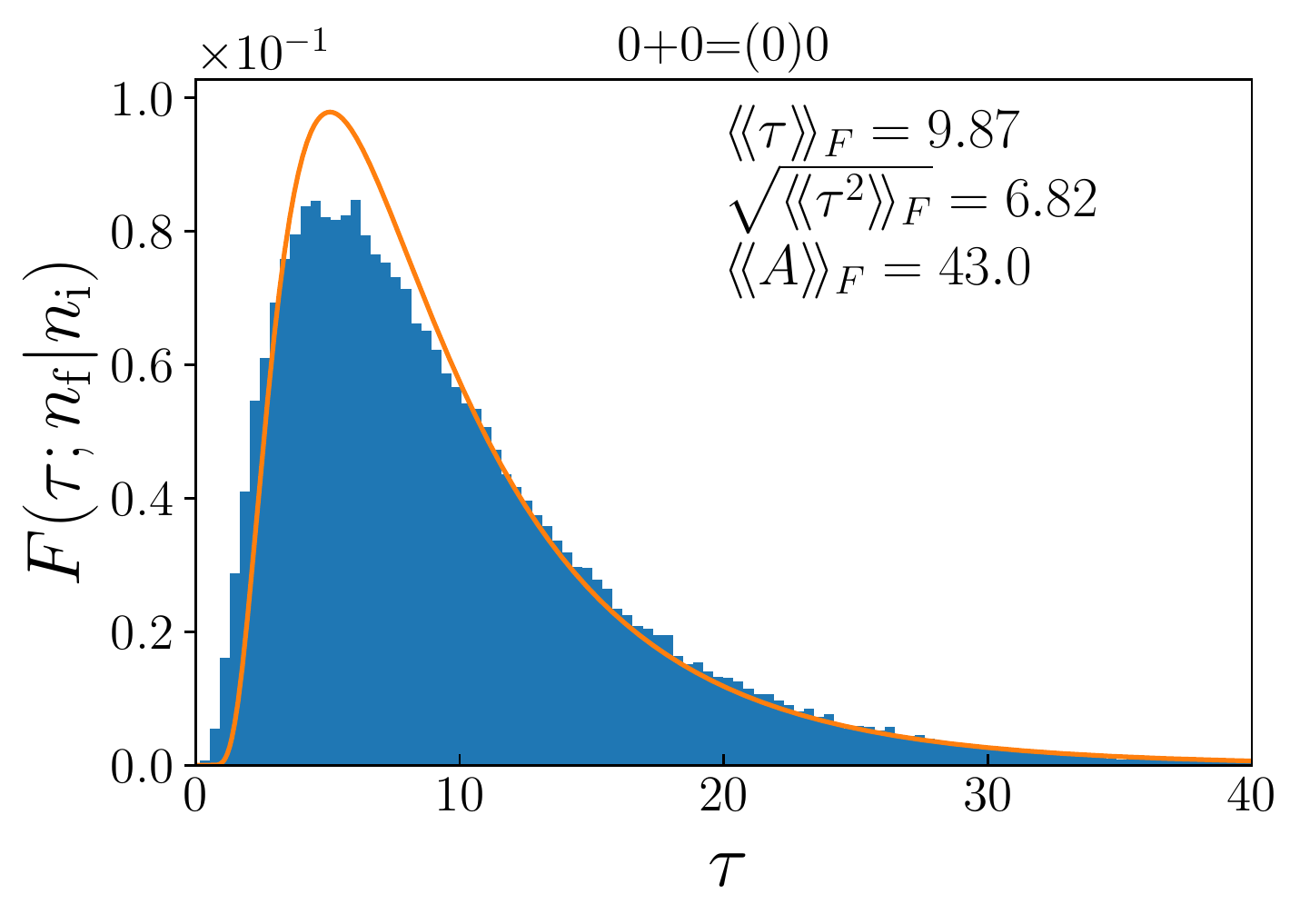}
\caption{Computation time distribution for $0+0=(0)0$ with $\gamma^\pm=10$. 
The solid line indicates the inverse Gaussian distribution (\ref{eqn:inverse_gauss_tau}) with $W=1$. 
}
\label{fig:lis_dat_HA_0p0f10b10_fptd}
\end{center}
\end{figure}

The number of jumps, i.e., the activity $A$, and the first-passage time, i.e., the computation time $\tau$, differ for each sample. 
Figure~\ref{fig:lis_dat_HA_0p0f10b10_fptd} shows the histogram of the computation time over $10^5$ runs. 
The solid line indicates the inverse Gaussian distribution [Eq.~(\ref{eqn:inverse_gauss_tau})] with one reset $W=1$. 
It well approximates the histogram, which implies that output detection drives the system in the non-equilibrium steady-state.

\begin{figure}[ht]
\begin{center}
\includegraphics[width=0.8 \columnwidth]{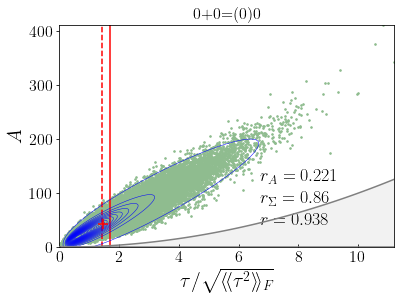}
\caption{Scatter plot of computation time and activity for the computation process $0+0=(0)0$ with $\gamma^\pm=10$. 
Solid curves (in blue) indicate the contour plot of the approximate expression (\ref{eqn:inverse_gauss_tau_a}) with $W=1$. 
The cross (in red) indicates the averages $( \langle \! \langle \tau \rangle \! \rangle_F/\sqrt{\langle \! \langle \tau^2 \rangle \! \rangle_F}, \langle \! \langle A \rangle \! \rangle_F)$.
The gray area depicts the regime above the kinetic bound $r_A > 1$. 
The solid vertical line indicates the mixed bound $\sqrt{ \langle \! \langle \Sigma^{{\rm tot}}_{{\rm bi}} \rangle \! \rangle_F/2 + \langle \! \langle A_{{\rm uni}} \rangle \! \rangle_F }$. 
The dashed vertical line indicates $\sqrt{ \langle \! \langle A_{\rm uni} \rangle \! \rangle_F  } = \sqrt{2}$. 
}
\label{fig:lis_dat_HA_0p0f10b10_tau_a}
\end{center}
\end{figure}

Figure~\ref{fig:lis_dat_HA_0p0f10b10_tau_a} shows the scatter plot of computation time and activity. 
The contour plot shows the approximate joint probability distribution (\ref{eqn:inverse_gauss_tau_a}) with $W=1$. 
The two quantities are approximately linearly correlated, i.e., the correlation coefficient is $r \approx 0.94$. 
The cross indicates the averages $(\langle \! \langle \tau \rangle \! \rangle_F/\sqrt{\langle \! \langle \tau^2 \rangle \! \rangle_F},\langle \! \langle A \rangle \! \rangle_F) = (S/N,\langle \! \langle A \rangle \! \rangle_F)$. 
The shaded area satisfies 
$\tau / \sqrt{ \langle \! \langle \tau^2 \rangle \! \rangle_F } \geq \sqrt{A}$. 
The cross is above the shaded area, which means that the SNR is below the kinetic bound $r_A < 1$. 
The solid vertical line indicates the mixed bound, 
\begin{align*}
\sqrt{ \langle \! \langle \Sigma^{{\rm tot}}_{{\rm bi}} \rangle \! \rangle_F/2 + \langle \! \langle A_{{\rm uni}} \rangle \! \rangle_F } \, ,
\end{align*}
see Eq.~(\ref{eqn:tur_mix}). 
The cross is on the left side of the line, which means that the SNR is below the mixed bound $r_\Sigma < 1$. 
The mixed bound is tighter than the kinetic bound $r_A \approx 0.2 < r_{\Sigma}  \approx 0.9 <1$. 
The dashed vertical line indicates $\sqrt{ \langle \! \langle A_{\rm uni} \rangle \! \rangle_F  } = \sqrt{2}$, which provides a reasonable estimate of the SNR.

\begin{figure}[ht]
\begin{center}
\includegraphics[width=0.9 \columnwidth]{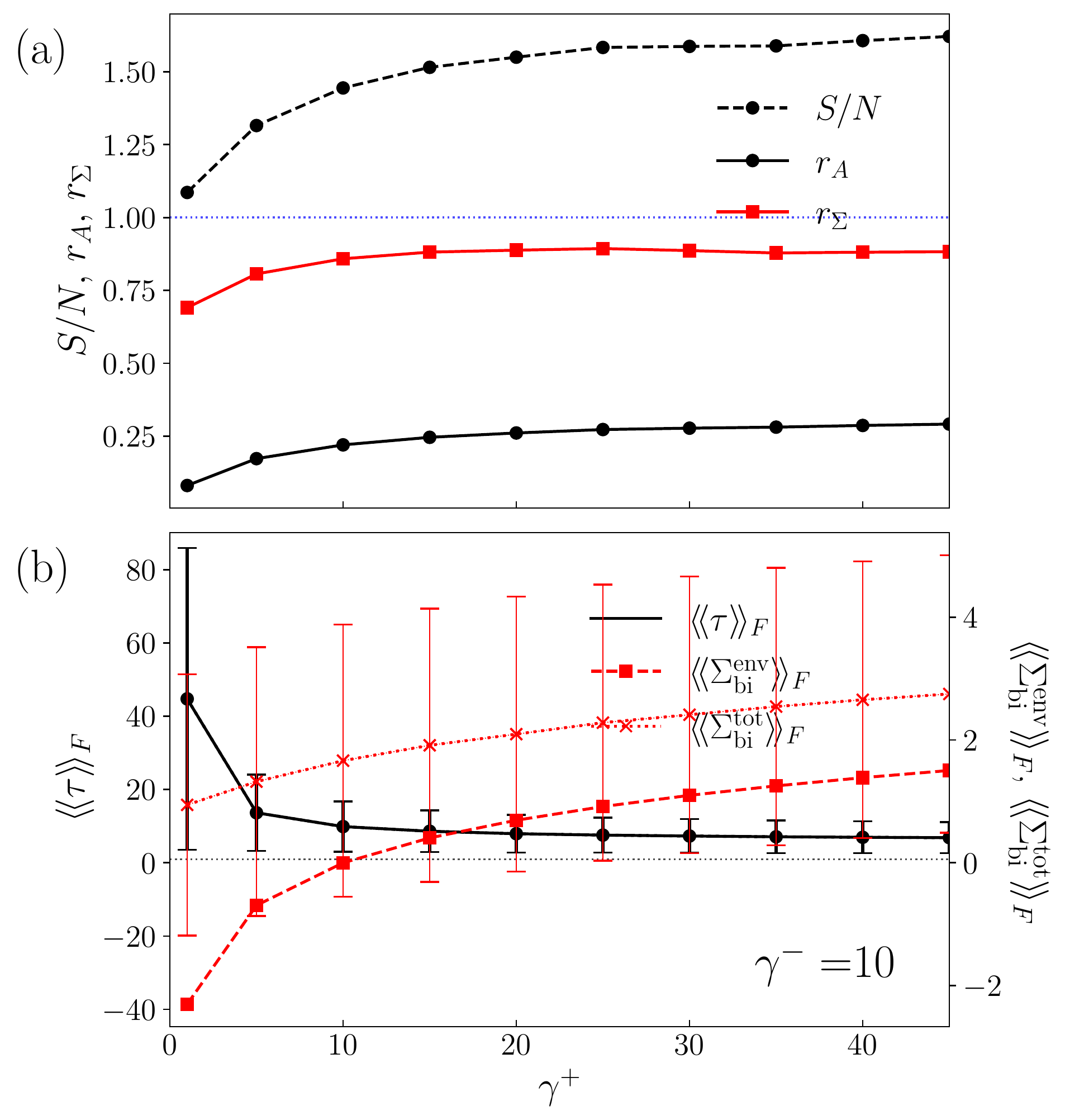}
\caption{
(a) SNR and the kinetic and mixed bounds versus the forward transition rate of CJoin for $0+0=(0)0$. 
The backward transition rate is fixed as $\gamma^-=10$. 
The two marked solid lines indicate $r_A$ and $r_\Sigma$.
(b) Average computation time, total entropy production and environment entropy production. 
Error bars indicate the standard deviation. 
}
\label{fig:plot_lis_gamma_dat_sn_r_t_hHA0p0fb10}
\end{center}
\end{figure}

Fig.~\ref{fig:plot_lis_gamma_dat_sn_r_t_hHA0p0fb10} (a) shows the SNR as well as the kinetic and the mixed bounds versus the CJoin forward transition rate $\gamma^+$ for $0+0=(0)0$. 
The backward transition rate is fixed at $\gamma^-=10$. 
The SNR (dashed line) increases with an increase in the forward transition rate and saturates above $\gamma^+/\gamma_- \approx 1$. 
Both $r_A$ and $r_\Sigma$ (the marked solid lines) increase and become saturated. 
Especially, the mixed bound approaches its maximum value $r_\Sigma \approx 1$.

The marked solid line in panel (b) of Fig.~\ref{fig:plot_lis_gamma_dat_sn_r_t_hHA0p0fb10} shows the mean computation time. 
It decreases and saturates around $\gamma_\pm \approx \gamma_- \gg \Gamma$. 
Saturation happens, because in each cell a token spends time $1/\Gamma=1$ to search for the correct place of the intended computation path, which cannot be shortened by increasing the forward transition rate of a CJoin. 
The marked dashed and dotted lines indicate the mean environment entropy production and the mean total entropy production, respectively. 
They increase logarithmically in $\gamma_+/\gamma_-$. 
Therefore, a forward transition rate that is much bigger than the backward transition rate imposes a cost without improving the computation time. 

For $\gamma_+ < \gamma_-$, negative environment entropy production, i.e., the cooling of the environment, is realized because there is a finite probability of reaching the final state by chance due to thermal fluctuations. 
This result does not contradict the second law of thermodynamics. 
The average total entropy production is always positive. 
The situation is similar to Maxwell's demon experiment~\cite{Toyabe2010} in which net work is extracted from thermal fluctuations by measurement and feedback. 
The stochastic total entropy associated with bidirectional transitions is separated into $\Sigma^{\rm tot}_{\rm bi}=\Sigma^{\rm env}_{\rm bi}+\Sigma^{\rm sys}_{\rm bi}$. 
We define the stochastic entropy $\Sigma^{\rm sys}_{\rm bi}$ in Eq.~(\ref{eqn:path_ent_pro_uni}), which fluctuates with each run. 
The error bars in panel (b) indicate the standard deviation.

At $\gamma_+=\gamma_-$, there is no environment entropy production. 
Figure \ref{fig:lis_dat_HAf10b10_t_sn_r_h} shows (a) the SNR and the kinetic and mixed bounds, (b) the mean computation time 
and (c) the mean total entropy production and mean environment entropy production. 
The horizontal axis indicates four computation processes. 
Lines connect data points belonging to the same reachability graph. 
Their initial and final states are summarized in Table~\ref{tab:ha_initial_final}. 
Figure~\ref{fig:lis_dat_HAf10b10_t_sn_r_h} (c) confirms that the environment entropy production is zero and the total entropy production is positive. 
No substantial differences exist among the four computation processes in all panels. 
We discuss the reason for this in the next section.

\subsection{Computation speed and cell size}
\label{sec:cscs}

The computation speed is qualitatively estimated from the size of the input cells of a gate module, $g$ [see Fig.~\ref{fig:blockdiag2in2out_module} (a)]. 
If the size of a cell is large, a token has to spend a long time to find a place connecting to a proper CJoin. 
The forward transition probability is roughly proportional to, 
\begin{align}
\frac{1}{ \left| P \left( { \alpha^{\rm in}_g }_{*} \right) \right| \left| P \left( { \beta^{\rm in}_g }_{*} \right) \right| } \, , \label{eqn:cscs_forward}
\end{align}
where the wildcard $*$ stands for 0 or 1. 

The probability of a backward transition of $g$ is roughly proportional to $| P( { \alpha^{\rm out}_g }_{*} ) |^{-1} | P( { \beta^{\rm out}_g }_{*} ) |^{-1}$. 
If $\alpha^{\rm out}_g$ ($\beta^{\rm out}_g$) is an output terminal of the whole circuit, the probability that one of two tokens is detected is roughly proportional to $| P( { \alpha^{\rm out}_g }_{*} ) |^{-1}$ ($| P( { \beta^{\rm out}_g }_{*} ) |^{-1}$). 
Once a token is detected, the backward transition is forbidden. 
Since, 
\begin{align*}
\frac{1}{| P ( { \alpha^{\rm out}_g }_{*} ) | | P( { \beta^{\rm out}_g }_{*} ) |} &< \frac{1}{| P( { \alpha^{\rm out}_g }_{*} ) |} \, ,  \frac{1}{| P( { \beta^{\rm out}_g }_{*} ) |} \, ,
\end{align*}
the detection of the token suppresses the backward transition. 
The detection prevents the current state from going astray into potentially available states on extraneous branches of the reachability graph.

\begin{figure}[ht]
\begin{center}
\includegraphics[width=0.9 \columnwidth]{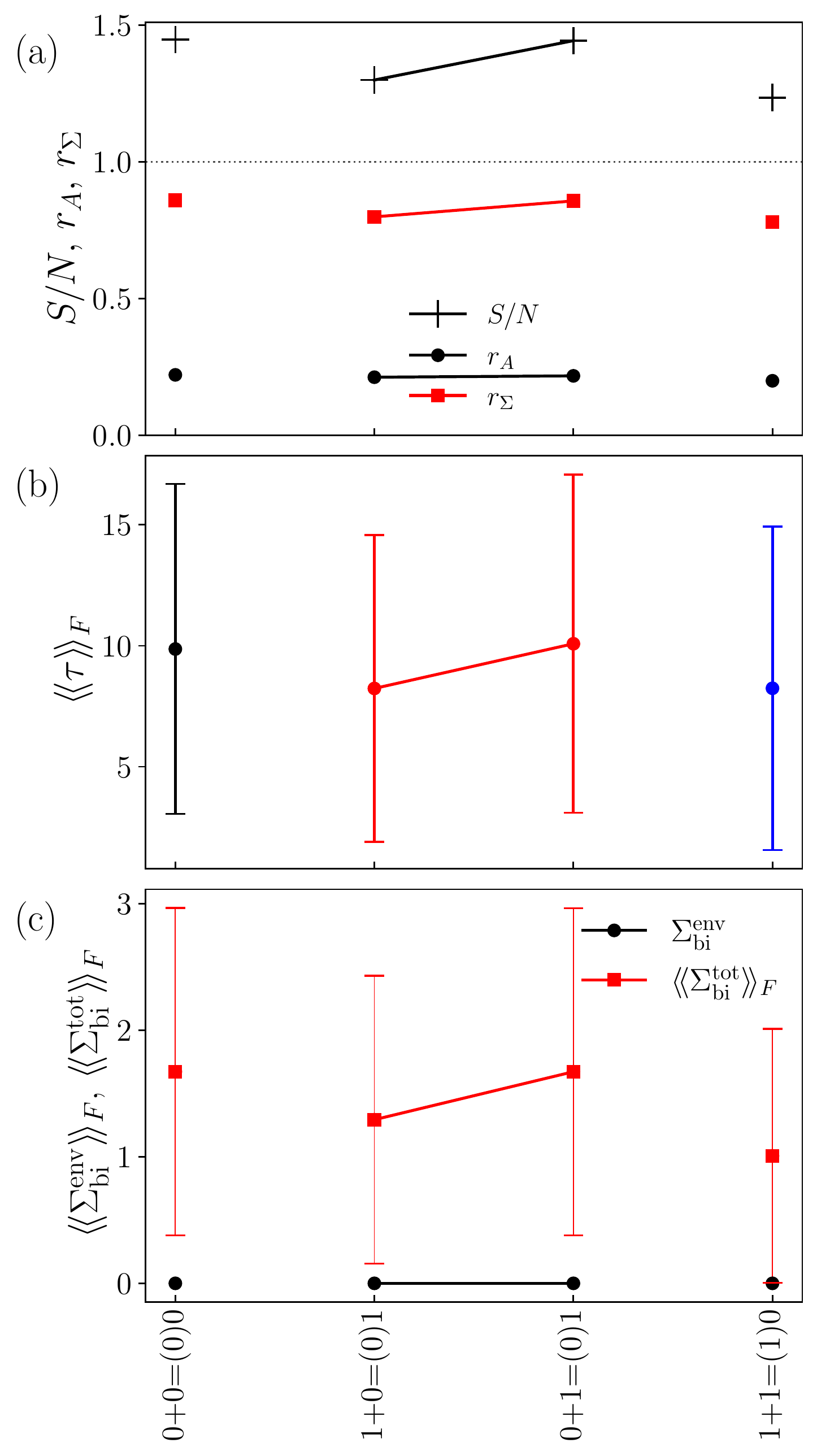}
\caption{
(a) SNR and the kinetic and mixed bounds, (b) average computation time, and (c) the average total entropy production and the average environment entropy production for four inputs with $\gamma^\pm=10$.
Error bars indicate the standard deviation. 
Lines connect data points belonging to the same reachability graph. 
}
\label{fig:lis_dat_HAf10b10_t_sn_r_h}
\end{center}
\end{figure}

For an isolated HA, $\alpha^{\rm in}_{\rm HA}$ and $\beta^{\rm in}_{\rm HA}$ are the input terminals of the whole circuit. 
Then, each Boolean input variable is encoded by a cell consisting of one Hub and thus $\left| P \left( { \alpha^{\rm in}_{\rm HA} }_{*} \right) \right|^{-1} \left| P \left( { \beta^{\rm in}_{\rm HA} }_{*} \right) \right|^{-1}=3^{-2}$. 
Therefore, the computation time is independent of the inputs, which explains why there are no substantial differences among the four computation processes of HA in Fig.~\ref{fig:lis_dat_HAf10b10_t_sn_r_h}.

\subsection{Addition of 2-digit binary numbers}

\begin{figure}[ht]
\begin{center}
\includegraphics[width=0.8 \columnwidth]{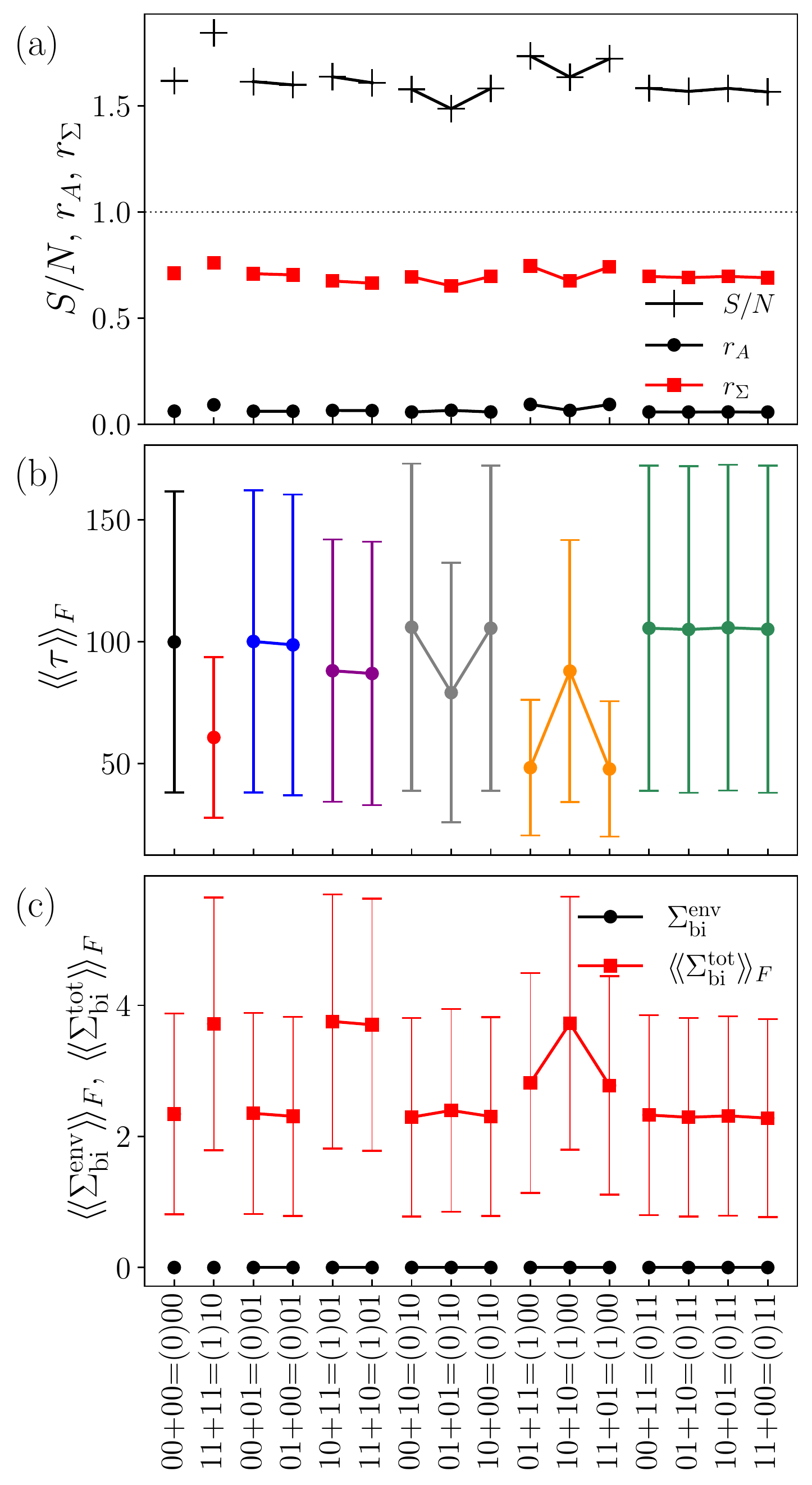}
\caption{
(a) SNR and the kinetic and mixed bounds, (b) average computation time, and (c) the average entropy productions for 16 inputs. 
Error bars indicate the standard deviation. 
Lines connect data points belonging to the same reachability graph. 
Parameters are set at $\gamma^\pm=10$. 
}
\label{fig:lis_dat_HAFAf10b10_t_sn_r_h}
\end{center}
\end{figure}

Figure~\ref{fig:lis_dat_HAFAf10b10_t_sn_r_h} (a) shows the SNR and the kinetic and mixed bounds for 16 computation processes. 
Their initial and final states are summarized in Table~\ref{tab:hafa_initial_final} of Appendix~\ref{sec:ifms}. 
The forward and backward transition rates are fixed at $\gamma^\pm=10$. 
For 16 computation processes, the SNR is slightly better than the SNR of the HA [see Fig.~\ref{fig:lis_dat_HAf10b10_t_sn_r_h} (a)].
As anticipated, the kinetic bound is looser,
but on the other hand the mixed bound still provides a reasonable estimation. 

The computation times in Fig.~\ref{fig:lis_dat_HAf10b10_t_sn_r_h} (b) differ substantially for different inputs. 
We consider that extraneous branches of reachability graphs (Fig.~\ref{fig:reachabilityHAFA}) are not relevant to the computation time,
because the detections of tokens prevent them from going astray into the extraneous branches, as we discussed in the previous section, Sec.~\ref{sec:cscs}.

The bottlenecks of computation are the cells encoding zero-carries, $c_{0,0}$, $c'_{0,0}$ and $c''_{0,0}$, possessing the large preimage (Appendix~\ref{sec:preima}), 
\begin{align*}
\left| f_{\rm HA}^{-1}(\cup,0) \right| = \left| \{ (0,0),(0,1),(1,0) \} \right|=3 \, .
\end{align*}
In fact, the number of zero-carries, or the number of one-carries $c_0+c'_1+c''_1$ summarized in Table \ref{tab:full_truth_table_two_binary_digit_sum}, qualitatively explains the computation time in Fig.~\ref{fig:lis_dat_HAFAf10b10_t_sn_r_h} (b), in which there are roughly three different computation times. 
The shortest computation times correspond to the reachability graph Fig.~\ref{fig:reachabilityHAFA} (b), and the $01+11$ and $11+01$ processes of Fig.~\ref{fig:reachabilityHAFA} (f), in which there is one zero-carry, $c_0+c'_1+c''_1=2$. 
The intermediate computation times correspond to the reachability graph in Fig.~\ref{fig:reachabilityHAFA} (d), the $01+01$ process in Fig.~\ref{fig:reachabilityHAFA} (e) and the $10+10$ process in Fig.~\ref{fig:reachabilityHAFA} (f), in which there are two zero-carries, $c_0+c'_1+c''_1=1$. 
The longest computation times correspond to the reachability graphs in Fig.~\ref{fig:reachabilityHAFA} (a), (c), (g) and the $00+10=10+00$ process in Fig.~\ref{fig:reachabilityHAFA} (e), in which there are three zero-carries, $c_0+c'_1+c''_1=0$. 

The approximate computation time is estimated only by the truth values of the carries. 
The forward transition probability of DOR is approximately proportional to 
$|P(c_1')|^{-1} |P(c_1'')|^{-1} = ( 2 | f^{-1}_{\rm HA}( \cup,  c_1')|+1)^{-1} \times ( 2 | f^{-1}_{\rm HA}( \cup,  c_1'')|+1)^{-1}$
and is dominated by the truth values of $c_1'$ and $c_1''$. 
The forward transition probability of HAb is approximately proportional to 
$|P(c_0)|^{-1} |P(z_1')|^{-1} = ( 2 | f^{-1}_{\rm HA}( \cup,  c_0)|+1)^{-1} \times ( 2 | f^{-1}_{\rm HA}( z_1', \cup)|+1)^{-1} = ( 2 | f^{-1}_{\rm HA}( \cup,  c_0)|+1)^{-1} \times 5^{-1}$ and is dominated by the truth values of $c_0$. 
Since the forward transition probability of the other gates, HA and HAa, 
$|P(x_0)|^{-1} |P(y_0)|^{-1} = |P(x_1)|^{-1} |P(y_1)|^{-1} = 3^{-2}$, 
are independent of their inputs, 
only the truth values of carries dominate the computation time. 
In Sec.~\ref{sec:Time_scale_separation}, we provide a more quantitative analysis. 

Another notable feature in Fig.~\ref{fig:lis_dat_HAFAf10b10_t_sn_r_h} (b) is that for $00+10=01+01=10+00$ ($01+11=10+10=11+01$), one computation time is shorter (longer) than the other two,
even though the three inputs belong to the same reachability graph, Fig.~\ref{fig:reachabilityHAFA} (e) [Fig.~\ref{fig:reachabilityHAFA} (f)]. 
For example, the computation time for $10+10$ differs from the other two, $01+11$ and $11+01$. 
The reason can be understood from the structure of the reachability graph in Fig.~\ref{fig:reachabilityHAFA} (f). 
Two main paths are merging at the 4th step. 
The initial states of $01+11$ and $11+01$ are $|2 \rangle$ and $|1 \rangle$, and they are the leaves of the same branch. 
The initial state of $10+10$ is $|3 \rangle$, which is the leaf of the other branch.

\subsection{Time scale separation}
\label{sec:Time_scale_separation}

For $\gamma^\pm \ll \Gamma =1$, the token is locally equilibrated inside each cell before an available CJoin fires. 
In this regime, we can coarse-grain over places belonging to the same cell and derive the coarse-grained master equation among 4-token logical states (Appendix~\ref{app:Time_scale_separation}). 
The transition state diagrams are derived from the reachability graphs in Fig.~\ref{fig:reachabilityHAFA}, by introducing arcs representing the backward transitions. 
Suppose the forward transition $(|m' \rangle \leftarrow |m \rangle) \in {\mathcal E}$ happens by annihilating two tokens in the cells, $u^{(m)}$ and $v^{(m)}$, and by creating two tokens in the cells, $u^{(m')}$ and $v^{(m')}$. 
Then the forward and backward transition rates of the coarse-grained master equation are, 
\begin{align}
\tilde{ L }_{m',m} = \frac{ \gamma^+ }{ D^{(m)} } \, , \;\;\;\;
\tilde{ L }_{m,m'} = \frac{ \gamma^- }{ D^{(m')} } \, , \label{eqn:coarse_tra_rat}
\end{align}
where the number of available states is, 
\begin{align}
D^{(m)} = \left| P \left( u^{(m)} \right) \right| \left| P \left( v^{(m)} \right) \right| \, . 
\end{align}
At the coarse-grained level, the logarithm of the ratio between the forward and backward transition rates is
\begin{align*}
\ln \frac{ \tilde{ L }_{m',m} }{ \tilde{ L }_{m,m'} } = \ln \frac{ \gamma^+ }{ \gamma^- } + \ln \frac{ D^{(m')} }{ D^{(m)} } \, . 
\end{align*}
The first and second terms of the right-hand side are compatible with Eqs.~(\ref{eqn:feynmann1}) and (\ref{eqn:feynmann2}), respectively.

\begin{figure}[ht]
\begin{center}
\includegraphics[width=1 \columnwidth]{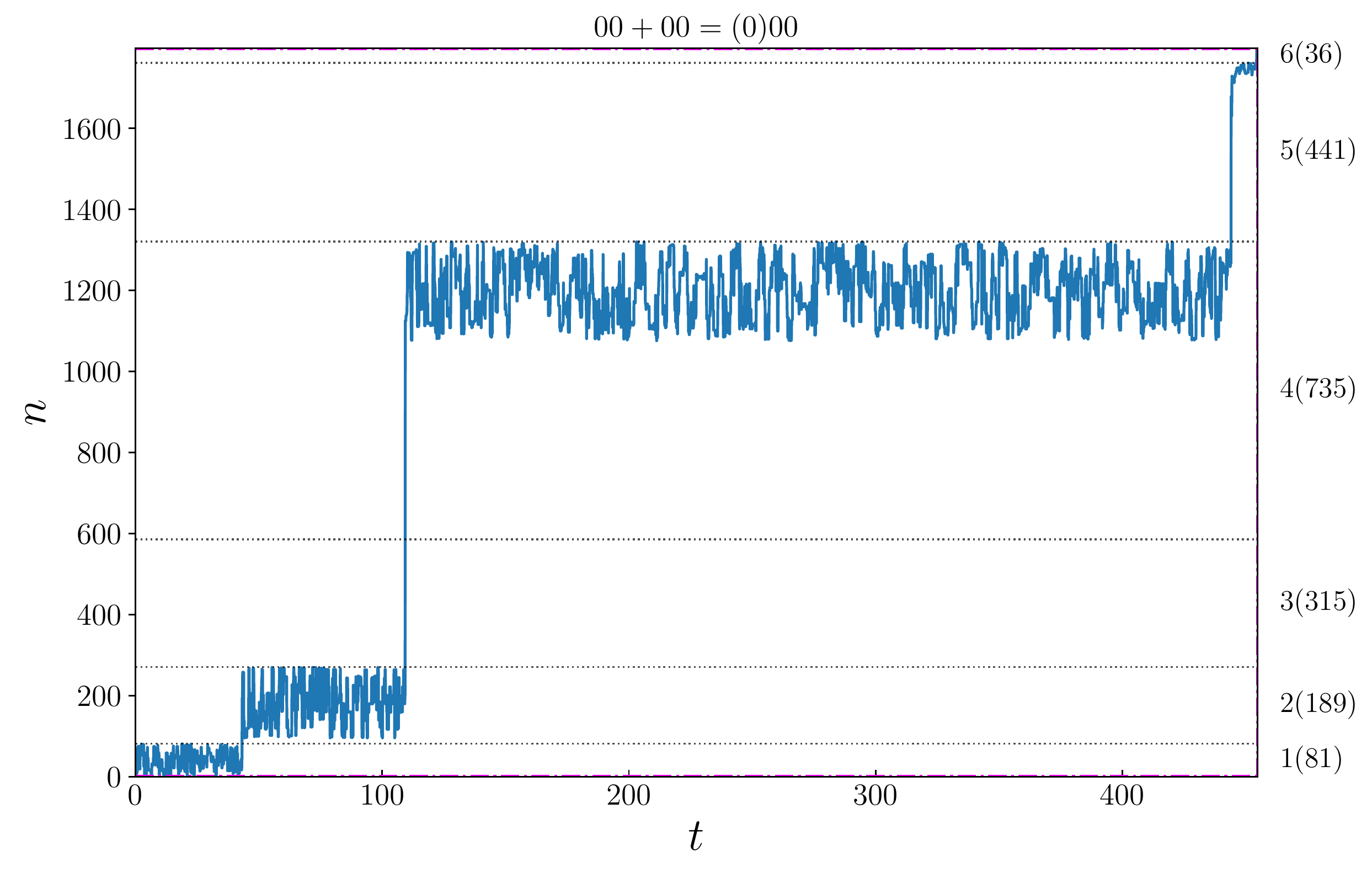}
\caption{Stochastic trajectory for $00+00=(0)00$ for $\gamma^+=0.1$ and $\gamma^-=0.001$. 
The left vertical axis indicates the serial numbers of all states, according to Eq.~(\ref{eqn:serial_number}). 
The horizontal dot-dashed lines indicate the initial and final states, $n_{\rm i}=1$ and $n_{\rm f}=1797$.
The right vertical axis denotes the index of the $4$-token logical states, $|m \rangle$, and the number of $4$-token fine grained states, $NU_{4}^{(m)}$, as $m(NU_{4}^{(m)})$. 
The horizontal dotted lines separate different $4$-token logical states. 
}
\label{fig:lis_dat_HAFA00p00f01b0001_path}
\end{center}
\end{figure}

Figure~\ref{fig:lis_dat_HAFA00p00f01b0001_path} shows a single stochastic trajectory for $00+00=(0)00$. 
The forward and backward transition rates obey $\gamma^- \ll \gamma^+ \ll \Gamma$. 
Dotted horizontal lines separate different 4-token logical states. 
The transitions between two 4-token logical states happen after many jumps within each 4-token logical state. 
The backward transitions are suppressed since $\gamma^- \ll \gamma^+$. 
In the logical states $|4 \rangle= \left|z_{0,0},z'_{1,0}, c_{0,0},c'_{1,0} \right \rangle$ and $|5 \rangle= \left|z_{0,0},z_{1,0}, c'_{1,0},c''_{1,0} \right \rangle$, not all fine-grained states participate for equilibration, since tokens in the output cells $z_{0,0}$ and $z_{1,0}$ are detected by Ratchets. 

\begin{figure}[ht]
\begin{center}
\includegraphics[width=0.9 \columnwidth]{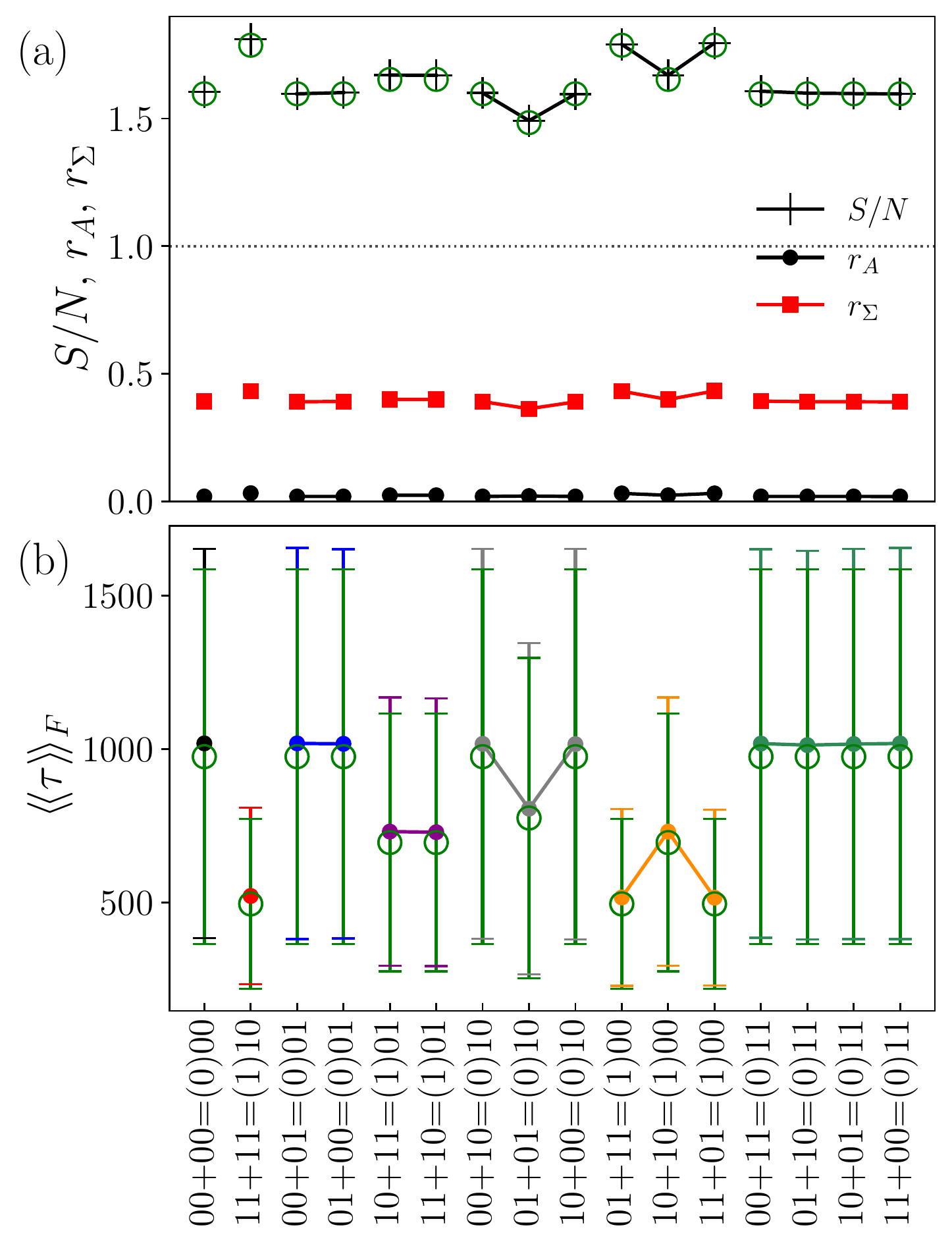}
\caption{
(a) SNR and the kinetic and mixed bounds and (b) average computation time for 16 inputs. 
Lines connect data points belonging to the same reachability graph. 
Error bars indicate the standard deviation. 
Open circles indicate approximate results, Eqs.~(\ref{eqn:cg_average}) and (\ref{eqn:cg_noise}), see also Table~\ref{tab:n3_n4_SN} in Appendix~\ref{app:Time_scale_separation}. 
Parameters are set to $\gamma^+=0.1$ and $\gamma^-=0.001$. 
}
\label{fig:lis_dat_HAFAf01b0001_appr_t_sn_r}
\end{center}
\end{figure}

Figure \ref{fig:lis_dat_HAFAf01b0001_appr_t_sn_r} (a) shows the SNR and the kinetic and mixed bounds for 16 computation processes. 
Since the activity counts the number of jumps within the 4-token logical state, which does not contribute to logical operations, $r_A$ becomes negligible. 

Figure \ref{fig:lis_dat_HAFAf01b0001_appr_t_sn_r} (b) shows average computation times. 
Although amplitudes are different, they resemble those in Fig.~\ref{fig:lis_dat_HAFAf10b10_t_sn_r_h} (b). 
For $\gamma^- \ll \gamma^+ \ll \Gamma$, the average and variance are expressed as, 
\begin{align}
\gamma_+ \langle \! \langle \tau \rangle \! \rangle_F =& \sum_{j=1}^4 D_{j} - \frac{D_{1} D_{2}}{D_{1}+D_{2}} \, , \label{eqn:cg_average} \\
\gamma_+^2 \langle \! \langle \tau^2 \rangle \! \rangle_F =& \sum_{j=1}^4 { D_{j} }^2 - 3 \left( \frac{D_{1} D_{2}}{D_{1}+D_{2}} \right)^2  \, , \label{eqn:cg_noise}
\end{align}
where, 
\begin{align}
D_{1} =& D_{2} = \left| P(x_{0,*}) \right| \left| P(y_{0,*}) \right| = \left| P(x_{1,*}) \right| \left| P(y_{1,*}) \right| \, ,  \label{eqn:D1_D2} \\
D_{3} =& \left| P(z'_{1,*}) \right| \left| P(c_{0,*}) \right| \nonumber \\ =& \left( 2 \left| f_{\rm HA}^{-1}(z'_{1},\cup) \right|+1 \right) \left( 2 |f_{\rm DOR}^{-1}(\cup, c_{0}) |+1 \right) \, , \label{eqn:D3} \\
D_{4} =& \left| P(c'_{1,*}) \right| \left| P(c''_{1,*}) \right| \nonumber \\  =& \left( 2 \left| f_{\rm HA}^{-1}(\cup, c''_{1}) \right|+1 \right) \left( 2 \left| f_{\rm HA}^{-1}(\cup, c'_{1}) \right|+1 \right) \, . \label{eqn:D4}  
\end{align}
Here the wildcard $*$ stands for the truth values of the corresponding computation, see Table~\ref{tab:full_truth_table_two_binary_digit_sum}. 
For the first and second steps, we have $D_{1}=D_{2}=3 \times 3=9$. 
The other numbers, $D_3$ and $D_4$, are summarized in Table~\ref{tab:n3_n4_SN} in Appendix \ref{app:Time_scale_separation}. 
Open circles in Fig.~\ref{fig:lis_dat_HAFAf01b0001_appr_t_sn_r} indicate analytic results. 
They approximate the numerical results well, which confirms that the computation time is dominated by the size of cells. 

The approximate SNR satisfies the following inequality, 
\begin{align}
\frac{S}{N} = \frac{ D_{1}/2 + \sum_{j=2}^{4} D_{j} }{ \sqrt{ ( D_{1}/2)^2 + \sum_{j=2}^{4} { D_{j} }^{2}  } } \leq \sqrt{4}=2 \, . 
\end{align}
It is derived from the Cauchy-Schwartz inequality. 
The SNR is bounded from above by the square root of the number of firings of CJoins, which is the number of tokens~\cite{Pal2021,Aldous1987}.

\subsection{Circuit without Ratchets} 
\label{sec:wdio}

This section considers the error-free detection and resetting conducted only at the final state. 
In other words, we remove the Ratchets from all output terminals of the circuit in Fig.~\ref{fig:DOR_HAFA} (b).

\begin{figure}[ht]
\begin{center}
\includegraphics[width=0.9 \columnwidth]{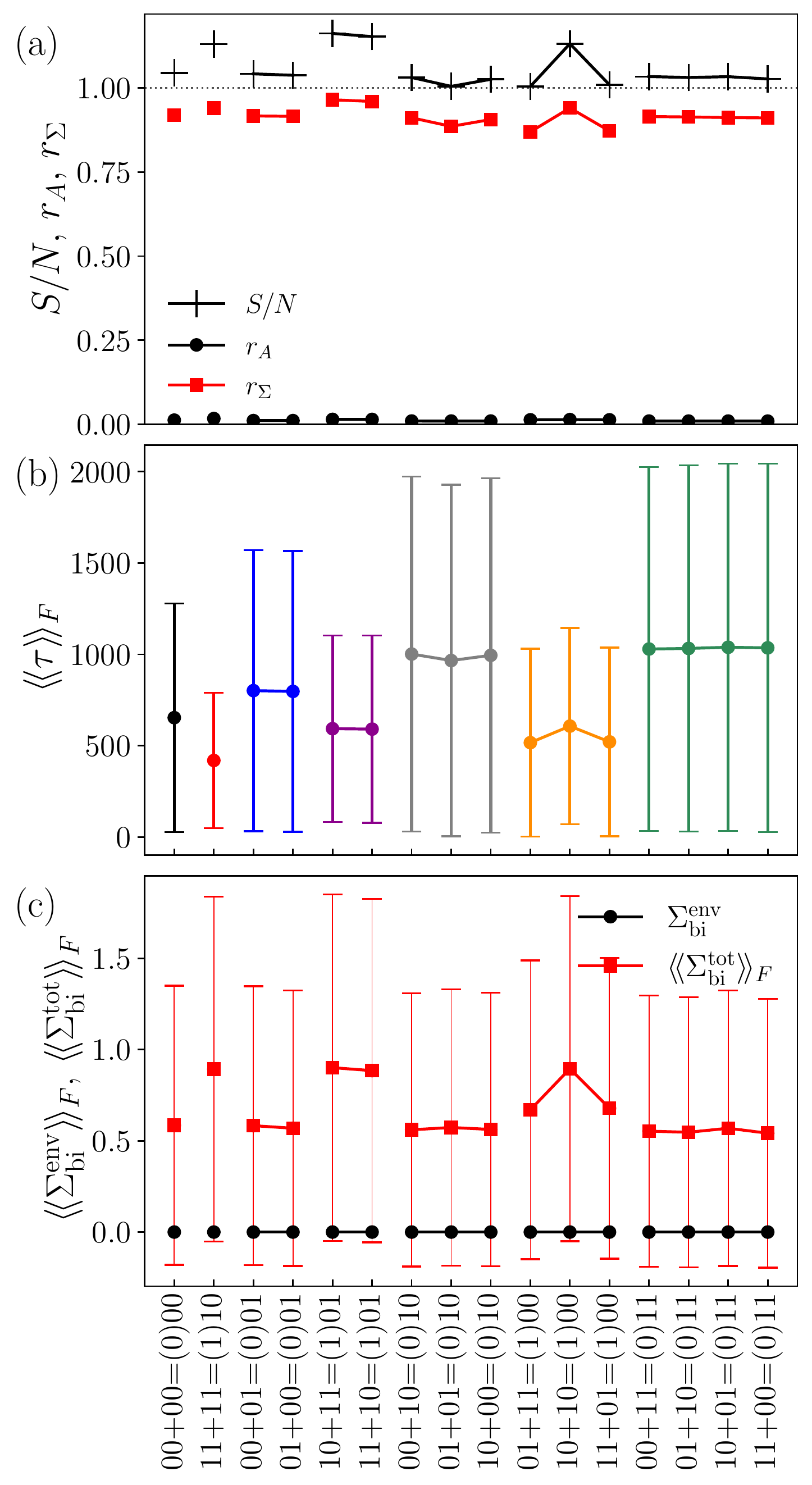}
\caption{
(a) SNR, kinetic bound and mixed bound, (b) computation time and (c) entropy productions for 16 computation processes for the circuit in Fig.~\ref{fig:DOR_HAFA} (b) with all Ratchets removed from the output terminals. 
Parameters are set at $\gamma^\pm=10$. 
}
\label{fig:lis_dat_HAFAf10b10NON_t_sn_r}
\end{center}
\end{figure}

Figure \ref{fig:lis_dat_HAFAf10b10NON_t_sn_r} (a) shows that the SNR approaches to 1. 
The mixed bound provides a good estimation for all computation processes of $r_\Sigma \approx 1$. 
On the other hand, the kinetic bound is negligible. 
By comparing Fig.~\ref{fig:lis_dat_HAFAf10b10NON_t_sn_r} (b) and Fig.~\ref{fig:lis_dat_HAFAf10b10_t_sn_r_h} (b), we observe that the computation slows down significantly. 
In addition, the differences in computation times observed among the three computation processes belonging to the same reachability graphs, $00+10=01+01=10+00$ and $01+11=10+10=11+01$ in Fig.~\ref{fig:lis_dat_HAFAf10b10NON_t_sn_r} (b), become unnoticeable. 
Figure~\ref{fig:lis_dat_HAFAf10b10NON_t_sn_r} (c) demonstrates that the average total entropy productions are reduced as compared with Fig.~\ref{fig:lis_dat_HAFAf10b10_t_sn_r_h} (c). 

These features make sense since the system reaches the uniform equilibrium distribution. 
The multi-token state equilibrates by diffusing over the whole state space, since the resetting occurs only at a single state in the whole state space. 
In fact, the Shannon entropy evaluated from the numerically obtained distribution probability in Eq.~(\ref{eqn:dis_pro_gillespie}) is close to its maximum value. 
For example, for the Shannon entropy of the computation process $00+11=(0)11$, we have
$H( \{ p_j \} )= - \sum_{j=1}^{\Omega} p_j \ln p_j \approx 7.98$, 
which is close to $\ln |\Omega| \approx 7.99$ with $|\Omega|+1=2940$.

The statistics of resets would be similar to that of emissions of particles from a particle reservoir. 
Namely, the resetting events are uncorrelated and Poisson distributed. 
Then the Fano factor $\langle \! \langle w^2 \rangle \! \rangle / \langle \! \langle w \rangle \! \rangle \approx 1$~\cite{Blanter2000} and thus $S/N \approx \sqrt{W}$ from Eq.~(\ref{eqn:snr_wald}). 
The computation time is inversely proportional to the reset current, 
$\langle \! \langle w \rangle \! \rangle = \sum_{j \in {\it \Gamma}^- {\rm f} } \Gamma^{\rm f}_{{\rm f},j} p_j^{\rm st} \approx \Gamma |{\it \Gamma}^- {\rm f}|/(|\Omega|+1) $, which depends only on the size of the state space $|\Omega|$ and the indegree $|\delta^- {\rm f}|=|{\it \Gamma}^- {\rm f}|$, which is the number of arcs negatively incident to the node $|{\rm f}\rangle$. 
The computation time is independent of the structure of the reachability graph. 
Thus computation times for the processes belonging to the same reachability graph become the same as observed in Fig.~\ref{fig:lis_dat_HAFAf10b10NON_t_sn_r} (b).

The average activity rate for the uniform distribution $p^{\rm st}_j \approx 1/|\Omega|$ is expressed in terms of the coarse-grained master equation as (see Appendix \ref{sec:sup_wdio}), 
\begin{align}
\langle \! \langle a \rangle \! \rangle \approx \sum_{m' \neq m} \tilde{L}_{m',m} \, p_{(m)}^{\rm eq} + \Gamma \sum_m \sum_{\ell=1}^N \frac{ \left| T \left( x_\ell^{(m)} \right) \right| }{\left| P \left( x_\ell^{(m)} \right) \right|}  \, ,  \label{eqn:a_uniform_dist_coagra}
\end{align}
where
$p_{(m)}^{\rm eq}=NU_N^{(m)}/(|\Omega|+1)$ 
is the equilibrium distribution probability. 
The second term of the right-hand side of Eq.~(\ref{eqn:a_uniform_dist_coagra}) is the activity rate associated with the local equilibration in each cell. 
Since the kinetic bound is $\sqrt{\langle \! \langle A\rangle \! \rangle_F } \approx \sqrt{ \langle \! \langle a \rangle \! \rangle \langle \! \langle \tau \rangle \! \rangle_F }$, it overestimates the SNR for large $\langle \! \langle \tau \rangle \! \rangle_F$, as observed in Fig.~\ref{fig:lis_dat_HAFAf10b10NON_t_sn_r} (a).

The magnitude of the total entropy production is small, as shown in Fig.~\ref{fig:lis_dat_HAFAf10b10NON_t_sn_r} (c),
because for the uniform distribution, the system entropy production due to the absorption $\langle \! \langle \Sigma^{{\rm abs}}_{{\rm uni}} \rangle \! \rangle_F \approx W \ln p^{\rm st}_{\rm f} \approx -W \ln |\Omega|$ is almost compensated by the entropy production due to the insertion $\langle \! \langle \Sigma^{{\rm ins}}_{{\rm uni}} \rangle \! \rangle_F \approx -W \ln p^{\rm st}_{\rm i} \approx W \ln |\Omega|$ as $\langle \! \langle \Sigma^{{\rm sys}}_{{\rm uni}} \rangle \! \rangle_F = \langle \! \langle \Sigma^{{\rm abs}}_{{\rm uni}} \rangle \! \rangle_F + \langle \! \langle \Sigma^{{\rm ins}}_{{\rm uni}} \rangle \! \rangle_F \approx 0$. 
In addition, since $\gamma^+=\gamma^-$, the environment entropy production is also zero $\langle \! \langle \sigma_{\rm bi}^{\rm env} \rangle \! \rangle=0$. 
Therefore, $\langle \! \langle \Sigma^{{\rm tot}}_{{\rm bi}} \rangle \! \rangle_F \approx 0$, and the mixed bound is the square root of the activity of unidirectional transitions, i.e., the number of resets, as $\sqrt{ \langle \! \langle A_{\rm uni} \rangle \! \rangle_F} \approx \sqrt{W}$. 
This results in $r_\Sigma \approx 1$ [see Fig.~\ref{fig:lis_dat_HAFAf10b10NON_t_sn_r} (a)].

\begin{figure}[ht]
\begin{center}
\includegraphics[width=0.8 \columnwidth]{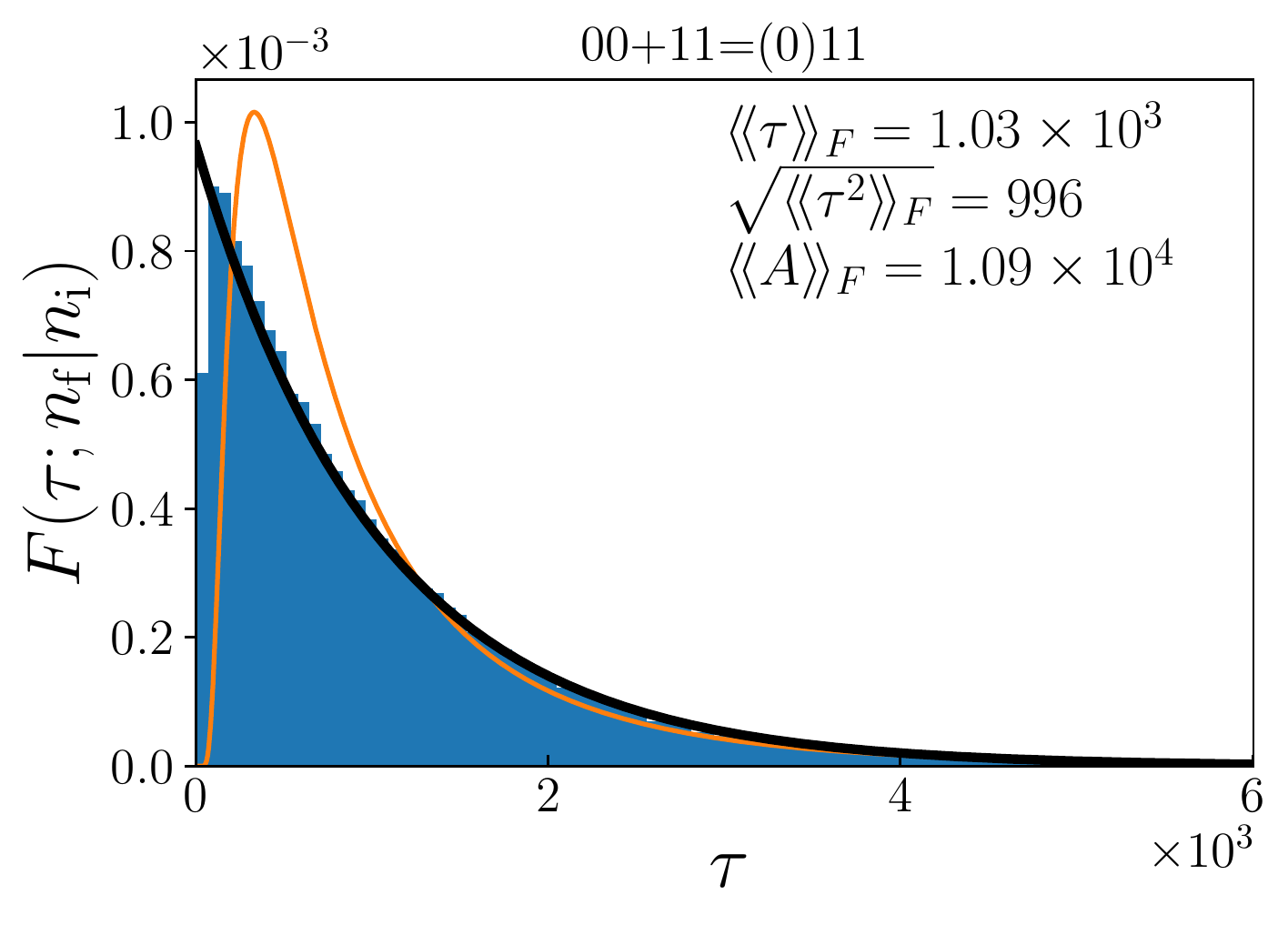}
\caption{
Probability distributions of computation time for $00+11=(0)11$ with $\gamma^\pm=10$ for $W=1$. 
The exponential distribution (thick solid line), given by Eq.~(\ref{eqn:exp_dist}), fits the histogram well. 
The solid line indicates the inverse Gaussian distribution (\ref{eqn:inverse_gauss_tau}). 
}
\label{fig:lis_dat_HAFA_NON_00p11f10b10_fptd}
\end{center}
\end{figure}

Figure \ref{fig:lis_dat_HAFA_NON_00p11f10b10_fptd} shows the computation time distribution of $00+11=(0)11$ with $W=1$. 
The histogram deviates from the inverse Gaussian distribution (solid line) and is well fitted by the exponential distribution (thick solid line):
\begin{align}
F(\tau) = \frac{1}{\langle \! \langle \tau \rangle \! \rangle_F} \exp \left( - \frac{\tau}{\langle \! \langle \tau \rangle \! \rangle_F} \right) \, . 
\label{eqn:exp_dist}
\end{align}
The exponential distribution is typical for the waiting time distribution of shot noise~\cite{VanKampen2007}, which also supports that the resetting events are Poisson distributed. 
The inverse Gaussian distribution is realized in the steady-state after many resets, as shown by Fig.~\ref{fig:lis_dat_HAFA_NONlp5_00p11f10b10_fptd} of Appendix \ref{sec:sup_wdio}. 

\section{Discussion}
\label{sec:discussion}

Error-free token detections and resets are unidirectional and drive the system out of equilibrium. 
For the RTM, the state space is one-dimensional, confining the dimensions of diffusion. 
Therefore, the error-free resets induce a strongly non-equilibrium distribution. 
As a consequence, the magnitude of the system entropy production due to the absorption $\langle \! \langle \Sigma^{{\rm abs}}_{{\rm uni}} \rangle \! \rangle_F \approx -2 W \ln |\Omega|$ [Eq.~(\ref{eqn:s_abs_RTM})] is twice as large as the logarithms of the size of the state space, $W \ln |\Omega|$. 
On the other hand, the entropy production due to insertion is also finite, $\langle \! \langle \Sigma^{{\rm ins}}_{{\rm uni}} \rangle \! \rangle_F \approx W \ln |\Omega|$ [Eq.~(\ref{eqn:s_ins_RTM})]. 
Therefore, the resetting entropy production is $\langle \! \langle \Sigma^{{\rm sys}}_{{\rm uni}} \rangle \! \rangle_F = \langle \! \langle \Sigma^{{\rm abs}}_{{\rm uni}} \rangle \! \rangle_F + \langle \! \langle \Sigma^{{\rm ins}}_{{\rm uni}} \rangle \! \rangle_F \approx - W \ln |\Omega|<0$. 
This means that the resets in computation cycles effectively erase information from the system. 
Note that the situation is different from the previous works~\cite{Norton2013,Strasberg2015}, which assume the system is reset to a zero entropy state in each computation cycle.

For token-based Brownian circuits, the direction of diffusion is less confined since the tokens spread over cells to search for the intended computation path. 
Especially, when Ratchets at the output terminals are absent (see Sec.~\ref{sec:wdio}), the uniform equilibrium distribution $p^{\rm st}_j \approx |\Omega|^{-1}$ is approached. 
In such a case, the entropy production due to the absorption $\langle \! \langle \Sigma^{{\rm abs}}_{{\rm uni}} \rangle \! \rangle_F \approx - W \ln |\Omega|$ is almost compensated by the entropy production due to the insertion $\langle \! \langle \Sigma^{{\rm ins}}_{{\rm uni}} \rangle \! \rangle_F \approx W \ln |\Omega|$. 
Therefore, in the course of many computation cycles, the resets only slightly reduce information.

%
%

Sec.~\ref{sec:wdio} demonstrates that the uniform equilibrium distribution is approached in a circuit without Ratchets. 
Similarly, the equilibrium state will be approached if we would allow an error in the detection modeled by a finite resetting rate $\Gamma_{\rm reset}$. 
In this case, the resetting entropy production would be reduced at the cost of error, which results in a longer computation time. 
The error-free detection of outputs is not a unique protocol for Brownian computation. 
Another possible protocol is to drive the system so that the output state turns into a unique low-energy state within a given tolerated error probability.
For this protocol, the speed limit for a classical system~\cite{Shiraishi2018,Nakajima2021} would be applicable to estimate its performance. 

%
%

Adding $N/2$-digit binary numbers is achieved by constructing an $N$-token circuit that cascades one HA and $N/2-1$ FAs. 
The computation time of such a circuit is dominated by internal HA and DOR modules. 
It is deduced from Eqs.~(\ref{eqn:cg_average}), (\ref{eqn:D3}) and (\ref{eqn:D4}) that
\begin{align*}
\gamma_+ \langle \! \langle \tau \rangle \! \rangle_F \approx & \sum_{j=1}^{N/2}
\left| P \left( c_{j-1, *} \right) \right| \left| P \left( z'_{j, *} \right) \right| \nonumber + \left| P(c''_{j, *}) \right| \left|P(c'_{j, *}) \right| \\
\approx & \sum_{j=1}^{N/2} \left( 2 |f_{\rm DOR}^{-1}(\cup, c_{j-1}) |+1 \right) \left( 2 \left| f_{\rm HA}^{-1}(z'_{j},\cup) \right|+1 \right) \\ &+ \left( 2 \left| f_{\rm HA}^{-1}(\cup, c''_{j}) \right|+1 \right) \left( 2 \left| f_{\rm HA}^{-1}(\cup, c'_{j}) \right|+1 \right) \, .
\end{align*}
This expression implies that the computation time roughly scales linearly in $N$. 
In addition to internal cells encoding zero-carries of HAs, the cells encoding $z=\tilde{z}=1$ of DORs are sized as (Appendix~\ref{sec:preima}), 
\begin{align*}
\left| f_{\rm DOR}^{-1}(1,\cup) \right| =& \left| f_{\rm DOR}^{-1}(\cup,1)  \right|  = \left| \{ (0,1), (1,0), (1,1) \}  \right| =3 \, .
\end{align*}
Therefore, they are also the bottlenecks of computation. 

For the addition of $N/2$-digit numbers, the number of unidirectional transitions equals the number of tokens $N$, and the mixed bound is roughly estimated as $\sqrt{A_{\rm uni} }= \sqrt{W N}$. 
This bound would be sufficiently tight for the following reason. 
The transition state diagram for the computation process adding $N/2$-digit numbers is effectively an $N$-site one-dimensional chain.
Roughly, its transition rate matrix is described by Eq.~(\ref{eqn:mod_Liouvill_FiniteRevTM}) with the replacements, $\gamma^+ \to \tilde{\gamma}^+$, $\gamma^- \to 0$, and $|\Omega| \to N$. 
Then the CGF is $\Lambda_0(\chi)=\tilde{\gamma}^+ (e^{i \chi/{N}}-1)$, from which the Fano factor is calculated as $\langle \! \langle w^2 \rangle \! \rangle/\langle \! \langle w \rangle \! \rangle=1/N$. 
The suppression of the Fano factor means that the reset current noise behaves like that of a macroscopic conductor~\cite{Shimizu1992} due to inelastic scattering~\cite{Shimizu1992,Utsumi2003,Utsumi2006}. 
Then from Eq.~(\ref{eqn:snr_wald}), the SNR is $S/N= \sqrt{W N}$. 
We speculate that the prescribed bound overestimates the SNR, 
because, from the analogy between the finite one-dimensional chain and the array of large diffusive quantum dot~\cite{Golubev2004}, the Fano factor is approximately $1/3$, which results in a smaller SNR $\sqrt{3W} \approx 1.7 \sqrt{W}$

At the same time, we speculate that the SNR would not be too small in general,
because, from the quantum transport viewpoint, the enhancement of the Fano factor requires non-trivial mechanisms, such as the bunching-like process~\cite{Belzig2005,Takase2021}.

Note that our model does not account for the spacial length of wires. 
For example, in the cascade connection of HA and FA in Fig.~\ref{fig:DOR_HAFA} (b), a wire which encodes $c_1'=1$, is spatially long. 
Therefore, a token needs to spend more time diffusing over this wire. 
However, each wire is replaced with a single place in the SPN model. 
Therefore, the sojourn time is independent of the length of the wire.

In the present work, we adopt the stochastic thermodynamics of resetting~\cite{Fuchs2016}. 
The information thermodynamics for the situation where the final state is fixed but the measurement time is uncertain~\cite{RibezziCrivellari2019,Garrahan2021} may be helpful for a transparent interpretation of the thermodynamic cost of Brownian computing.

\section{Conclusion}
\label{sec:summary}

We analyze the computation time distribution of the token-based Brownian circuit.
The computation can be completed within a finite time with zero or even negative environment entropy production by paying the thermodynamic cost necessary for output detections and resets. 
We analyze the mixed bound of the signal-to-noise ratio of the computation time, which is the square root of the number of token detections and resets, plus half of the entropy production related to the thermodynamic cost. 

The magnitude of the system entropy production for token-based Brownian computers tends to be small,
because the diffusion in multi-token state space is less confined: the tokens are designed to search randomly for correct computation paths in the multi-token state space, which leads to a uniform distribution of the equilibrium. 
Consequently, the entropy production due to absorption and insertion compensate each other. 
Then the mixed bound is approximately the squire root of the number of tokens, which reasonably estimates the signal-to-noise ratio.

The results mentioned above are contrasted with those on the logically reversible Brownian Turing machine, in which the resetting cost increases logarithmically in the size of the state space. 
Accordingly, the mixed bounds overestimate the signal-to-noise ratio, 
because the diffusion in the state space is confined to a one-dimensional chain, and consequently, the error-free resets strongly drive the system out of equilibrium.

For token-based Brownian computers, the detection of output tokens helps the multi-token state stay on the reachability graph's intended computation path. 
Therefore, the extraneous branches in the reachability graph are irrelevant to the computation time. 
Rather, the sizes of the internal cells dominate the computation time. 
The token-based Brownian circuit meets the physical prerequisite that the amount of information encoded in the state of a finite system is bounded. 
Consequently, the size of a cell encoding the output of a gate module depends on the size of its preimage. 
For this reason, the internal cells encoding zero-carries of half-adders are the bottlenecks of token-based Brownian adders.

\begin{acknowledgments}
We thank Teijiro Isokawa, Satoshi Nakajima, Hiroki Okada, Takahiro Sagawa, Yoshishige Suzuki, Kazutaka Takahashi, Yoshimichi Teratani, and Yasuhiro Tokura for valuable discussions. 
This work was supported by JSPS KAKENHI Grants No. 18KK0385, No. 20H01827, No. 	20H02562 and No. 20H05666, and JST, CREST Grant Number JPMJCR20C1, Japan. 

\end{acknowledgments}

\begin{appendix}

\section{One-tape Turing-machine}
\label{sec:RTM}

A one-tape Turing machine (TM) consists of a read-write head, a finite control, and a two-way infinite tape, see Fig.~\ref{fig:RTMTcopy}~\cite{Morita2017}. 
It is formally defined as a 6-tuple $T=(Q,S,q_{\rm s},F,b,\delta)$, where 
$Q$ and $S$ are finite sets of states and tape symbols, $q_{\rm s} \in Q$ is the initial state, $F \subseteq Q$ is the set of final states, and $b \in S$ is a blank symbol. 
The internal state $q$ is an element of $Q$ or the initial (start) state $q_{\rm s}$ or final (halt) state $q_{\rm h} \in F$. 
Each tape cell contains a symbol drawn from a finite set of symbols $S$. 
The TM operates according to prescribed transition rules expressed in the quintuple form $\delta \subseteq (Q \times S \times S \times \{-1,0,+1 \} \times Q)$,
represented by
\begin{align}
\left[ p,s,s',d, q \right] \in \delta \, . \label{eqn:quintuple}
\end{align}
This means that if the finite control is in state $p \in Q$ and the tape head reads a symbol $s \in S$ from the tape cell currently being scanned by the head, 
the TM overwrites the tape cell by the symbol $s' \in S$, shifts the head in a direction depending on $d \in \{-1,0,+1 \}$, and it updates the state to $q' \in Q$. 
The direction into which the tape head shifts is to the left if $d = -1$ and to the right if $d=1$. 
The tape head's position is kept unchanged if $d=0$.

A deterministic TM possesses a unique successor and a reversible TM (RTM) possesses a unique predecessor. 
For any pair of distinct quintuples
$[p_1,s_1,s_1',d_1,q_1], [p_2,s_2,s_2',d_2,q_2] \in \delta$, 
$T$ is a deterministic TM 
if $p_1=p_2$ implies $s_1 \neq s_2$. 
$T$ is an RTM, if $q_1=q_2$ implies $(d_1=d_2) \wedge (s_1' \neq s_2')$. 
For an RTM, it is possible to construct the inverse RTM $T^{-1}$, which undoes the computation performed by $T$, by starting from a final state of $T$, tracing the computation process of $T$ backward, and arriving at the initial state of $T$. 
Let $T=(Q,S,q_{\rm s}, \{ q_{\rm h} \},b,\delta)$ be an RTM, where there is no quintuple in $\delta$ whose fifth element is $q_{\rm s}$.
The inverse RTM is $T^{-1}=(Q,S,q_{\rm h}, \{ q_{\rm s} \}, b, \hat{\delta})$, 
where 
\begin{align}
\hat{\delta} = \left \{ [q,t,s,-d,p] \middle|  [p,s,t,d,q] \in \delta \right \} \, .
\end{align}

The following is an example of an RTM~\cite{Morita2017} that copies a unary number given at the left of a delimiter $a$ to the right of the delimiter: 
$T_{\rm copy}=(Q_{\rm copy}, S_{\rm copy}, q_{\rm s},\{ q_{\rm h} \} , b,\delta_{\rm copy})$, 
where 
$Q_{\rm copy}= \{ q_{\rm s},q_{\rm f},q_{\rm n},q_{\rm r},q_{\rm h} \}$, $S_{\rm copy}= \{ b,1,a,m \}$ and, 
\begin{align}
\delta_{\rm copy} = \{ & [q_{\rm s},1,m,+,q_{\rm f}], [q_{\rm s},a,a,0,q_{\rm h}], [q_{\rm f},b,1,+,q_{\rm n}], \nonumber \\ & 
[q_{\rm f},1,1,+,q_{\rm f}], [q_{\rm f},a,a,+,q_{\rm f}], 
[q_{\rm n},b,b,-,q_{\rm r}], \nonumber \\ & 
[q_{\rm r},1,1,-,q_{\rm r}], [q_{\rm r},a,a,-,q_{\rm r}], [q_{\rm r},m,1,+,q_{\rm s}] \} \, . \label{eqn:T_copy}
\end{align}
Figure \ref{fig:RTMTcopy} shows the step-by-step snapshots of the computation process of $T_{\rm copy}$. 
The tape contains a finite sequence of tape symbols sandwiched by semi-infinite sequences of blanks. 
The instantaneous description (ID) specifies the computational configuration given by the tape contents, the head state, and the head position. 
For example, the tape content and the ID of Fig.~\ref{fig:RTMTcopy} $|1 \rangle$ are $1 a$ and $q_{\rm s}1a$. 
The set of all IDs of $T_{\rm copy}$ is 
${\rm ID}(T_{\rm copy})=\{ 
q_{\rm s} 1 a, 
m q_{\rm f} a, 
m a q_{\rm f}, 
m a 1 q_{\rm n}, 
m a q_{\rm r} 1, 
m q_{\rm r} a 1, 
q_{\rm r} m a 1, 
1 q_{\rm r} a 1, 
1 q_{\rm h} a 1
\}$. 
For the master equation, we assign serial numbers to these states as, 
\begin{align*}
|1 \rangle =& | q_{\rm s} 1 a \rangle = | {\rm i} \rangle \, , \\
|2 \rangle =& | m q_{\rm f} a \rangle \, , \\
|3 \rangle =& | m a q_{\rm f} \rangle \, , \\
|4 \rangle =& | m a 1 q_{\rm n} \rangle \, , \\
|5 \rangle =& | m a q_{\rm r} 1 \rangle \, , \\
|6 \rangle =& | m q_{\rm r} a 1 \rangle \, , \\
|7 \rangle =& | q_{\rm r} m a 1 \rangle \, , \\
|8 \rangle =& | 1 q_{\rm s} a 1 \rangle \, , \\
|9 \rangle =& | 1 q_{\rm h} a 1 \rangle = | {\rm f} \rangle \, .
\end{align*}
Then the state space except for the final state is $\Omega = \left\{ |n \rangle \middle| n=1,\cdots , |\Omega| \right \}$, where $|\Omega|=8$.

\section{Mixed bound in steady-state}
\label{sec:tur_oneway}

The mixed bound at an arbitrary measurement time is derived in Ref.~\onlinecite{Pal2021}. 
Here we derive the mixed bound in the limit of a long measurement time. 
For this purpose, we take the resetting (\ref{eqn:reset_ham}) into account to ensure the steady-state with finite reset current and take $\Gamma_{\rm reset} \to \infty$. 
Suppose the state changes from $|\omega_k \rangle$ to $|\omega_{k+1} \rangle$ at time $t=t_k$, where $(|\omega_{k+1} \rangle \leftarrow |\omega_{k} \rangle) \in E_{\rm bi} \cup E_{\rm uni}$. 
Such a transition is induced by $\Gamma_{\omega_{k+1},\omega_{k}}$. 
The stochastic trajectory with $N$ transitions is described by the initial time $t_0=0$ and $N$ successive times at which transitions happen, 
\begin{align}
t^{N+1}=t_0 \, t_1 \cdots t_N \, , \label{eqn:tN}
\end{align}
where $t_0 < t_1 < \cdots < t_N$, 
and the successive states are, 
\begin{align}
\omega^{N+1}=\omega_0  \, \omega_1 \cdots \omega_N \, . \label{eqn:omegaN}
\end{align}
The measurement time is fixed as $t_{N+1}=\tau > t_{N}$. 
For each trajectory, the total sojourn time in the state $|\omega \rangle$ and the number of transitions from $|\omega' \rangle$ to $|\omega \rangle$ are, 
\begin{align}
\tau_{\omega} \left( t^{N+1} \right) =& \sum_{k=0}^{N} (t_{k+1}-t_k) \delta_{\omega_k, \, \omega} \, , \label{eq:sojourn} \\
W_{ \omega \leftarrow \omega' } \left( \omega^{N+1} \right) =& \sum_{k=0}^{N-1} \delta_{\omega_{k+1} , \, \omega} \, \delta_{\omega_{k}, \, \omega'} \, , \label{eq:jump}
\end{align}
where $|\omega \rangle, \, |\omega' \rangle \in \Omega$. 

The joint probability distribution for the empirical current $j_{i,j}=(W_{i \leftarrow j}-W_{j \leftarrow i})/\tau$ and the empirical density $n_j=\tau_j/\tau$, can be derived, e.g. by exploiting the FCS for current and dwell-time~\cite{Utsumi2007}. 
We skip derivations and begin with an obvious extension of level 2.5 large deviations~\cite{Gingrich2016,Barato2015,GingrichRotskoff2017}, which discuss the bidirectional processes, to include unidirectional processes. 
In the limit of a long measurement time, 
\begin{align*}
P_\tau ( \{ j_{i,j} \} , \{ n_{j} \} ) \approx \exp \left[ \tau  I ( \{ j_{i,j} \} , \{ n_{j} \} ) \right] \, , 
\end{align*}
where the rate function is, 
\begin{align}
I ( \{ j_{i,j} \} , \{ n_{j} \} ) &= \frac{1}{2} \sum_{(i \leftarrow j) \in E_{\rm bi}} \tau_{i,j} \Psi_{\rm bi} \left( \frac{ j_{i,j} }{ \tau_{i,j} } , \frac{ \tilde{j}_{i,j} }{ \tau_{i,j} } \right)  \nonumber \\ & + \sum_{(i \leftarrow j) \in E_{\rm uni}} \Gamma_{i,j} n_j \Psi_{\rm uni} \left( \frac{j_{i,j}}{\Gamma_{i,j} n_j} \right) \, ,
\end{align}
whereby 
$\tau_{i,j} = 2 \sqrt{\Gamma_{i,j} \Gamma_{j,i} n_i n_j}$, 
and, 
\begin{align}
\Psi_{\rm bi}(x,y) =&  \sqrt{ 1 + x^2 }  - \sqrt{ 1 + y^2 } \nonumber \\
 &- x \left( \sinh^{-1} x -  \sinh^{-1} y  \right) \, , \\
\Psi_{\rm uni}(x) =& x-1-x \ln x \, .
\end{align}
In the steady-state, the current is conserved at each node $|j \rangle \in \Omega$, i.e.,
\begin{align}
\sum_{i \in {\it \Gamma}^+ {j} - {\it \Gamma} {j} } j_{i,j} - \sum_{i \in {\it \Gamma}^- {j} - {\it \Gamma} {j} } j_{j,i} + \sum_{i \in {\it \Gamma} {j} } j_{i,j} = 0  \, , \label{eqn:current_conservation}
\end{align}
where the sets of nodes connected by an edge with tail $j$ and head $j$ are, respectively,
\begin{align}
{\it \Gamma}^+ {j} =& \left \{  i \in \Omega \middle | \Gamma_{i,j} >0  \wedge j \in \Omega \right \} \, , \\
{\it \Gamma}^- {j} =& \left \{  i \in \Omega \middle | \Gamma_{j,i} >0  \wedge j \in \Omega \right \} \, . 
\end{align}
An element of the set ${\it \Gamma}^+ {j} \cap {\it \Gamma}^- {j} = {\it \Gamma} {j} $ is a node connected to one edge with head $j$ and one edge with tail $j$, which corresponds to the bidirectional process. 

The stochastic reset current is, 
\begin{align}
w = \sum_{j \in  {\it \Gamma}^- {\rm f}  } j_{ {\rm i} ,j} \, , \label{eqn:def_w}
\end{align}
where the `final' state $| {\rm f} \rangle$ is identical to the `initial' state $| {\rm f} \rangle = | {\rm i} \rangle$. 
Note that it is not the final state of the sequence (\ref{eqn:omegaN}), $| {\rm f} \rangle \neq | \omega_N \rangle$. 
The probability distribution of reset current is obtained from the contraction, 
\begin{align}
\tau^{-1} \ln P_\tau ( w ) =& \sup_{ j_{i,j} \in J(w) , n_j \in P } I ( \{ j_{i,j} \} , \{ n_{j} \} ) \, , \label{eqn:pdf_w}
\end{align}
where the supremum is taken under the constraints Eq.~(\ref{eqn:def_w}), and the current conservation in Eq.~(\ref{eqn:current_conservation}), 
\begin{align}
J(w)=\left \{ j_{k,\ell} \in {\mathbb R} \middle | 
k,\ell \in \Omega \wedge  {\rm Eq. \, (\ref{eqn:current_conservation})}  \wedge {\rm Eq. \, (\ref{eqn:def_w})}  \right \}  \, , \label{eqn:set_jw}
\end{align}
and the normalization, 
\begin{align}
P=\left \{ n_k \in {\mathbb R}_+ \middle | k \in \Omega \wedge \sum_{j \in \Omega } n_{j} =1  \right \}  \, . \label{eqn:set_p}
\end{align}
By observing that $n_j=p_j^{\rm st}$ and 
$j_{i,j} = j_{i,j}^{\rm st} w/\langle \! \langle w \rangle \! \rangle$
satisfy the constraints, the following inequality is derived~\cite{Garrahan2017}, 
\begin{align}
\frac{ \ln P_\tau ( w ) }{\tau} \geq & I \left( \left \{ j_{i,j}^{\rm st} \frac{w}{ \langle \! \langle w \rangle \! \rangle }  \right \} , \{ p_{j}^{\rm st} \} \right) = I^* \, . 
\end{align}
A lower bound for the bidirectional processes follows from the inequality~\cite{Gingrich2016,Gingrich2017}: 
\begin{align}
\Psi_{\rm bi}(x,y) &\geq - \frac{(x-y)^2}{2 y} \sinh^{-1} y \, . 
\end{align}
For the unidirectional processes, the following inequality holds:
\begin{align}
\Psi_{\rm uni}(x) \geq - \frac{(x-1)^2}{2} + \varphi(x) \, ,
\end{align}
where, $\varphi(x)=0$ for $x \geq 1$ and $\varphi(x)=(x-1)^3/2$ for $0 \leq x < 1$. 
Then, the rate function is, 
\begin{align}
I^* \geq & - \frac{ (w - \langle \! \langle w \rangle \! \rangle)^2 }{2 \langle \! \langle w \rangle \! \rangle^2} 
\left( \langle \! \langle \sigma_{\rm uni}^{\rm tot} \rangle \! \rangle/2 + \langle \! \langle a_{\rm uni} \rangle \! \rangle \right)
\nonumber \\ &
+\langle \! \langle a_{\rm uni} \rangle \! \rangle \varphi( w / \langle \! \langle w \rangle \! \rangle )
 \, . 
\end{align}
From the first and second terms of the right-hand side of the above inequality, we can deduce the variance and the skewness of an approximate probability distribution function. 
This inequality means that the width of the approximate probability distribution function is narrower than that of the actual probability distribution function. 
Therefore, the variance $\langle \! \langle w^2 \rangle \! \rangle$ is larger than or equal to 
$\langle \! \langle w \rangle \! \rangle^2 /\left( \langle \! \langle \sigma_{\rm uni}^{\rm tot} \rangle \! \rangle/2 + \langle \! \langle a_{\rm uni} \rangle \! \rangle \right)$. 
Consequently, a lower bound of the Fano factor is, 
\begin{align}
\frac{\langle \! \langle w^2 \rangle \! \rangle}{\langle \! \langle w \rangle \! \rangle} \geq 
\frac{ \langle \! \langle w \rangle \! \rangle }{ \langle \! \langle \sigma_{\rm uni}^{\rm tot} \rangle \! \rangle/2 + \langle \! \langle a_{\rm uni} \rangle \! \rangle } = 
\frac{ W }{ \langle \! \langle \Sigma_{\rm uni}^{\rm tot} \rangle \! \rangle_F/2 + \langle \! \langle A_{\rm uni} \rangle \! \rangle_F }\, . 
\end{align}
Here we used Eq.~(\ref{eqn:c_t}). 
By substituting this into Eq.~(\ref{eqn:snr_wald}), we obtain Eq.~(\ref{eqn:tur_mix}).

\section{Detailed calculations of Sec.~\ref{sec:fcs}}
\label{sec:First_passage_resettings}

The characteristic function of the first-passage time for $W$ resets is calculated from the definition Eq.~(\ref{eqn:fw_def}) as, 
\begin{align}
F_\chi(s) =& \int_0^\infty d \tau e^{-s \tau} \sum_{W=1}^\infty e^{i \chi W} F_W(\tau) \nonumber \\ =& \langle {\rm i}| \hat{V}(\chi) (s \hat{I} - \hat{L}^{\rm f} - \hat{V}(\chi))^{-1} |{\rm i} \rangle = \frac{e^{i \chi} F_1(s)}{1 - e^{i \chi} F_1(s)} \, , 
\label{eqn:cffw}
\end{align}
where the modified transition rate matrix (\ref{eqn:reset_ham}) is, 
\begin{align}
\hat{V}(\chi) = \Gamma_{\rm reset} (|{\rm i} \rangle \langle {\rm f}| e^{i \chi} - |{\rm f} \rangle \langle {\rm f}|) \, . \label{eqn:reset_ham_modify}
\end{align}
Here, 
\begin{align}
F_1(s) =& \Gamma_{\rm reset} \langle {\rm f} | ( s \hat{I} - \hat{L}^{\rm f} +\Gamma_{\rm reset} |{\rm f} \rangle \langle {\rm f}| )^{-1} |{\rm i} \rangle \\
=&
\frac{ \Gamma_{\rm reset} \langle {\rm f}| (s \hat{I} - \hat{L}^{\rm f})^{-1} |{\rm i} \rangle }{ 1 + \Gamma_{\rm reset} \langle {\rm f} |(s \hat{I} - \hat{L}^{\rm f})^{-1} |{\rm f} \rangle } \, ,
\end{align}
which results in Eq.~(\ref{eqn:cff1}) in the limit of $\Gamma_{\rm reset} \to \infty$. 

The probability distribution of the number of resets $W$ during the measurement time $\tau$ is expressed by the probability distributions of first-passage time $\tau$ for $W$ resets, Eq.~(\ref{eqn:fw_def}), as~\cite{Redner2001}, 
\begin{align}
P_\tau(W) =& \int_0^\tau d \tau' P_{\tau - \tau'}(0) F_W(\tau') \, ,
\label{eqn:pw} \\
P_\tau(0) =& \langle e | e^{ \hat{L}^{\rm f} \tau } | {\rm i} \rangle \, .
\label{eqn:p0}
\end{align}
Equation~(\ref{eqn:pw}) is solved by performing the Laplace transform, which results in Eq.~(\ref{eqn:cf_fpt}). 
 
The characteristic function of the probability distribution of the number of resets is, 
\begin{align}
P_\tau(\chi) =& \sum_{W=0}^\infty e^{i \chi W} P_\tau(W) = \langle e | e^{ \hat{L}^{\rm f}(\chi) \tau} |{\rm i} \rangle \, , \label{eqn:cf} \\
\hat{L}^{\rm f}(\chi) =& \hat{L}^{\rm f}+\hat{V}(\chi) \, . \label{eqn:mod_Liou}
\end{align}
In the limit of a long measurement time $\tau$, the characteristic function is approximately $\sum_{W=0}^\infty e^{i \chi W} P_\tau(W) \approx e^{\Lambda_0(\chi) \tau}$. 
Here $\Lambda_0(\chi)$ is given by Eq.~(\ref{eqn:cf_long_tau}) and is the eigenvalue of the modified transition rate matrix $\hat{L}^{\rm f}(\chi)$ with maximum real part. 
Equation (\ref{eqn:pws}) is obtained after the inverse Fourier transform of the Laplace transform of Eq.~(\ref{eqn:cf}), 
\begin{align}
P_s(W) = \int_{-\pi}^\pi \frac{d \chi}{2 \pi} e^{-i \chi W} \int_0^\infty d \tau e^{-s \tau} P_\tau(\chi) \, .
\end{align}

The eigenvalue of the modified transition rate matrix in the limit of $\Gamma_{\rm reset} \to \infty$ is derived as follows. 
The characteristic polynomial is, 
\begin{align}
p(\Lambda) =& {\rm det} \left( \Lambda \hat{I} -\hat{L}^{\rm f}(\chi) \right) 
={\rm det} \left[ \hat{Q} \left( \Lambda \hat{I} -\hat{V}(\chi) \right) \hat{Q}  \right] \nonumber \\ 
& \times {\rm det} \biggl[ \Lambda \hat{I}- \hat{P} \hat{L}^{\rm f} \hat{P} \nonumber \\ &- \hat{P} \hat{V}(\chi) \hat{Q} \left( \Lambda \hat{I}- \hat{Q} \hat{V}(\chi) \hat{Q} \right)^{-1} \hat{Q} \hat{L}^{\rm f} \hat{P}  \biggl] \nonumber \\
\approx& ( \Lambda + \Gamma_{\rm reset}) {\rm det} \left( \Lambda \hat{I}- \hat{ {L} }(\chi) \right) =0 \, ,
\end{align}
where the projection operators are, 
$\hat{Q} = |f \rangle \langle f|$ and $\hat{P} = \hat{I}-\hat{Q}$. 
The effective transition rate matrix $\hat{L}(\chi)$ is defined in Eq.~(\ref{eqn:eff_liouvill}). 

The counting field for the activity $\eta$ can also be included straightforwardly. 
The following discussion is not limited to the activity but apply to any stochastic observable $O$ measured by a counting field, including the environment entropy production. 
The scaled CGF expanded up to the second-order in the counting fields is, 
\begin{align}
\Lambda_0 (\chi,\eta) \approx {\bm C}_1^T {\bm x} + \frac{1}{2} {\bm x}^T {\bm C}_2 {\bm x} \, , \label{eqn:gauss_expansion} \;\;\;\; {\bm x} = \left[ \begin{array}{c} i \chi \\ i \eta \end{array} \right] \, ,
\end{align}
whereby the first and second cumulants are, 
\begin{align}
{\bm C}_1 = \left[ \begin{array}{c} \langle \! \langle w \rangle \! \rangle \\ \langle \! \langle a \rangle \! \rangle \end{array} \right]
\, , \;\;\;\;
{\bm C}_2 = \left[ \begin{array}{cc} \langle \! \langle w^2 \rangle \! \rangle & \langle \! \langle a w \rangle \! \rangle \\ \langle \! \langle w a \rangle \! \rangle & \langle \! \langle a^2 \rangle \! \rangle \end{array} \right] \, . 
\end{align}
Within the Gaussian approximation (\ref{eqn:gauss_expansion}), the cumulant generating function is, 
\begin{align}
\ln F_W(s,\eta) =& \ln \frac{P_W(s,\eta)}{P_0(s,\eta)} \nonumber \\
=& W \frac{ \langle \! \langle \tilde{w} \rangle \! \rangle - \sqrt{ \langle \! \langle \tilde{w} \rangle \! \rangle^2 + 2 \tilde{s} \langle \! \langle w^2 \rangle \! \rangle } }{\langle \! \langle w^2 \rangle \! \rangle} \, , \label{eqn:cgf_gauss} \\
\langle \! \langle \tilde{w} \rangle \! \rangle = & \langle \! \langle w \rangle \! \rangle+ \langle \! \langle w a \rangle \! \rangle i \eta \, , \nonumber \\
\tilde{s} = & s- \langle \! \langle a \rangle \! \rangle i \eta + \langle \! \langle a^2 \rangle \! \rangle (i \eta)^2/2 \, . \nonumber
\end{align}
Then the joint probability distribution of the first-passage time and activity is, 
\begin{align}
F_W(\tau, A) =& \int_0^{2 \pi} \frac{d \eta}{2 \pi} e^{- i \eta A} \int_{a-i \infty}^{a+ i \infty} \frac{d s}{2 \pi i} e^{s \tau} F_W(s,\eta) \, , \nonumber \\ \approx& \frac{W}{ \tau \, \sqrt{ {\rm det} \left( 2 \pi \tau {\bm C}_2 \right) } } \exp \left( - \frac{\delta {\bm J}^T {\bm C}_2^{-1} \delta {\bm J}}{2 \tau} \right) \, , \label{eqn:inverse_gauss_tau_a} 
\end{align}
where, 
\begin{align*}
\delta {\bm J} =&  \left[ \begin{array}{c} W - \tau \langle \! \langle w \rangle \! \rangle \\ A - \tau \langle \! \langle a \rangle \! \rangle \end{array} \right] \, . 
\end{align*}
Equation~(\ref{eqn:inverse_gauss_tau}) is obtained by performing the integral of Eq.~(\ref{eqn:inverse_gauss_tau_a}) over $A$. 

From the derivatives of CGF, Eq.~(\ref{eqn:cgf_gauss}), we obtain the cumulants, see Eq.~(\ref{eqn:joint_cumulants}): 
\begin{align}
\langle \! \langle \tau \rangle \! \rangle_F =& \frac{W}{ \langle \! \langle w \rangle \! \rangle } \, , \tag{ \ref{eqn:c_t} } \\
\langle \! \langle A \rangle \! \rangle_F =& \langle \! \langle a \rangle \! \rangle \langle \! \langle \tau \rangle \! \rangle_F \, .  \label{eqn:c_a} \\
\langle \! \langle \tau^2 \rangle \! \rangle_F =& W \frac{ \langle \! \langle w^2 \rangle \! \rangle }{ \langle \! \langle w \rangle \! \rangle^3 } \, , \label{eqn:c_tt} \\
\langle \! \langle A^2 \rangle \! \rangle_F =& W \frac{ \langle \! \langle w^2 \rangle \! \rangle \langle \! \langle a \rangle \! \rangle^2 + \langle \! \langle w \rangle \! \rangle^2 \langle \! \langle a^2 \rangle \! \rangle - 2 \langle \! \langle w \rangle \! \rangle \langle \! \langle a \rangle \! \rangle \langle \! \langle w a \rangle \! \rangle } { \langle \! \langle w \rangle \! \rangle^3 }
\, , \label{eqn:c_aa} \\
\langle \! \langle \tau A \rangle \! \rangle_F =& W \frac{ \langle \! \langle w^2 \rangle \! \rangle \langle \! \langle a \rangle \! \rangle - \langle \! \langle w \rangle \! \rangle \langle \! \langle w a \rangle \! \rangle }{ \langle \! \langle w \rangle \! \rangle^3 }  \label{eqn:c_ta} \, .
\end{align}
By substituting Eqs.~(\ref{eqn:c_t}) and (\ref{eqn:c_tt}) into Eq.~(\ref{eqn:snr}), we obtain Eq.~(\ref{eqn:snr_wald}).

\section{Detailed calculations of Sec.~\ref{sec:semi-infinite_reversible_TM}
 }
\label{sec:suppl_semi-infinite_reversible_TM}

Instead of Eq.~(\ref{eqn:revTM_liouvill}) with $|\Omega| \to \infty$, 
it is simpler to consider the infinite chain~\cite{Singh2019}, 
\begin{align}
\hat{L}_N(\xi,\eta) =& \sum_{n=-N/2+1 }^{N/2} \biggl( - (\gamma^+ + \gamma^-) |n \rangle \langle n| \nonumber \\ & + \gamma^+ e^{i \Delta \Sigma_{\rm bi}^{\rm env} \xi + i \eta} |n \rangle \langle n-1| \nonumber \\
&+ \gamma^- e^{-i \Delta \Sigma_{\rm bi}^{\rm env} \xi + i \eta} |n \rangle \langle n+1| \biggl) \, ,
\end{align}
and impose the periodic boundary condition $|N/2 + j \rangle = |-N/2+j \rangle$ ($j=0,1$). 
The discrete Fourier transform, 
\begin{align}
|\phi_{k_\ell} \rangle = \frac{1}{\sqrt{N}} \sum_{n=-N/2+1}^{N/2} e^{i k_\ell n} |n \rangle \, , \;\;\;\; k_\ell = \frac{2 \pi \ell}{N} \, , 
\end{align}
diagonalizes the transition rate matrix. 
By taking $N \to \infty$, we derive the propagator in Eq.~(\ref{eqn:cff1}), 
\begin{align}
\langle n'|( s \hat{I} - \hat{L}_N)^{-1} |n \rangle &= \int_{-\pi}^\pi \frac{d k}{2 \pi} \frac{e^{i (n-n') k}}{s- \Lambda (\xi,\eta;k)} \, , \label{eqn:gf} 
\end{align}
where 
\begin{align}
\Lambda(\xi,\eta;k) =& \gamma^+ (e^{i k + i \xi \Sigma_{\rm bi}^{\rm env} + i \eta}-1) \nonumber \\ &+ \gamma^- (e^{-i k - i \xi \Delta \Sigma_{\rm bi}^{\rm env} + i \eta}-1) \, , 
\label{eqn:scgf_bidirectional_Poisson}
\end{align}
is the cumulant generating function of the bidirectional Poisson form or the Skellam distribution.  
By substituting Eq.~(\ref{eqn:gf}) into Eq.~(\ref{eqn:cff1}), we obtain Eq.~(\ref{eqn:cgf_infinite_TM}). 
The first and second cumulants are, 
\begin{align}
\langle \! \langle \tau \rangle \! \rangle_F =& \frac{m}{\gamma^+ - \gamma^-} \, , \tag{ \ref{eqn:infiniteTM_ave_tau} }
\\
\langle \! \langle A \rangle \! \rangle_F =& (\gamma^+ + \gamma^-) \langle \! \langle \tau \rangle \! \rangle_F \, , \label{eqn:infiniteTM_ave_A} 
\\
\langle \! \langle \tau^2 \rangle \! \rangle_F =& \frac{\gamma^+ + \gamma^-}{m (\gamma^+ - \gamma^-)}  \langle \! \langle \tau \rangle \! \rangle_F^2 \, , \label{eqn:infiniteTM_var_tau} 
\\
\langle \! \langle A^2 \rangle \! \rangle_F =& 4 \gamma^+ \gamma^- \langle \! \langle \tau^2 \rangle \! \rangle_F \, , \label{eqn:infiniteTM_var_A}
\\
\langle \! \langle A \tau \rangle \! \rangle_F =& \frac{ \langle \! \langle A^2 \rangle \! \rangle_F }{\gamma^+ + \gamma^-} \, . \label{eqn:infiniteTM_cross}
\end{align}

\section{Dyson series, Gillespie's algorithm and stochastic entropy}
\label{sec:Gillespies_algorithm}

For the first-passage time problem, a stochastic trajectory with $N$ transitions is described by the successive times at which the transitions happen, Eq.~(\ref{eqn:tN}), as well as by the successive states, Eq.~(\ref{eqn:omegaN}), with the following constraints: 
($\mathrm{i}$) The transitions are induced by the transition rate matrix $\hat{L}^{\rm f}$ in Eq.~(\ref{eqn:full_Liouvillian}), and thus a transition satisfies $(|\omega_{k+1} \rangle \leftarrow |\omega_{k} \rangle) \in E_{\rm bi}^{\rm f} \cup E_{\rm uni}^{\rm f}$, 
where the sets of directed edges $E_{\rm uni}^{\rm f}$ and $E_{\rm bi}^{\rm f}$ are described by Eq.~(\ref{eqn:E_uni}) and Eq.~(\ref{eqn:E_bi}) with the replacements, 
$\Gamma_{i,j} \to \Gamma_{i,j}^{\rm f}$ and $\Omega \to \Omega \cup \{ |{\rm f} \rangle \}$. 
($\mathrm{ii}$) At the initial time $t_0=0$, the system is in the initial state $|\omega_0 \rangle =|{\rm i} \rangle$. 
($\mathrm{iii}$) The first-passage time is introduced as $t_N=\tau$. 
Namely, at $t_N=\tau$, the final state $|\omega_N \rangle =|{\rm f} \rangle$ is reached for the first time, and the successive states, according to Eq.~(\ref{eqn:omegaN}), satisfy $|\omega_j \rangle \neq |{\rm f} \rangle$ for $j=0,\cdots,N-1$. 
The summation of Eq.~(\ref{eq:sojourn}) is performed over $\sum_{k=0}^{N-1}$.

The path probability is, 
\begin{align}
Q \left( \omega^{N+1}, t^{N+1} \right) = \prod_{k=0}^N P(\omega_{k+1}|\omega_{k}) P(\omega_{k};t_{k+1} -t_{k}) \, , \label{eqn:path_prob}
\end{align}
where the sojourn times in the state $\omega$ are exponentially distributed: 
\begin{align}
P(\omega;t) = \gamma_{\omega}^{\rm f} e^{- \gamma_{\omega}^{\rm f} t} \, ,
\;\;\;\;
\gamma^{\rm f}_{n} = \sum_{n' (\neq n)} \Gamma^{\rm f}_{n',n} \, .
\end{align}
The conditional probability for the transition from $|\omega \rangle$ to $|\omega' \rangle$ is 
\begin{align}
P(\omega'|\omega) = \frac{\Gamma_{\omega',\omega}^{\rm f}}{\gamma_\omega^{\rm f}} \, . 
\end{align}
The characteristic function is expressed in a Dyson series~\cite{Bagrets2003,Utsumi2007} or a path ensemble form~\cite{Esposito2010}:
\begin{align}
F_1(\tau,A) &= \sum_{N=0}^\infty \sum_{ \omega^{N+1} }
 \int_0^\tau dt_{N-1} \int_0^{t_{N-1}} dt_{N-2} \cdots \int_0^{t_2} dt_{1} 
\nonumber \\ &\times Q \left( \omega^{N+1}, t^{N+1} \right) \delta \left( A -A\left( \omega^{N+1} \right) \right) \, , \label{eqn:F_1_Dyson}
\end{align}
where we accounted for the activity, 
\begin{align}
A \left( \omega^{N+1} \right) = N \, . 
\end{align}
It is straightforward to extend the above derivations to the transition rate matrix $\hat{L}^{\rm f}+\hat{V}$ and derive Eq.~(\ref{eqn:fw_def}). 

For each trajectory, the total sojourn time according to Eq.~(\ref{eq:sojourn}), and the number of transitions, Eq.~(\ref{eq:jump}), are defined for the states $|\omega \rangle \in \Omega \cup \{ | {\rm f} \rangle \}$.
The path environment entropy production is
\begin{align}
\Sigma^{\rm env}_{\rm bi} \left( \omega^{N+1} \right) = \sum_{ (\omega \leftarrow \omega') \in E_{\rm bi}^{\rm f} } W_{ \omega \leftarrow \omega' } \left( \omega^{N+1} \right) \ln \frac{ \Gamma_{\omega, \, \omega'}}{\Gamma_{\omega', \, \omega}} \, . 
\end{align}
In stochastic thermodynamics~\cite{Seifert2012,VandenBroeck2015,Fuchs2016}, the system entropy for each stochastic trajectory, i.e., the stochastic entropy, is introduced. 
We define the stochastic entropies associated with unidirectional and bidirectional transitions as, 
\begin{align}
\Sigma^{\rm sys}_{\rm uni} \left( \omega^{N+1},t^{N+1} \right) =& \sum_{n=0}^{N-1} \ln \frac{ p_{\omega_{n}} (t_{n+1}-0) }{ p_{\omega_{n+1}} (t_{n+1}+0) }
\nonumber \\ & \times \phi_{E_{\rm uni}}( (\omega_{n+1} \leftarrow \omega_{n})  ) \, , \label{eqn:path_ent_pro_uni} \\
\Sigma^{\rm sys}_{\rm bi} \left( \omega^{N+1},t^{N+1} \right) =& \ln \frac{p_{\omega_{0}} (t_{0})}{p_{\omega_{N}} (t_{N})} - \Sigma^{\rm sys}_{\rm uni} \left( \omega^{N+1},t^{N+1} \right) \, , \label{eqn:path_ent_pro_bi} 
\end{align}
where the indicator function is $\phi_E(x)=1$ for $x \in E$ and $\phi_E(x)=0$ for $x \notin E$. 
The distribution probability $p_\omega(t)$ at time $t$ is obtained by solving the master equation with a given initial distribution probability~\cite{Seifert2012}. 
In the present paper, we approximate it by the steady-state distribution probability, $p_\omega(t) \approx p_\omega^{\rm st}$.

Gillespie's algorithm numerically generates sequences (\ref{eqn:tN}) and (\ref{eqn:omegaN}) obeying the path probability (\ref{eqn:path_prob})~\cite{Gillespie1976,Gillespie1977}. 
For each sample $(t^N,\omega^N)$, the quantities $\tau$, $\tau_\omega$, $\Sigma_{\rm bi}^{\rm env}$, $\Sigma_{\rm uni(bi)}^{\rm sys}$ and $A$ are evaluated. 
For example, to obtain the average and the variance of the sojourn time, we first generate $N_{\rm sample}$ sequences $t_1^{N+1}, t_2^{N+1}, \cdots, t_{N_{\rm sample}}^{N+1}$ numerically. 
Then their sample average and sample variances are, 
\begin{align}
\bar{\tau}_\omega =& \frac{1}{N_{\rm sample}} \sum_{j=1}^{N_{\rm sample}} \tau_{\omega} \left( t_j^{N+1} \right) \, , \\
S^2 =& \frac{1}{N_{\rm sample}} \sum_{j=1}^{N_{\rm sample}} \left( \tau_{\omega} \left( t_j^{N+1} \right) - \bar{\tau}_\omega \right)^2 \, , 
\end{align}
representing the numerical values of the average $\langle \! \langle \tau \rangle \! \rangle_F$ and the variance $\langle \! \langle \tau^2 \rangle \! \rangle_F$. 
The distribution probability is 
\begin{align}
p_\omega = \frac{ \bar{\tau}_{\omega} }{ \sum_{\omega'} \bar{\tau}_{\omega'} } \, . \label{eqn:dis_pro_gillespie}
\end{align}
For our case, Eq.~(\ref{eqn:dis_pro_gillespie}) approximates the steady-state distribution probability, $p_\omega^{\rm st} \approx p_\omega$. 
Due to error-free resetting, the distribution probability of the final state is set to be that of the initial state $p_{{\rm f}}=p_{{\rm i}}$. 
Therefore, the first term of the right-hand side of Eq.~(\ref{eqn:path_ent_pro_bi}) vanishes.

\section{Rayleigh-Schr\"odinger perturbation theory}
\label{sec:RSPT} 

Suppose ${\bm L}^{\rm f}$ is a $ (|\Omega|+1) \times (|\Omega|+1)$ transition rate matrix corresponds to Eq.~(\ref{eqn:full_Liouvillian}). 
Then the matrix representation of Eq.~(\ref{eqn:eff_liouvill}) becomes,
\begin{align}
{\bm L}(\chi) = {\bm P} {\bm L}^{\rm f} {\bm P}^T + {\bm Q}' {\bm L}^{\rm f} {\bm P}^T e^{i \chi} \, . 
\end{align}
Here, we introduce $|\Omega| \times (|\Omega|+1)$ projection matrices ${\bm P}$ and ${\bm Q}'$, 
\begin{align}
{\bm P}^T =& ( {\bm u}_1, \cdots, \overset{n_{\rm f}-1}{ {\bm u}_{n_{\rm f}-1} }, \overset{n_{\rm f}}{ {\bm u}_{n_{\rm f}+1} } , \cdots, \overset{|\Omega|}{ {\bm u}_N } ) \, , \\
{\bm Q}^{\prime \, T} =& ( {\bm 0}, \cdots, \overset{n_{\rm i}-1}{ {\bm 0} }, \overset{n_{\rm i}}{ {\bm u}_{n_{\rm f}} }, \overset{n_{\rm i}+1}{ {\bm 0} } , \cdots, \overset{|\Omega|}{ {\bm 0} } ) \, , 
\end{align}
where
$({\bm u}_k)_\ell = \delta_{k,\ell}$ is a $|\Omega|+1$ component unit vector and $( {\bm 0} )_\ell = 0$ is a $|\Omega|+1$ component zero vector. 

The counting field $\chi$ counts the number of resets $W$. 
We attach the phase factor $e^{i \eta}$ to all off-diagonal components of the transition rate matrix, 
\begin{align}
\left [ {\bm L}(\chi,\eta) \right]_{i,j} = \left[ {\bm L}(\chi) \right]_{i,j} e^{i \eta} \, , \;\;\;\; (i \neq j ) \, ,
\end{align}
to count the number of jumps, i.e., the activity.

To numerically calculate the first and second cumulants, we adopt Rayleigh-Schr\"odinger perturbation theory~\cite{FlindtPRB2010,Takase2021}. 
The basic idea is to solve the eigenvalue equation, 
\begin{align}
\left( { {\bm L} }(\chi,\eta) - \Lambda_0(\chi,\eta) \right) {\bm v}_0(\chi,\eta) =0 \, , \label{eqn:eigenvalue_equation}
\end{align}
by applying perturbation theory. 
Here $\Lambda_0$ is the eigenvalue with the maximum real part satisfying $\Lambda_0(0,0)=0$. 
The zero eigenvector is normalized to satisfy ${\bm v}_0(0,0)={\bm p}^{\rm st}$. 
First, we perform the following expansions: 
\begin{align}
{ {\bm L} }(\chi,\eta) =& \sum_{k=0}^\infty \sum_{\ell=0}^\infty \frac{(i \chi)^k (i \eta)^\ell}{k ! \ell !} { {\bm L} }^{(k,\ell)} \, , \\
\Lambda_0(\chi,\eta) =& \sum_{k=0}^\infty \sum_{\ell=0}^\infty \frac{(i \chi)^k (i \eta)^\ell}{k ! \ell !} \langle \! \langle w^k a^\ell \rangle \! \rangle \, ,\\
{\bm v}_0(\chi,\eta) =& \sum_{k=0}^\infty \sum_{\ell=0}^\infty \frac{(i \chi)^k (i \eta)^\ell}{k ! \ell !} {\bm v}_0^{(k,\ell)} \, .
\end{align}
Then we compare the left-hand side and right-hand side of Eq.~(\ref{eqn:eigenvalue_equation}) order by order~\cite{JJSakurai,FlindtPRB2010,Takase2021}, 
giving, 
\begin{align}
\langle \! \langle w \rangle \! \rangle = {\bm e}_{|\Omega|}^T { {\bm L} }^{(1,0)} {\bm p}^{\rm st} \, , \;\;\;\;
\langle \! \langle a \rangle \! \rangle = {\bm e}_{|\Omega|}^T { {\bm L} }^{(0,1)} {\bm p}^{\rm st} \, , 
\end{align}
\begin{align}
\langle \! \langle w^2 \rangle \! \rangle =& {\bm e}_{|\Omega|}^T \left( { {\bm L} }^{(2,0)} - 2 { {\bm L} }^{(1,0)} {\bm R} { {\bm L} }^{(1,0)} \right) {\bm p}^{\rm st} \, , \\
\langle \! \langle a^2 \rangle \! \rangle =& {\bm e}_{|\Omega|}^T \left( { {\bm L} }^{(0,2)} - 2 { {\bm L} }^{(0,1)} {\bm R} { {\bm L} }^{(0,1)} \right) {\bm p}^{\rm st} \, , \\
\langle \! \langle w a \rangle \! \rangle =& {\bm e}_{|\Omega|}^T \left( { {\bm L} }^{(1,1)} - { {\bm L} }^{(1,0)} {\bm R} { {\bm L} }^{(0,1)} \right. 
\nonumber \\
& \left. - { {\bm L} }^{(0,1)} {\bm R} { {\bm L} }^{(1,0)} \right) {\bm p}^{\rm st} \, , 
\end{align}
where ${ {\bm e}_{n} } =(1,\cdots,1)^T $ is an $n$-component vertical vector with $1$ for all entries. 
The pseudo-inverse satisfies 
${\bm R} { {\bm L} }^{(0,0)}  = \bm{I} - {\bm p}^{\rm st} {\bm e}_{|\Omega|}^T$.
Setting $\chi=\eta=0$ and omitting $(0,0)$,
we obtain the explicit form in our numerical calculation as, 
${\bm R} = \sum_{i,j \neq 0} {\bm v}_i g_{i,j} \tilde{\bm v}_j^T/\Lambda_j$,
whereby 
${\bm L}^T \tilde{\bm v}_j=\Lambda_j \tilde{\bm v}_j$
and 
$\sum_{j} g_{i,j} \tilde{\bm v}_j^T {\bm v}_k = \delta_{i,k}$. 
To obtain this from, we used $\tilde{\bm v}_0 = {\bm e}_{|\Omega|}$ and $\tilde{\bm v}_0^T {\bm v}_j = \tilde{\bm v}_j^T {\bm v}_0 = 0$ for $j \neq 0$.

\section{Stochastic Petri net}
\label{sec:SPN}

A stochastic Petri net (SPN)~\cite{Murata1989} is a quintuple $SPN=(P,T,F,M_0, {\it \Gamma} )$ for which  
\begin{enumerate}
\item
$P= \{ p_1,p_2, \cdots, p_{|P|} \}$ is a finite set of places. 

\item
$T= \{ t_1,t_2, \cdots, t_{|T|} \}$ is a finite set of transitions. 

\item
$F \subseteq (P \times T) \cup (T \times P)$ is a finite set of arcs representing the flow relation. 
$I=\{ (p_{i_1}  \to  t_{i_1}),(p_{i_2}  \to  t_{i_2}), \cdots (p_{ i_{|I|} }  \to  t_{ i_{|I|} }) \} \subseteq P \times T$ is a set of input arcs and $O =\{ (t_{i_1} \to p_{i_1}),(t_{i_2} \to p_{i_2}), \cdots (t_{ i_{|O|} } \to p_{ i_{|O|} }) \} \subseteq T \times P$ is a set of output arcs. 

\item
$M_0 = \{ M_0({p_1}) , M_0({p_2}), \cdots, M_0({p_{|P|}}) \} $ is an initial marking. 

\item
$ {\it \Gamma} = \{ \gamma(t_1), \gamma(t_2), \cdots , \gamma(t_{|T|}) \} $ is a set of firing rates associated with the transitions. 

\end{enumerate}
The sets of places and transitions satisfy $P \cup T = \emptyset$ and $P \cap T \ne \emptyset$. 

A Petri net is visualized as a bipartite graph consisting of two kinds of nodes, {\it places} and {\it transitions}, drawn as circles and bars, respectively, see Figs.~\ref{fig:Circuit_elements} (b-1), (b-2) and (b-3). 
An arc, or a directed edge, is represented by a directed arrow connecting a place and a transition. 
The forward (backward) incidence matrix with the matrix elements $w(p \leftarrow t)$ ($w(t \leftarrow p)$), each of which equals the number of arcs from $t$ ($p$) to $p$ ($t$), characterizes the structure of the graph.  
The marking of the Petri net is updated according to its transition (firing) rule. 
A firing of an enabled transition $t$ removes $w(t \leftarrow p)$ tokens from each input place $p$ of $t$, and adds $w(p' \leftarrow t)$ tokens to each output place $p'$ of $t$. 
For the stochastic Petri net, transitions are governed by the Doi-Peliti Hamiltonian Eq.~(\ref{eqn:Doi_Peliti_Ham}). 
The SPN of the token-based Brownian circuit is {\it pure}, i.e., it does not contain self-loops.

\section{Doi-Peliti formalism of SPN}
\label{sec:DPH}

The state with $n$ tokens is defined as $|n \rangle = (\hat{a}^\dagger )^n |0 \rangle$, where $|0 \rangle$ is the vacuum state, $\hat{a} |0 \rangle = 0$. 
In this convention, 
$ \hat{a}^\dagger |n \rangle = |n+1 \rangle$ and 
$ \hat{a} |n \rangle = n |n-1 \rangle$. 
The left and right eigen vectors are orthogonal 
$\langle n | n' \rangle = n! \delta_{n,n'}$~\cite{Kamenev2002,Kamenevbook2011}. 
Let the set of places be $P=\{p_1,p_2, \cdots, p_{|P|} \}$. 
The state is specified by the {\it markings}~\cite{Murata1989}, 
\begin{align}
M(P) = \{ M(p_1),M(p_2),\cdots, M(p_{|P|}) \} \, , 
\end{align}
where $M(p)$ is the number of tokens contained in the place $p$. 
In the Doi-Peliti formalism, this corresponds to the Fock state (the number state), 
\begin{align}
|M(p_1),M(p_2),\cdots, M(p_{|P|}) \rangle = \prod_{p \in P} \left( \hat{a}_{p}^\dagger \right)^{M(p)} |0 \rangle  \, . \label{eqn:fock}
\end{align}
The SPNs of token-based Brownian circuits in the present paper are {\it safe} or {1-bounded}~\cite{Peper2013,Lee2016}, i.e. at most one token is contained in each place, $M(p)=0,1$. 
In addition, the circuit elements conserve the number of tokens. 
Therefore, for a system with $N \leq {|P|}$ tokens, 
$N$ places are singly occupied, and the other places are empty. 
In such a case, it is convenient to specify the $N$-token state with the set of occupied places $Q=\{ q_1, q_2, \cdots, q_N \}$ as, 
\begin{align}
|q_1, q_2, \cdots q_N \rangle = \prod_{\ell = 1}^N \hat{a}_{q_\ell}^\dagger |0 \rangle  \, .
\end{align}
We adopt this representation in Eq.~(\ref{eqn:Ntokenstate}).

The stochastic dynamics is characterized by sets of transitions $T= \{ t_1,t_2, \cdots, t_{|T|} \}$ with transition rates $ {\it \Gamma} = \{ \gamma(t_1), \gamma(t_2), \cdots , \gamma(t_{|T|}) \} $. 
In the Doi-Peliti formalism, the modified transition rate matrix (the Doi-Peliti Hamiltonian) of $SPN=(P,T,F,M_0, {\it \Gamma} )$, see Appendix~\ref{sec:SPN}, is, 
\begin{align}
\hat{H}( \{ \hat{a}_i^\dagger \}, \{ \hat{a}_i \} ; \xi, \eta) =& \sum_{j=1}^{|T|} \gamma(t_j) \left( \hat{f}_j(\xi)^\dagger e^{i \eta + i \xi \Delta \Sigma_{\rm bi}^{\rm env}(t_j) } \hat{b}_j(\xi) \right. \nonumber \\ & \left. - \hat{b}_j(0)^\dagger \hat{b}_j(0) \right)  \, , \label{eqn:Doi_Peliti_Ham} \\ 
\hat{f}_j(\xi) =& \prod_{i=1}^{|P|} \left( \hat{a}_i \hat{N}_i^{i \xi} \right)^{w(p_i \leftarrow t_j )} \, , \label{eqn:f_DP} \\
\hat{b}_j(\xi) =& \prod_{i=1}^{|P|} \left( \hat{a}_i \hat{N}_i^{i \xi} \right)^{w(t_j \leftarrow p_i)} \, , \label{eqn:b_DP}
\end{align}
where $\hat{N}_i=\hat{a}_i^\dagger \hat{a}_i$ is the number operator of tokens in the place $p_i$. 
Here, $w(p_i \leftarrow t_j )$ is the matrix element of the forward incidence matrix, while $w(t_i \leftarrow p_i)$ is the matrix element of the backward incidence matrix. 
If the transition $t_{\tilde{j}}$ is the time reversal of the transition $t_j$, where $\hat{f}_{\tilde{j}}=\hat{b}_j$, the environment entropy production is 
\begin{align}
\Delta \Sigma_{\rm bi}^{\rm env}(t_j) = \ln \frac{\gamma(t_j)}{\gamma(t_{\tilde{j}})} \, . 
\end{align}
The factor $\hat{N}_i^{i \xi}$ in Eqs.~(\ref{eqn:f_DP}) and (\ref{eqn:b_DP}) is introduced to preserve the symmetry of the fluctuation relation~\cite{Golubev2011}, $\Gamma_{n,n'}(\xi)=\Gamma_{n',n}(-\xi+i)$. 
Although this factor is crucial to be thermodynamically consistent, it can be omitted if the Petri net is safe. 
If the time-reversal of the transition $t_j$ is absent, the transition is unidirectional. 
Such a process is associated with a Ratchet, and we do not assign a counting field $\xi$ to it, or equivalently, we set $\Delta \Sigma_{\rm bi}^{\rm env}(t_j) = 0$. 
Note that the modified transition rate matrix satisfies the normalization~\cite{Kamenev2002}
$\hat{H}( \{ \hat{a}_i^\dagger=1 \}, \{ \hat{a}_i \}; \xi=0, \eta=0 ) = 0$.

\section{Fibers and preimages}
\label{sec:preima}

The fibers and preimages~\cite{set_theory_1} of $f_{\rm HA}$ and $f_{\rm DOR}$ are summarized as follows: 
\begin{align}
f_{\rm HA}^{-1}(0,0) =& \{ (0,0) \}  \, , \\
f_{\rm HA}^{-1}(1,0) =& \{ (0,1) , (1,0) \}  \, , \\
f_{\rm HA}^{-1}(0,1) =& \{ (1,1) \}  \, .
\end{align}
\begin{align}
f_{\rm HA}^{-1}(0,\cup) =& \{ (0,0),(1,1) \} \, , \label{eqn:preimag_HA_1} \\
f_{\rm HA}^{-1}(1,\cup) =& \{ (0,1),(1,0) \} \, , \\
f_{\rm HA}^{-1}(\cup,0) =& \{ (0,0),(0,1),(1,0) \} \, , \\
f_{\rm HA}^{-1}(\cup,1) =& \{ (1,1) \} \, . \label{eqn:preimag_HA_4}
\end{align}
\begin{align}
f_{\rm DOR}^{-1}(0,0) =& \left \{ (0,0) \right \} \, , \\
f_{\rm DOR}^{-1}(1,1) =& \left \{ (0,1), (1,0), (1,1) \right \} \, .
\end{align}
\begin{align}
f_{\rm DOR}^{-1}(0,\cup) =& f_{\rm DOR}^{-1}(\cup,0) = \{ (0,0) \} \, , \label{eqn:preimag_DOR_1} \\
f_{\rm DOR}^{-1}(1,\cup) =& f_{\rm DOR}^{-1}(\cup,1)  = \{ (0,1), (1,0), (1,1) \} \, . \label{eqn:preimag_DOR_2}
\end{align}

\section{Serial number for reachable fine-grained multi-token states}
\label{sec:Ser_num_for_fin_gra_mul_tok_sta}

In the following, we focus on one of the separate blocks of the full transition rate matrix corresponding to a specific reachability graph. 
The $N$-token logical state is specified with a serial number $m$, 
\begin{align}
|m \rangle = \left| x_1^{(m)}, \cdots , x_N^{(m)} \right \rangle \, . \label{eqn:mthNtokenlogicalstate}
\end{align}
Suppose there are $M$ states reachable from the initial $N$-token logical state $| m_{\rm i} \rangle$. 
The sets of nodes and arcs of the reachability graph are, 
\begin{align}
{\mathcal R}( | m_{\rm i} \rangle ) =& \left \{ |1 \rangle \, , \cdots \, , |M \rangle \right \} \, , \label{eqn:set_RE} \\
{\mathcal E}(| m_{\rm i} \rangle) =& \{ ( |m'_1 \rangle \leftarrow |m_1 \rangle ),  \cdots, ( | m'_{|{\mathcal E}|} \rangle \leftarrow | m_{|{\mathcal E}|} \rangle ) \} \, , \label{eqn:edge_RA}
\end{align}
where $| m_{\rm i} \rangle \in {\mathcal R}( | m_{\rm i} \rangle )$. 
For example, for panel (g) in Fig.~\ref{fig:reachabilityHAFA}, 
$N=4$, $M=13$ and the initial $N$-token logical state is $m_{\rm i}=1, 2, 3$, or $4$. 

\begin{figure}[ht]
\begin{center}
\includegraphics[width=0.6 \columnwidth]{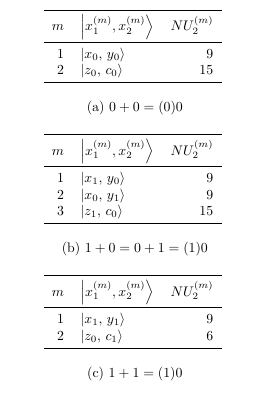}
\caption{ Reachable 2-token logical states and $NU_2^{(m)}$ for the three reachability graphs of Fig.~\ref{fig:reachabilityHA}. }
\label{fig:HAtable}
\end{center}
\end{figure}

\begin{figure*}[ht]
\begin{center}
\includegraphics[width=2 \columnwidth]{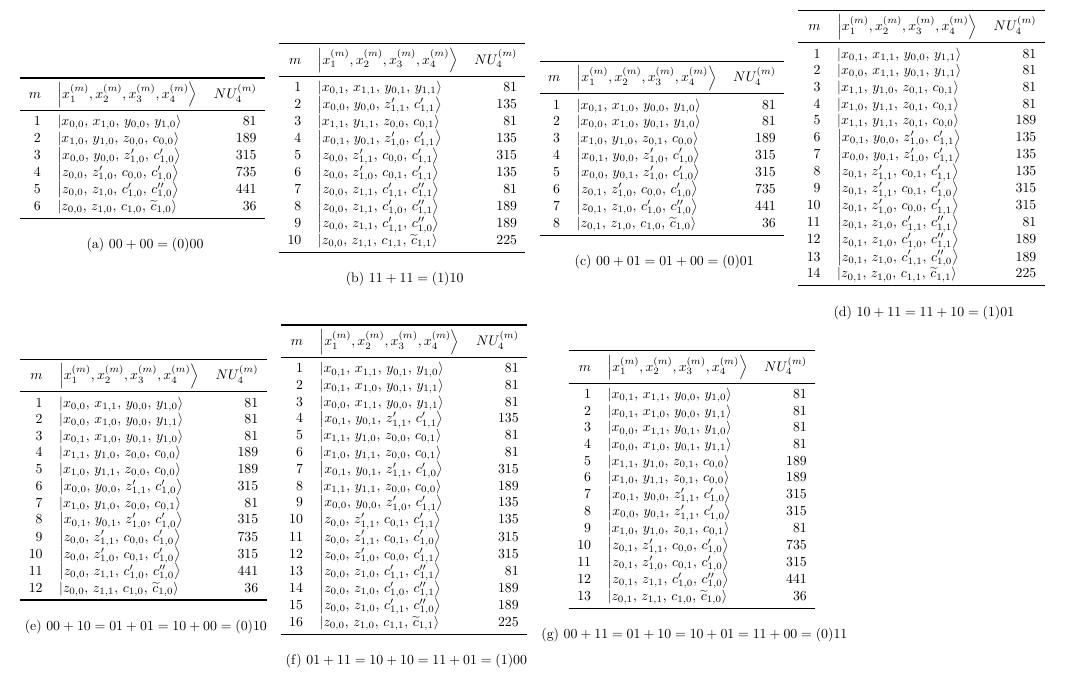}
\caption{Reachable 4-token logical states for seven reachability graphs. 
The variable $m$ denotes the serial number of the 4-token logical state in each reachability graph. 
$NU_4^{(m)}$ is the number of fine-grained states in the 4-token logical state $|m \rangle$, see Eq.~(\ref{eqn:num_l_tokenstates_of_mth_state}). 
}
\label{tab:HAFAtable}
\end{center}
\end{figure*}

Let $\Omega$ be the set of all $N$-token fine-grained states except for the final $N$-token fine-grained state $|{\rm f} \rangle$. 
Each logical state $x$, or each cell $x$, consists of $\left| P \left (x \right) \right|$ places. 
The number of fine-grained states of the last $\ell$ tokens of the $m$th $N$-token logical state (\ref{eqn:mthNtokenlogicalstate}) is 
\begin{align}
NU_{\ell}^{(m)} = \prod_{i=1}^{\ell} \left| P \left (x _{N+1-i}^{(m)} \right) \right| \, , \;\;\;\;  (\ell = 1, \cdots, N) \, . \label{eqn:num_l_tokenstates_of_mth_state}
\end{align}
In the tables in Fig.~\ref{fig:HAtable}, we summarize the numbers of 2-token fine-grained states $NU_2^{(m)}$ corresponding to the 3 reachability graphs in Fig.~\ref{fig:reachabilityHA}. 
Tables in Fig.~\ref{tab:HAFAtable} summarize the numbers of 4-token fine-grained states $NU_4^{(m)}$ corresponding to the 7 reachability graphs in Fig.~\ref{fig:reachabilityHAFA}. 

The serial number $n$ of an $N$-token fine grained state Eq.~(\ref{eqn:Ntokenstate}),
\begin{align}
|n \rangle =& \left| x_1^{(m)}, \cdots , x_N^{(m)} \right \rangle \left| \sigma_1^{(m)}, \cdots , \sigma_N^{(m)} \right \rangle \, , \label{eqn:state_serial_number} 
\end{align}
where
$\sigma_j^{(m)} = 0, \cdots, \left| P  \left(x_j^{(m)} \right) \right|-1$, 
is 
\begin{align}
n = 1+ \sum_{i=0}^{m-1} NU_N^{(i)} + \sum_{\ell=1}^{N}  \sigma_{\ell}^{(m)} NU_{\ell-1}^{(m)} \, . 
\label{eqn:serial_number}
\end{align}
Here we define $NU_N^{(0)}=0$ and $NU_0^{(m)}=1$. 
In total, there are 
\begin{align}
|\Omega|+1 = \sum_{i=1}^{M} NU_N^{(i)} \, , 
\end{align}
fine-grained $N$-token states. 
The sequence of serial numbers belonging to the $N$-token logical state $|m \rangle$ is, 
\begin{align}
E_N(m)= \left ( NU_N^{(m-1)}+1, \cdots , NU_N^{(m)} \right ) \, . \label{eqn:set_mth_logical_state}
\end{align}
We denote the serial numbers for the initial and final states as $n_{\rm i}$ and $n_{\rm f}$ with 
$| n_{\rm i} \rangle =| {\rm i} \rangle$ and $| n_{\rm f} \rangle =| {\rm f} \rangle$. 
In the basis of the states (\ref{eqn:state_serial_number}), we construct the $(|\Omega|+1) \times (|\Omega|+1)$ transition rate matrix.

\section{Transition rate matrix of safe and token conserving SPN}
\label{sec:transition_rate_matrix}

We construct the transition rate matrix starting from the reachability graph $({\mathcal R},{\mathcal E})$. 
First we extend an element $(|m' \rangle \leftarrow |m \rangle) \in {\mathcal E}$ to the following pair of triples: 
\begin{align}
& \left( \left( |m' \rangle, (u^{(m')},r^{(m')}), (v^{(m')},s^{(m')}) \right) 
\right.   
\nonumber \\ & 
\left. 
\leftarrow \left( |m \rangle, (u^{(m)},r^{(m)}), (v^{(m)},s^{(m)}) \right) \right) \, . \label{eqn:tran_rule_element}
\end{align}
Each triple consists of $N$-token logical states $|m \rangle (|m' \rangle)$, two pairs of cell indices and place indices, $(u^{(m)},r^{(m)})$ and $(v^{(m)},s^{(m)})$ ( $(u^{(m')},r^{(m')})$ and $(v^{(m')},s^{(m')})$ ), where two tokens are annihilated (created). 
The element (\ref{eqn:tran_rule_element}) specifies a forward transition from, 
\begin{align}
&|\cdots, x_k^{(m)}=u^{(m)}, \cdots, x_\ell^{(m)}=v^{(m)}, \cdots \rangle \nonumber \\
&\times |\cdots, \sigma_k^{(m)}=r^{(m)}, \cdots, \sigma_\ell^{(m)}=s^{(m)}, \cdots \rangle \, ,
\end{align}
to, 
\begin{align}
&|\cdots, x_{k'}^{(m')}=u^{(m')}, \cdots, x_{\ell'}^{(m')}=v^{(m')}, \cdots \rangle \nonumber \\
&\times |\cdots, \sigma_{k'}^{(m')}=r^{(m')}, \cdots, \sigma_{\ell'}^{(m')}=s^{(m')}, \cdots \rangle \, ,
\end{align}
by annihilating $k$th and $\ell$th tokens of the $N$-token logical state $|m \rangle$ and creating $k'$th and $\ell'$th tokens of the $N$-token logical state $|m' \rangle$ . 
The other $N-2$ tokens stay at the same places as, 
\begin{widetext}
\begin{align}
\left( {\nu_1^{(m)}}, \cdots, {\nu_{k-1}^{(m)}}, {\nu_{k+1}^{(m)}}, \cdots, {\nu_{\ell-1}^{(m)}}, {\nu_{\ell+1}^{(m)}}, \cdots, {\nu_N^{(m)}} \right)
=&
\left( {\nu_1^{(m')}}, \cdots, {\nu_{k'-1}^{(m')}}, {\nu_{k'+1}^{(m')}}, \cdots, {\nu_{\ell'-1}^{(m')}}, {\nu_{\ell'+1}^{(m')}}, \cdots, {\nu_N^{(m')}} \right) \, ,  \label{eqn:sum_cond}
\end{align}
\end{widetext}
where $\nu_j^{m}= \left( x_j^{(m)},\sigma_j^{(m)} \right)$. 
Therefore, the element (\ref{eqn:tran_rule_element}) specifies, 
\begin{align*}
\prod_{j (\neq k,\ell)} \left| P \left( x_j^{(m)} \right) \right| = \prod_{j (\neq k',\ell')} \left|P \left( x_j^{(m')} \right) \right| \, 
\end{align*}
matrix elements. 

Matrix elements of the transition rate matrix ${\bm L}^{\rm f}$ are arranged according to the serial number (\ref{eqn:serial_number}). 
The transition rate matrix is divided into $M$ sub-matrices, with $NU_{N}^{(m)}$ components ($m=1,\cdots, M$). 
The block-diagonal sub-matrix with rows and columns $E_N(m)$, Eq.~(\ref{eqn:set_mth_logical_state}), describes transitions within the $m$th $N$-token logical states, 
\begin{align}
{{\bm L}^{\rm f}}^{E_N(m)}_{E_N(m)}= {\bm H}^{(m)} - { {\bm \gamma}^{\rm f} }^{(m)} \, . 
\label{eqn:block_diagonal}
\end{align}
The first term of the right-hand side of the equation is, 
\begin{align}
{\bm H}^{(m)}=& \sum_{\ell=1}^{N} {\bm I}_{ NU_{1,\ell-1}^{(m)}} \otimes {\bm H} \left( x_\ell^{(m)} \right) \otimes {\bm I}_{ NU_{\ell+1,N}^{(m)} } \, , 
\label{eqn:tra_rat_blo_dia_dec}
\\
{\bm I}_{ NU_{\ell,\ell'}^{(m)} } =& \left \{ \begin{array}{cc} \bigotimes_{j=\ell}^{\ell'} {\bm I}_{NU \left( x_j^{(m)} \right) } & (\ell ' \geq \ell) \\ 1 & (\ell'=\ell-1) \end{array} \right. \, , 
\end{align}
where, ${\bm I}_n$ is a $n \times n$ unit matrix for $n \in {\mathbb Z}_+ $. 
Here ${\bm H} ( x )$ means the matrix representation of the Doi-Peliti Hamiltonian. 
For Eq.~(\ref{eqn:Ham_nhub}), 
\begin{align*}
{\bm H}_{n} =& {\bm H}_{n-1} \oplus {\bm 0}_2 + {\bm 0}_{2n-2} \oplus {\bm H}_1
\, , \\
{\bm H}_1 =& \Gamma \left( \begin{array}{ccc} -2 &  1 &  1 \\  1 & -2 &  1 \\  1 &  1 & -2 \end{array} \right) \, ,
\end{align*}
where $\oplus$ stands for the matrix direct sum and ${\bm 0}_n$ is a $n \times n$ zero matrix. 
For Eqs.~(\ref{eqn:Ham_0hub}), (\ref{eqn:Ham_Ratchet}) and (\ref{eqn:terminal_Ratchet}), 
\begin{align*}
{\bm H}_0 = \Gamma \left( \begin{array}{cc} -1 &  1 \\ 1 & -1 \end{array} \right) \, , \;\;\;\;
{\bm H}_0^{\rm R} = \Gamma \left( \begin{array}{cc} -1 & 0 \\ 1 & 0 \end{array} \right) \, , \\
{\bm H}_n^{\rm R}={\bm H}_n- {\bm 0}_{2 (n-1)} \oplus \Gamma (1,1,-2)^T (0,0,1)
\, .
\end{align*}

A block off-diagonal component is associated with the transitions induced by a CJoin. 
Then, sub-matrices indicating forward and backward transitions are generated from the element (\ref{eqn:tran_rule_element}). 
The forward transition is described by the $NU_N^{(m')} \times NU_N^{(m)}$ sub-matrix with rows $E_N(m')$ and columns $E_N(m)$: 
\begin{align}
{{\bm L}^{\rm f}}^{E_N(m')}_{E_N(m)} =& \gamma^+ \sum_{  {\rm Eq. \, (\ref{eqn:sum_cond}) } } {\bm u}^{(m')} { {\bm u}^{(m)} }^T
\, , \label{eqn:tra_rat_blo_off_f}
\end{align}
where ${\bm u}^{(m)}$ and ${\bm u}^{(m')}$ are $NU_N^{(m)}$ and $NU_N^{(m')}$ dimensional unit vectors, respectively. 
They are given by, 
\begin{widetext}
\begin{align*}
{\bm u}^{(m)} &= {\bm u}(x_1^{(m)})_{\sigma_1^{(m)}} \otimes \cdots \otimes 
{\bm u}(x_{k}^{(m)}=u^{(m)})_{\sigma_{k}^{(m)}=r^{(m)} } \otimes \cdots \otimes {\bm u}(x_{\ell}^{(m)}=v^{(m)})_{\sigma_{\ell}^{(m)}=s^{(m)} } \otimes \cdots \otimes {\bm u}(x_{N}^{(m)})_{\sigma_{N}^{(m)} } \, , 
\\
{\bm u}^{(m')} &= {\bm u}(x_1^{(m')})_{\sigma_1^{(m')}} \otimes \cdots \otimes 
{\bm u}(x_{k'}^{(m')}=u^{(m')})_{\sigma_{k'}^{(m')}=r^{(m')} } \otimes \cdots \otimes {\bm u}(x_{\ell'}^{(m')}=v^{(m')})_{\sigma_{\ell'}^{(m')}=s^{(m')} } \otimes \cdots \otimes {\bm u}(x_{N}^{(m')})_{\sigma_{N}^{(m')} } \, , 
\end{align*}
\end{widetext}
where ${\bm u}(x)_\sigma$ is a $|P(x)|$ dimensional vertical unit vector with 1 for its the $\sigma+1$st component. 
The summation is performed over place indexes of the other $N-2$ tokens, 
${\sigma_1^{(m)}}, \cdots, {\sigma_{k-1}^{(m)}}, {\sigma_{k+1}^{(m)}}, \cdots, {\sigma_{\ell-1}^{(m)}}, {\sigma_{\ell+1}^{(m)}}, \cdots, {\sigma_N^{(m)}}$ satisfying the condition Eq.~(\ref{eqn:sum_cond}).  
Similarly, the backward transition is described by the $NU_N^{(m)} \times NU_N^{(m')}$ sub-matrix with rows $E_N(m)$ and columns $E_N(m')$: 
\begin{align}
{{\bm L}^{\rm f}}^{E_N(m)}_{E_N(m')} = \gamma^-  \sum_{  {\rm Eq. \, (\ref{eqn:sum_cond}) } } {\bm u}^{(m)} { {\bm u}^{(m')} }^T \, . \label{eqn:tra_rat_blo_off_b}
\end{align}

The second term of the right-hand side of Eq.~(\ref{eqn:block_diagonal}) is a diagonal matrix related to the decay caused by the transition between different $N$-token logical states induced by the CJoin, 
\begin{align}
{{\bm \gamma}^{\rm f}}^{(m)}=& \sum_{m' (\neq m)}{\rm diag} \left( { {\bm e}_{NU_N^{(m')}} }^T {{\bm L}^{\rm f}}^{E_N(m')}_{E_N(m)} \right) \, . 
\end{align}

The matrix constructed in this way is sparse. 
Our program was written in Python.

\section{Initial and final states for $f_{\rm HA+FA}$}
\label{sec:ifms}

Table~\ref{tab:hafa_initial_final}  summarizes the initial and final 4-token logical state for 16 computation processes of $f_{\rm HA+FA}$. 
$|n_{\rm i} \rangle$ and $|n_{\rm f} \rangle$ are the initial and final states represented in the serial number (\ref{eqn:serial_number}). 

\begin{center}
\begin{table*}
\begin{tabular}{c c c l} 
  & $|n_{\rm i} \rangle $ & $ |n_{\rm f} \rangle$ & \\ \cmidrule{1-3}
$00+00=(0)00$  & $|1 \rangle = |x_{0,0},x_{1,0},y_{0,0},y_{1,0} \rangle |0,0,0,0 \rangle$ & $|1797 \rangle = |z_{0,0}, z_{1,0}, c_{1,0},\tilde{c}_{1,0} \rangle |2,2,1,1 \rangle$ & (a) \\ 
\rowcolor{gray!30}
$11+11=(1)10$  & $|1 \rangle = |x_{0,1},x_{1,1},y_{0,1},y_{1,1} \rangle |0,0,0,0 \rangle$ & $|1566 \rangle = |z_{0,0}, z_{1,1}, c_{1,1},\tilde{c}_{1,1} \rangle |2,2,4,4 \rangle$ & (b) \\ 
$00+01=(0)01$  & $|82 \rangle = |x_{0,0},x_{1,0},y_{0,1},y_{1,0} \rangle |0,0,0,0 \rangle$ & \multirow{2}{*}{$|2193 \rangle =|z_{0,1}, z_{1,0}, c_{1,0},\tilde{c}_{1,0} \rangle |2,2,1,1 \rangle$} & \multirow{2}{*}{(c)} \\ 
$01+00=(0)01$  & $|1 \rangle = |x_{0,1},x_{1,0},y_{0,0},y_{1,0} \rangle |0,0,0,0 \rangle$ & & \\ 
\rowcolor{gray!30}
$10+11=(1)01$  & $|82 \rangle =|x_{0,0},x_{1,1},y_{0,1},y_{1,1} \rangle |0,0,0,0 \rangle$ & & \\ 
\rowcolor{gray!30}
$11+10=(1)01$  & $|1 \rangle =|x_{0,1},x_{1,1},y_{0,0},y_{1,1} \rangle |0,0,0,0 \rangle$ & \multirow{-2}{*}{$|2232 \rangle =|z_{0,1}, z_{1,0}, c_{1,1},\tilde{c}_{1,1} \rangle |2,2,4,4 \rangle$} & \multirow{-2}{*}{(d)} \\ 
$00+10=(0)10$  & $|82 \rangle =|x_{0,0},x_{1,0},y_{0,0},y_{1,1} \rangle |0,0,0,0 \rangle$ & \multirow{3}{*}{$|2859 \rangle =|z_{0,0}, z_{1,1}, c_{1,0},\tilde{c}_{1,0} \rangle |2,2,1,1 \rangle$} & \multirow{3}{*}{(e)} \\ 
$01+01=(0)10$  & $|163 \rangle =|x_{0,1},x_{1,0},y_{0,1},y_{1,0} \rangle |0,0,0,0 \rangle$ & & \\ 
$10+00=(0)10$  & $|1 \rangle =|x_{0,0},x_{1,1},y_{0,0},y_{1,0} \rangle |0,0,0,0 \rangle$ & & \\ 
\rowcolor{gray!30}
$01+11=(1)00$  & $|82 \rangle =|x_{0,1},x_{1,0},y_{0,1},y_{1,1} \rangle |0,0,0,0 \rangle$ & &  \\ 
\rowcolor{gray!30}
$10+10=(1)00$  & $|163 \rangle =|x_{0,0},x_{1,1},y_{0,0},y_{1,1} \rangle |0,0,0,0 \rangle$ & & \\ 
\rowcolor{gray!30}
$11+01=(1)00$  & $|1 \rangle =|x_{0,1},x_{1,1},y_{0,1},y_{1,0} \rangle |0,0,0,0 \rangle$ & \multirow{-3}{*}{$|2628 \rangle =|z_{0,0}, z_{1,0}, c_{1,1},\tilde{c}_{1,1} \rangle |2,2,4,4 \rangle$} & \multirow{-3}{*}{(f)} \\ 
$00+11=(0)11$  & $|244 \rangle =|x_{0,0},x_{1,0},y_{0,1},y_{1,1} \rangle |0,0,0,0 \rangle$ & \multirow{4}{*}{$|2940 \rangle =|z_{0,1}, z_{1,1}, c_{1,0},\tilde{c}_{1,0} \rangle |2,2,1,1 \rangle$} & \multirow{4}{*}{(g)} \\ 
$01+10=(0)11$  & $|82 \rangle =|x_{0,1},x_{1,0},y_{0,0},y_{1,1} \rangle |0,0,0,0 \rangle$ & & \\ 
$10+01=(0)11$  & $|163 \rangle =|x_{0,0},x_{1,1},y_{0,1},y_{1,0} \rangle |0,0,0,0 \rangle$ & & \\ 
$11+00=(0)11$  & $|1 \rangle =|x_{0,1},x_{1,1},y_{0,0},y_{1,0} \rangle |0,0,0,0 \rangle$ & & \\ 
\end{tabular}
\caption{Initial and final 4-token fine-grained states of 16 computation processes of $f_{\rm HA+FA}$. }
\label{tab:hafa_initial_final}
\end{table*}
\end{center}

\section{Coarse-grained master equation}
\label{app:Time_scale_separation}

For $\gamma^\pm \ll \Gamma=1$, we separate the transition rate matrix into two parts ${\bm L}^{\rm f}= {\bm L}^{\rm f}_0 + {\bm L}^{\rm f}_1$, 
where ${\bm L}^{\rm f}_0$ and ${\bm L}^{\rm f}_1$ are proportional to $\Gamma$ and $\gamma^\pm$, respectively. 
In this section, we omit the superscript ${\rm f}$. 
The block $(m',m)$ component of the unperturbed and perturbative parts are $\left( {\bm L}_0 \right)_{m',m} = {\bm H}^{(m)} \delta_{m',m}$ and, 
\begin{align}
\left( {\bm L}_1 \right)_{m',m} =& \left \{ \begin{array}{cc} {\bm L}^{E_N(m')}_{E_N(m)} & (m' \neq m) \\ - {{\bm \gamma}}^{(m)} & (m'=m) \end{array} \right. \, .
\end{align}
The steady-state solution of the block $(m,m)$ component of the unperturbed part, 
\begin{align}
{\bm H}^{(m)} {{\bm p}^{\rm st}}^{(m)}  =0 \, , 
\end{align}
is written with the Kronecker product of equilibrium distributions of $N$ logical states as, 
\begin{align}
{{\bm p}^{\rm st}}^{(m)} = {\bm p}^{\rm st} \left( x^{(m)}_1 \right) \otimes \cdots \otimes {{\bm p}^{\rm st}} \left( x^{(m)}_N \right) \, ,
\end{align}
where, 
${{\bm p}^{\rm st}}(x) = {\bm e}_{\left| P(x) \right|} / \left| P(x) \right|$. 
Then we introduce the coarse-grained Green function, 
\begin{align}
G(m'|m;s) =& \left( {\bm u}_{m'} \otimes {\bm e}_{ NU_N^{(m')} } \right)^T
\left( s {\bm I}_{|\Omega|+1} - {\bm L} \right)^{-1} \nonumber \\ 
&\times \left( {\bm u}_m \otimes {{\bm p}^{\rm st}}^{(m)} \right) \, ,
\end{align}
where $({\bm u}_m)_\ell = \delta_{m,\ell} $ is an $M$-component unit vector. 
We approximate the block diagonal component of the unperturbed resolvent as, 
\begin{align}
\left( s {\bm I}_{NU_N^{(m)}} +{\bm H}^{(m)} \right)^{-1} \approx \frac{ {{\bm p}^{\rm st}}^{(m)} {{\bm e}_{NU_N^{(m)}}}^T }{s} \, , 
\end{align}
and account for only the zero eigenstate. 
Then the coarse-grained Green function is approximately 
\begin{align}
G(m'|m;s) \approx& \left( s {\bm I}_M - \tilde{ {\bm L} } \right)^{-1} \, , 
\\
\tilde{L}_{m',m} =& {{\bm e}_{NU_N^{(m')}}}^T \left( {\bm L}_1 \right)_{m',m} {{\bm p}^{\rm st}}^{(m)} \, .
\end{align}
By substituting the block off-diagonal component ${\bm L}^{E_N(m')}_{E_N(m)}$ given by Eqs.~(\ref{eqn:tra_rat_blo_off_f}) and (\ref{eqn:tra_rat_blo_off_b}) into $\left( {\bm L}_1 \right)_{m',m}$, we derive the coarse-grained forward and backward transition rates, Eq.~(\ref{eqn:coarse_tra_rat}). 
The diagonal component $m'=m$ is obtained as, 
\begin{align}
\tilde{ L }_{m,m} = - {{\bm e}_{NU_N^{(m)}}}^T {\bm \gamma}^{(m)} {{\bm p}^{\rm st}}^{(m)} = - \sum_{m' (\neq m)} \tilde{ L }_{m',m} \, .
\end{align}

In the following, we consider the $00+00$ process with $\gamma^-=0$. 
The transition state diagram is deduced from the reachability graph in Fig.~\ref{fig:reachabilityHAFA} (a). 
Non-vanishing matrix elements are, 
\begin{align*}
\tilde{L}_{2,1} & = \tilde{L}_{4,3} = \gamma^+/D_{1}  \, ,
\\
\tilde{L}_{3,1} & = \tilde{L}_{4,2} = \gamma^+/D_{2} \, ,
\\
\tilde{L}_{5,4} &= \gamma^+/D_{3} \, ,
\\
\tilde{L}_{6,5} &= \gamma^+/D_{4} \, ,
\end{align*}
where $D_1$, $D_2$, $D_3$ and $D_4$ are given by Eqs.~(\ref{eqn:D1_D2}), (\ref{eqn:D3}) and (\ref{eqn:D4}). 
The characteristic function of the first-passage time distribution from $| 1 \rangle = | m_{\rm i} \rangle$ to $| 6 \rangle = | m_{\rm f} \rangle $ is
\begin{align}
F(m_{\rm f}|m_{\rm i};s) =& \frac{G(m_{\rm f}|m_{\rm i};s)}{G(m_{\rm f}|m_{\rm i};s)} \nonumber \\
=& \frac{2 s+\gamma^+ (D_{1}+D_{2})}{s+\gamma^+ (D_{1}+D_{2})} \prod_{j=1}^{4} \frac{\gamma^+ D_{j}}{s+\gamma^+ D_{j}}  \, . \label{eqn:F4step}
\end{align}
From Eq.~(\ref{eqn:F4step}), the average and variance, Eqs.~(\ref{eqn:cg_average}) and (\ref{eqn:cg_noise}), are obtained. 
We can repeat similar calculations for the other computation processes. 
The results are summarized in Table~\ref{tab:n3_n4_SN}.

\begin{table}[htb]
\begin{tabular}{c  c c c c c l} 
 & $D_3$ & $D_4$ & $\gamma^+ \langle \! \langle \tau \rangle \! \rangle_F$ & $\gamma^+ \sqrt{ \langle \! \langle \tau^2 \rangle \! \rangle_F }$ & $S/N$ & \\ 
\cmidrule{1-6}
$00+00$ & 35 & 49 & 97.5 & 61.2 & 1.59 & (a) \\
\rowcolor{gray!30}
$11+11$  & 15 & 21 & 49.5 & 28.1 & 1.76 & (b)\\
$01+00$  & 35 & 49 & 97.5 & 61.2 & 1.59 & \\
$00+01$  & 35 & 49 & 97.5 & 61.2 & 1.59 & \multirow{-2}{*}{(c)} \\
\rowcolor{gray!30}
$11+10$  & 35 & 21 & 69.5 & 42.3 & 1.64 & \\
\rowcolor{gray!30}
$10+11$  & 35 & 21 & 69.5 & 42.3 & 1.64 &  \multirow{-2}{*}{(d)} \\
$10+00$  & 35 & 49 & 97.5 & 61.2 & 1.59 & \\
$01+01$  & 15 & 49 & 77.5 & 52.4 & 1.48 & \\
$00+10$  & 35 & 49 & 97.5 & 61.2 & 1.59 &  \multirow{-3}{*}{(e)} \\
\rowcolor{gray!30}
$11+01$  & 15 & 21 & 49.5 & 28.1 & 1.76 & \\
\rowcolor{gray!30}
$10+10$  & 35 & 21 & 69.5 & 42.3 & 1.64 & \\
\rowcolor{gray!30}
$01+11$  & 15 & 21 & 49.5 & 28.1 & 1.76 & \multirow{-3}{*}{(f)} \\
$11+00$  & 35 & 49 & 97.5 & 61.2 & 1.59 & \\
$01+10$  & 35 & 49 & 97.5 & 61.2 & 1.59 & \\
$10+01$  & 35 & 49 & 97.5 & 61.2 & 1.59 & \\
$00+11$  & 35 & 49 & 97.5 & 61.2 & 1.59 & \multirow{-3}{*}{(g)} \\
\end{tabular}
\caption{Numbers of available states, $D_3$ and $D_4$, average, variance and signal-to-noise ratio. 
The other numbers are $D_1=D_2=9$. 
}
\label{tab:n3_n4_SN}
\end{table}

\section{Supplemental discussions on Sec.~\ref{sec:wdio}}
\label{sec:sup_wdio}

To derive Eq.~(\ref{eqn:a_uniform_dist_coagra}), we substitute the uniform distribution $p^{\rm st}_j \approx 1/|\Omega|$ into Eq.~(\ref{eqn:act_rat}): 
\begin{align}
\langle \! \langle a \rangle \! \rangle =& \frac{1}{2} \sum_{(i \leftarrow j) \in E_{\rm bi}} \Gamma_{i,j} p_{j}^{\rm st} + \sum_{(i \leftarrow j) \in E_{\rm uni}} \Gamma_{i,j} p_{j}^{\rm st} \nonumber \\
 \approx& \frac{1}{2} \sum_{(i \leftarrow j) \in E_{\rm bi}} \frac{ \Gamma_{i,j}^{\rm f} }{ |\Omega|+1 } = \Gamma  \frac{ \sum_{m} N_{m,m} }{ |\Omega|+1 } 
\nonumber \\ &+ \gamma^+ \frac{ \sum_{m' > m }  N_{m',m} }{ |\Omega|+1 } + \gamma^- \frac{\sum_{m' < m } N_{m',m}}{ |\Omega|+1 } \, , \label{eqn:a_uniform_dist}
\end{align}
where $N_{m,m}$ is the number of positive matrix elements in ${{\bm L}^{\rm f}}^{E_N^{(m')}}_{E_N^{(m)}}$, see Eqs.~(\ref{eqn:tra_rat_blo_off_f}) and (\ref{eqn:tra_rat_blo_off_b}). 
By using 
$N_{m',m} = NU_N^{(m)}/ \left( \left| P \left( u^{(m)} \right) \right| \left| P \left(v^{(m)} \right) \right| \right)$ ($m \neq m'$) and 
$N_{m,m} = \sum_{\ell=1}^N \left| T \left( x^{(m)}_\ell \right) \right| NU_N^{(m)}/\left| P \left(x^{(m)}_\ell \right) \right|$, 
we obtain Eq.~(\ref{eqn:a_uniform_dist_coagra}). 

\begin{figure}[ht]
\begin{center}
\includegraphics[width=0.8 \columnwidth]{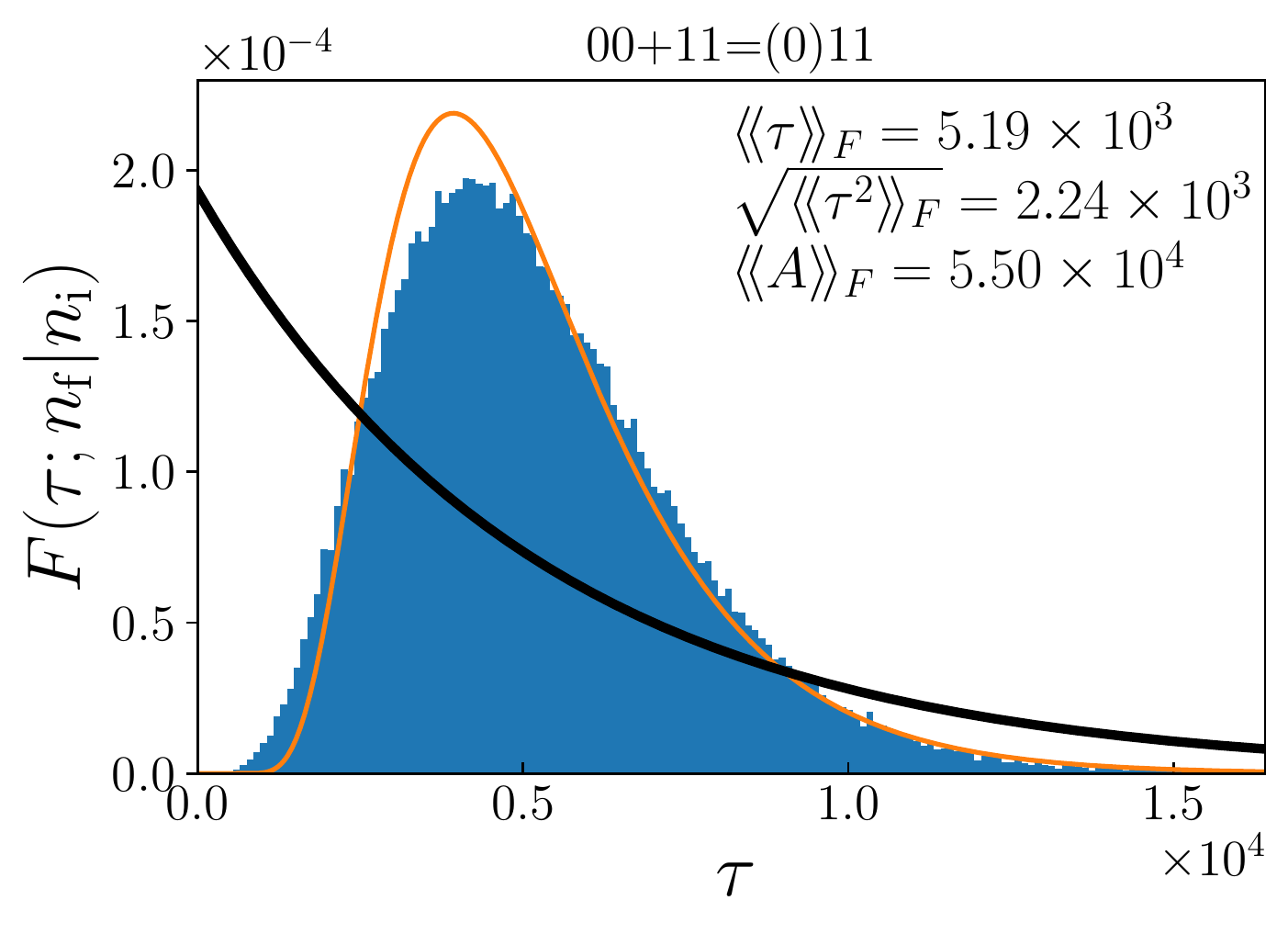}%
\caption{
Probability distributions of computation time for $00+11=(0)11$ with $\gamma^\pm=10$ and $W=5$. 
The inverse Gaussian distribution (\ref{eqn:inverse_gauss_tau}) closely approximates the histogram. 
The thick solid line indicates the exponential distribution in Eq.~(\ref{eqn:exp_dist}). 
}
\label{fig:lis_dat_HAFA_NONlp5_00p11f10b10_fptd}
\end{center}
\end{figure}

Figure \ref{fig:lis_dat_HAFA_NONlp5_00p11f10b10_fptd} shows the computation time distribution of $00+11=(0)11$ with $W=5$. 
It is well-approximated by the inverse Gaussian distribution (solid line). 
The exponential distribution (\ref{eqn:exp_dist}) deviates from the histogram (thick solid line). 

\begin{figure}[ht]
\begin{center}
\includegraphics[width=0.8 \columnwidth]{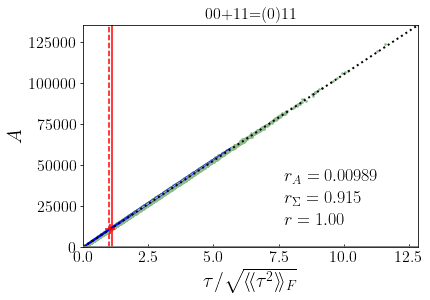}
\caption{Scatter plot of computation time and activity without Ratchets for $00+11=(0)11$ and $\gamma^\pm=10$. 
The dotted line indicates $\langle \! \langle a \rangle \! \rangle \tau$, 
with $\langle \! \langle a \rangle \! \rangle$ calculated from Eq.~(\ref{eqn:a_uniform_dist})
}
\label{fig:lis_dat_HAFA_00p11f10b10_tau_a}
\end{center}
\end{figure}

Figure~\ref{fig:lis_dat_HAFA_00p11f10b10_tau_a} shows the scatter plot of the computation time and the activity for $00+11=(0)11$. 
The linear correlation is observed to be similar to the RTM case [Eq.~(\ref{eqn:cor_coe_seminfTM})]. 
The dotted line indicates $\langle \! \langle a \rangle \! \rangle \tau$, 
with $\langle \! \langle a \rangle \! \rangle$ calculated from Eq.~(\ref{eqn:a_uniform_dist}) by setting $ |\Omega|+1 = 2940$, $ \sum_{m} N_{m,m} = 26214$ and $ \sum_{m' \neq m }  N_{m',m} = 488$. 
It closely fits the scatter plot. 

\clearpage

\end{appendix}


\begin{thebibliography}{99}
\bibitem{set_theory_1}
The number of elements of the set $\Omega$ is denoted as $|\Omega|$. 
%
The fibre of an element $y \in Y$ of the map $f:X \to Y$ is, 
\begin{align*}
f^{-1}(y) = \left \{ x \in X \middle | f(x)=y  \right \}  \, .
\end{align*}
The preimage of a set $A \subseteq Y$ is 
\begin{align*}
f^{-1}(A) 
= \left \{ x \in X \middle | f(x) \in A  \right \}  
= \bigcup_{y \in A} f^{-1}(y)  \, .
\end{align*}
\end{thebibliography}
\end{document}